\documentclass{article}

\usepackage{PRIMEarxiv}

\usepackage[utf8]{inputenc}
\usepackage[T1]{fontenc}
\PassOptionsToPackage{hyphens}{url}
\usepackage{hyperref}
\usepackage{url}
\usepackage{booktabs}
\usepackage{amsfonts}
\usepackage{amsmath}
\usepackage{nicefrac}
\usepackage{microtype}
\usepackage[numbers,sort&compress]{natbib}
\usepackage{fancyhdr}
\usepackage{graphicx}
\usepackage{pgfplots}
\pgfplotsset{compat=1.18}
\usetikzlibrary{arrows.meta}
\usepackage{array}
\usepackage{siunitx}
\usepackage{multirow}
\usepackage{xcolor}
\usepackage{fvextra}
\graphicspath{{media/}}

\title{Recipe-Matching, Not Equivalence}

\author{%
  Ali Habibullah\footnotemark[2]\ , Mohammad Alshiekh\footnotemark[1] \ \& Yazan Alshoibi\footnotemark[1] \\
  KAUST Academy \\
  Computer, Electrical \& Mathematical Sciences \& Engineering (CEMSE) \\
  King Abdullah University of Science and Technology (KAUST) \\
  Thuwal, Saudi Arabia \\
  \texttt{\{ali.habibullah, mohammad.shiekh, yazen.shaebi\}@kaust.edu.sa} \\
  \AND
  Salman Khan \\
  Visual Artificial Intelligence Laboratory \\
  Oxford Brookes University \\
  \texttt{salmankhan@brookes.ac.uk} \\
  \And
  Naeemullah Khan \\
  KAUST Academy \& CEMSE, KAUST \\
  Lady Margaret Hall, University of Oxford \\
  \texttt{naeemullah.khan@kaust.edu.sa}
}

\newcommand{\repourl}{\url{https://github.com/KAUST-Academy/recipe-matching-not-equivalence}}
\newcommand{\artifactsnote}{\footnote{Code, data, results and trained models: \repourl.}}

\begin{document}
\maketitle
{\renewcommand{\thefootnote}{\fnsymbol{footnote}}%
 \footnotetext[1]{Equal contribution.}%
 \footnotetext[2]{Corresponding author.}}
\begin{abstract}
MathNet-Retrieve tests whether a retriever can find, for a math problem, a document that states the same problem. It builds those documents itself: an LLM under one fixed prompt writes each gold document and its near-miss distractors, and LLM judges filter the result. We call that procedure the benchmark's \emph{recipe}, training on pairs built the same way \emph{recipe-matching}, and ask how much score it buys beyond the ability the benchmark claims to test. Two models trained from the same base on the same number of rows under the same settings differ only in the training file: pairs written with the benchmark's published prompt by a different vendor's LLM and judge, or pairs a computer-algebra system verified with no LLM anywhere. The first beats the second by $45$ R@1 points on the easy tier. As far as a non-LLM paraphrase control can tell, half to two thirds of that gap comes from the training pairs being LLM-written at all: LLM rewrites under two prompts unrelated to the benchmark, with the verified model's negatives, recover $30$ and $22$ of the $45$ points, while back-translated paraphrases with the same negatives recover almost none. The remaining $15$ to $25$ points, depending on the comparison, appear only under the benchmark's own prompt, and they vanish on real duplicates no generator wrote, the same problem printed in two languages. The hard tier rewards the recipe's pair structure, a deep rewrite against a minimal-edit near-miss: LLM rewrites alone score zero on it, attaching negatives unlocks it, and every kind of negative that does so costs cross-language points against the same rewrites trained alone; the training sets that score highest on it discriminate near-misses no LLM wrote worse than plain LLM rewrites with verified negatives. MELD also moves when a model trains on pairs built its way, without losing retention; on SABER-Math the registered attack fails, and the one gain, from training on its LLM-written summaries, is small but holds at a matched budget; only on MathNet-Retrieve could we pin an inversion, benchmark score up and real retention down, to one edit of a training file. We release the generator-free duplicate evaluations, the near-miss test and three trained models\artifactsnote.
\end{abstract}


\section{Introduction}
\label{sec:intro}

Finding a math problem that is the same problem written differently matters for spotting duplicates and for handing solved problems to a reasoning model. Many benchmarks now test retrieval over math and other reasoning problems~\citep{su2025bright,alshammari2026mathnet,georgiev2026saber,ye2026meld}, and more of them are built by an LLM: it writes the queries or the gold documents and decides what counts as relevant. When every gold document comes out of one fixed LLM procedure, a model can score well by learning what that procedure produces instead of the skill the benchmark says it tests. We call such gains \emph{recipe-matching}. Unlike an ordinary shortcut, a quirk of one dataset, this one is built into how the benchmark is made, so it carries over when a different LLM runs the same procedure and needs no access to the exact prompt (Section~\ref{sec:related}).

MathNet-Retrieve~\citep{alshammari2026mathnet} is a good place to study this. Its three tiers use the same corpus and the same queries, and differ only in how heavily the gold document is disguised. Every gold is a Gemini-3-flash paraphrase of its query, together with three near-misses, small edits of the query that change the problem, and it is kept only when Gemini-3-flash and GPT-5 both judge it equivalent. The hard tier is presented as deep mathematical equivalence, yet at rank~$1$ nobody solves it: every model in the authors' evaluation scores almost zero R@1, and no public model we tested exceeds $0.01$. That floor is built into the tier's design rather than a sign of a hard skill: the near-misses are tiny edits of the query, so any embedder that goes by surface wording ranks them above the heavily disguised gold, and for existing models R@5 and R@10 say more (Appendix~\ref{app:results}).

To find out what that tier rewards, we ran a controlled experiment on Qwen3-Embedding-0.6B~\citep{qwen3embedding}: two training sets of exactly $6{,}145$ rows built from the same source problems and trained under the same settings, so the only difference is the training file. In the \emph{verified arm}, a computer algebra system proves each positive equivalent to its source and refutes each minimal-edit negative with a counterexample. In the \emph{recipe arm}, the pairs are generated with the benchmark's own Appendix-F prompt and kept or dropped by an LLM judge alone, which is how the benchmark itself was built, except that our generator is Qwen3-32B rather than Gemini-3-flash and one judge decides where the benchmark requires two to agree. Both arms carry the same item-level contamination~\citep{dekoninck2024constat,yao2024crosslingual}, about a third of their rows anchoring on a benchmark query; it accounts for about one point of the gap, which survives two corrected-gate retrainings (Section~\ref{sec:setup}, Appendix~\ref{app:contamination}).

The recipe arm beats the verified arm on every tier (Table~\ref{tab:controlled}): over eight seeds it leads by $45.33$ easy-tier R@1 points and reaches $9.42$ on the hard tier, where the verified arm sits at zero. A third model, the \emph{recipe-free reference}, trains on LLM rewrites from a prompt unrelated to the benchmark plus the verified arm's negatives; it recovers $29.85$ of the $45.33$ points and matches the recipe arm on the hard tier, a second unrelated prompt recovers $22.15$, and back-translated paraphrases, which no LLM wrote, recover only $1.76$ with the same negatives. Half to two thirds of the gap is therefore bought by LLM-written pairs of any kind, and $15.47$ to $24.59$ points, depending on the comparison, need the benchmark's own prompt (Section~\ref{sec:gaming}). A ladder of rewriter prompts and a factorial over positives and negatives (Section~\ref{sec:mechanism}) show that the easy tier rewards LLM-written text in general, the \emph{genre} of LLM rewrites, whichever prompt wrote them. The hard tier rewards the \emph{pair structure} the recipe produces, a deep rewrite set against a minimal-edit near-miss: rewrites alone or near-misses alone score nothing, and restatements from an unrelated prompt set against the recipe's own near-misses reach $26.97$ over three seeds, nearly three times the recipe arm.

To see whether the lead means anything real, we test on duplicates no LLM wrote: the same problem printed in two languages, and the same problem reprinted in one. On the cross-language duplicates the recipe arm's easy-tier lead over the verified arm shrinks from $45.33$ to under ten points; the drop from benchmark to real data is $35.59$ points, and $27.24$ of it is recipe-specific, since the reference, with no recipe wording, loses far less. On same-language reprints the recipe arm holds no measurable lead over the verified arm and is within noise of the reference (Section~\ref{sec:ood}).

We tried the same attack on two other benchmarks. MELD~\citep{ye2026meld} never published the prompt behind its evaluation set; we reconstructed the procedure from its description and ran it through a different vendor's LLM, and training a $0.6$B model on the result more than triples its R@1 on MELD's pairs. That model also improves, rather than worsens, on cross-language duplicates, so most of its gain is the task itself being learned. SABER-Math~\citep{georgiev2026saber} has organic documents, the attack we registered against them fails, and the only movement is a small gain, robust to a matched budget, from training on the LLM-written summaries its pipeline mines pairs from (Section~\ref{sec:ood}).

The model trained on rewrites from an unrelated prompt with no negatives is at the top on cross-language duplicates, above the untrained base, yet scores zero on the hard tier; one edit to its training, attaching verified negatives, lifts it off that floor and costs it $15.79$ cross-language points. We release the evaluations, near-miss tests, three trained models and recommendations to benchmark builders.

\section{Related Work}
\label{sec:related}

\paragraph{Math and reasoning retrieval.}
Our target is MathNet-Retrieve~\citep{alshammari2026mathnet}, which asks whether an embedder can match a competition problem to an LLM-written version of the same problem at three levels of disguise. Among neighbouring benchmarks, SABER-Math~\citep{georgiev2026saber} labels human-written problems automatically from start to finish, BRIGHT~\citep{su2025bright} gathers reasoning-heavy queries from real forum posts and curated problem sets, and \citet{wei2026rirsurvey} survey the field. Among retrievers built for reasoning, ReasonIR~\citep{shao2025reasonir} and RaDeR~\citep{das2025rader} train on synthetic queries, the MathLeap embedders come with MELD, a diagnostic for mathematical equivalence~\citep{ye2026meld}, and Rank1~\citep{weller2025rank1} reranks with test-time reasoning; RaDeR, ReasonIR and MathLeap are our $7$--$8$B baselines, none above $0.01$ hard-tier R@1 (Table~\ref{tab:leaderboard-full}).

\paragraph{The blind spot of item-level contamination detection.}
Most contamination research looks for leaked test items. ConStat~\citep{dekoninck2024constat} checks whether a model does better on the benchmark than on reworded copies of it, \citet{yao2024crosslingual} show that a model still remembers a test item after translation, and \citet{yang2023rephrased} show that reworded copies of test items in the training data slip past $n$-gram filters. In all of these the reworded text copies a \emph{benchmark item}, which an item-level detector can catch; our training data rewords public corpus problems. Task contamination~\citep{li2024taskcontamination} and ConStat's synthetic references look past single items to a distribution, and recipe-matching is of that kind, except that the distribution is the benchmark's own writing procedure, which anyone can regenerate rather than leak. What carries over is a style of writing, invisible to item-level checks; leakage explains $1.24$ of the $45$ points (Appendix~\ref{app:contamination}).

\paragraph{Shortcut learning at the procedure level.}
Recipe-matching is a relative of shortcut learning, where a model passes a test by picking up accidental patterns in the data: natural language inference (NLI) models that answer from the hypothesis alone and other crowdworker artefacts~\citep{gururangan2018artifacts,poliak2018hypothesis}, HANS~\citep{mccoy2019hans}, and the framing of \citet{geirhos2020shortcut}. Our style probe is, mechanically, a supervised detector of machine-written text, unlike the zero-shot DetectGPT~\citep{mitchell2023detectgpt}. The usual fix, adversarial filtering, discards the answers or items a shortcut model finds easy~\citep{zellers2019hellaswag,lebras2020aflite}, but it cannot remove a property of how the benchmark was written, which lives in no particular item. Recipe-matching is also not an LLM judge favouring its own outputs~\citep{panickssery2024selfpref,zheng2023judging}, since our generator differs from the benchmark's in vendor and model family, so the model is not recognising itself. What remains is a construct-validity failure~\citep{raji2021everything,bowman2021fixbench}, the benchmark measuring something other than what it claims, here in retrieval evaluation, where BEIR and MTEB~\citep{thakur2021beir,muennighoff2023mteb} set the standard. On MathNet-Retrieve the gaming comes with forgetting of real retrieval, and Appendix~\ref{sec:forgetting} compares two remedies for it, weight interpolation~\citep{wortsman2022wiseft} and replay~\citep{chaudhry2019tiny}.

\section{Experimental Setup}
\label{sec:setup}

\paragraph{Terms and design.}
Five terms recur (Table~\ref{tab:terms}, Appendix~\ref{app:mechanism}): the \emph{recipe} is how a benchmark builds itself, the prompts and filters under which an LLM writes part of it; the \emph{template} is the exact wording of that prompt; the \emph{genre} is the style of LLM-written rewrites, whichever prompt produced them; the \emph{pair structure} is a deep LLM rewrite as the positive against a minimal-edit near-miss as the negative; and a \emph{surface} is the part of a benchmark its LLM writes, the part an attacker can imitate. Table~\ref{tab:design} (Appendix~\ref{app:mechanism}) shows every training set as a combination of positives and negatives, and it is where the four models the paper keeps returning to get their names: the \emph{verified arm} trains on computer-algebra positives and negatives; the \emph{recipe arm} on positives and near-misses written with the benchmark's own prompt; \emph{D4} on LLM restatements from a prompt unrelated to the benchmark, with no negatives; and the \emph{reference} on those same D4 restatements with the verified arm's negatives attached. D1 to D4 are four rewriter prompts at increasing distance from the benchmark's own, the ladder Section~\ref{sec:mechanism} climbs. Every model is scored three ways: on the benchmark's tiers, on real duplicates no LLM wrote, and with an $n$-gram probe that asks whether its similarity scores follow the generator's writing style.

\paragraph{Task and metric.}
A retriever embeds queries and documents and ranks the corpus by cosine similarity; each query has one gold document, and R@$k$ is the share of queries whose gold lands in the top $k$. All models are trained with the same in-batch contrastive loss, under which each hard negative of a training row becomes its own example; the loss, the notation and the formal statement of the recipe-matching gain are in Appendix~\ref{app:setup}.

\paragraph{Benchmark and harness.}
MathNet-Retrieve has no training split; it is a test set only. Its three tiers share one corpus of $117{,}088$ documents and the same $15{,}000$ queries, and differ only in which Gemini-3-flash rewrite counts as the gold document. We report R@$k$ for $k \in \{1,5,10\}$. Our evaluation code is a sentence-transformers~\citep{reimers2019sentencebert} reimplementation that reproduces the benchmark authors' published all-mpnet-base-v2 easy-tier result. One check does not pass: our best Qwen3-Embedding-4B setting gives $11.96$ easy R@1 where the authors report $14.76$, and a fourteen-setting sweep narrows the gap to $1.57$ points without closing it (Appendix~\ref{app:setup}).

\paragraph{Two supervision arms at a matched budget.}
The verified arm is built from MathNet's own public $27{,}817$-problem corpus with no LLM involved: each positive an in-place symbolic edit that SymPy certifies equivalent, each negative a minimal edit that a numeric counterexample refutes, and a manual audit of sampled records found $20/20$ correct (Appendix~\ref{app:setup}). The recipe arm copies the benchmark's own procedure: the Appendix-F rewriter prompt, near-verbatim, run through Qwen3-32B-AWQ, with a single LLM judge deciding what to keep and no symbolic check; the judge kept $95.8\%$ of what the generator wrote. Because the benchmark uses two judges where we used one, a second judge from another vendor later re-checked every pair and accepted $86\%$ of the positives; training on the rows both judges admit moves the arm by $+2.33$ easy, $-1.84$ hard and $+8.56$ cross-language points (Appendix~\ref{app:mechanism}). Both arms draw on the same $7{,}089$ source problems, hold exactly $6{,}145$ rows, share every hyperparameter (Table~\ref{tab:hparams}, Appendix~\ref{app:setup}) and pick their checkpoints on their own dev splits; each trains with eight seeds ($42$--$49$), and Table~\ref{tab:ablations} replicates on other backbones and at $4$B. The budget matches sources, rows and settings, not paraphrase depth: recipe positives are full-text LLM paraphrases, verified ones edits that leave the prose untouched, and Section~\ref{sec:mechanism} separates what paraphrase buys from what the recipe buys.

Matched \emph{rows} are not matched \emph{signal}: the recipe arm's file has $1.7\times$ as many negatives per row, and since every negative becomes its own training example, the same $6{,}145$ rows give that arm more contrastive examples and more optimiser steps (Eq.~\ref{eq:loss}, Appendix~\ref{app:setup}). We therefore built a second verified arm matched to the recipe arm on source problems, on the number of negatives and on the dev split. It scores within noise of the first verified arm, so the ${\sim}45$-point gap is not the recipe arm training more; we predicted this before running it (pre-registered; Table~\ref{tab:ablations}). A verified arm with no cap at all, with more rows, sources and examples, does no better (Appendix~\ref{app:setup}).

\paragraph{Real-duplicate evaluations and statistics.}
To test on equivalent problems no LLM wrote, we use official reprints: the same olympiad problem printed in the booklets of different countries in different languages. Mining found $754$ such pairs, $85$--$90\%$ of them genuine on inspection. Their $393$ non-English problems serve as queries against the public corpus, and a hit counts only if the model retrieves the same problem in a \emph{different} language. The mining leans on shared formulas, which favours formula-matching models, yet both trained arms still score below the untrained base on every slice (Table~\ref{tab:xling-slices}). A strong multilingual embedder already scores above $90$ here, BM25 only $10.69$ (Appendix~\ref{app:results}), and the set has no near-miss distractors, so there is no recipe to match; what a drop here can still mix in is lost cross-lingual alignment, since both arms train on English alone, which the same-language set controls. The same graph's single-language clusters give that set, $498$ reprint queries, of which the $125$ with reworded copies and problems in no training file are the primary slice (Appendix~\ref{app:setup}). Every gap between arms carries a $95\%$ nested-bootstrap interval over queries, or over clusters on the duplicate sets, and over the eight seeds (Appendix~\ref{app:setup}).

\paragraph{Contamination.}
The exclusion lists compared exact text and missed corpus copies of benchmark queries that differ only by a stray OCR header, so about $33\%$ of \emph{both} arms' rows anchor on a benchmark query's own statement. The leak is the same size per row in both arms and slightly larger in the arm that loses; the gap holds within about a point on every clean slice of queries; netting each arm's gain on its leaked queries puts $1.24 \pm 0.08$ of the $45.33$ points on leakage; and both arms retrained under a corrected filter (eight seeds) read $43.11$ and $44.45$ (Table~\ref{tab:cleangate}, Appendix~\ref{app:contamination}).

\section{Recipe-Matching Games the Benchmark}
\label{sec:gaming}
\begin{table}[t]
\centering
\caption{The main experiment. Columns: the untrained base, the two matched-budget arms and the recipe-free \emph{reference} (D4 restatements with the verified arm's negatives, no benchmark prompt anywhere), each over eight seeds (mean $\pm$ std), then the recipe arm's lead over the verified arm and over the reference with $95\%$ nested-bootstrap intervals (Section~\ref{sec:setup}). Rows: the benchmark's own tiers first, then real duplicates that no LLM wrote (cross-language reprints, scored \emph{strictly}: the query's own corpus entry is masked and only a reprint in another language counts as a hit; and the $125$-query same-language slice of Appendix~\ref{app:setup}), then the difference-in-differences (DiD), the benchmark lead minus the real-duplicate lead. Bold marks the largest value in a row; on the duplicate rows it is the untrained base. Seed-$42$ breakdown: Table~\ref{tab:controlled-full}.}
\label{tab:controlled}
\scriptsize
\setlength{\tabcolsep}{2pt}
\renewcommand{\arraystretch}{0.90}
\begin{tabular}{lcccccc}
\toprule
Evaluation & Base & Verified & Reference & Recipe & Recipe $-$ verified & Recipe $-$ reference \\
\midrule
\multicolumn{7}{l}{\emph{Benchmark tiers}} \\
Easy (R@1)   & 8.32 & 17.05 $\pm$ 0.79 & 46.91 $\pm$ 1.25 & \textbf{62.38 $\pm$ 0.85} & 45.33 $[44.25, 46.41]$ & 15.47 $[14.11, 16.73]$ \\
Medium (R@1) & 1.78 & 3.14 $\pm$ 0.46  & 4.22 $\pm$ 0.25  & \textbf{8.84 $\pm$ 0.36}  & 5.70 $[5.18, 6.24]$    & 4.62 $[4.13, 5.15]$ \\
Hard (R@1)   & 0.00 & 0.16 $\pm$ 0.06  & \textbf{10.30 $\pm$ 1.15} & 9.42 $\pm$ 0.61  & 9.26 $[8.71, 9.85]$    & $-0.88$ $[-1.85, 0.10]$ \\
Hard (R@5)   & 4.08 & 8.35 $\pm$ 1.13  & \textbf{60.05 $\pm$ 1.22} & 57.20 $\pm$ 0.82 & 48.84 $[47.95, 49.74]$ & $-2.86$ $[-4.01, -1.67]$ \\
\multicolumn{7}{l}{\emph{Real duplicates, no LLM involved}} \\
Cross-language (strict R@1) & \textbf{84.73} & 56.52 $\pm$ 4.05 & 78.02 $\pm$ 0.91 & 66.26 $\pm$ 2.31 & 9.74 $[3.96, 15.71]$ & $-11.77$ $[-17.33, -6.09]$ \\
Same-language (R@1) & \textbf{62.40} & 51.30 $\pm$ 2.44 & 39.20 $\pm$ 1.28 & 46.30 $\pm$ 1.89 & $-5.00$ $[-13.42, 3.35]$ & 7.10 $[-2.34, 17.04]$ \\
\multicolumn{7}{l}{\emph{Benchmark lead minus real-duplicate lead}} \\
DiD, easy $-$ cross-language & & & & & 35.59 $[29.51, 41.47]$ & 27.24 $[21.40, 32.93]$ \\
DiD, easy $-$ same-language & & & & & 50.33 $[41.85, 58.75]$ & 8.37 $[-1.14, 18.16]$ \\
\bottomrule
\end{tabular}
\end{table}

\paragraph{The recipe arm leads the verified arm on every tier; we read the lead as a decomposition.}The reference column of Table~\ref{tab:controlled} does the splitting: the recipe arm minus the reference is the part of each gap that is specific to the benchmark's own prompt, and the reference minus the verified arm is the part any LLM rewrites would buy. Two tiers need care. On the medium tier the recipe arm's lead over the verified arm is no larger on the benchmark than on real duplicates, so against that arm the tier shows no inflation (Section~\ref{sec:ood}). On the hard tier the verified arm scores near zero, so a comparison against it says nothing; the recipe arm is read against the reference instead (Section~\ref{sec:mechanism}).

\paragraph{A better training file, or a file built like the test?}
Perhaps the recipe arm's training file is simply the better file, and a better file would score higher on any test. Two facts rule that out. First, a better file would help on real duplicates, the same problem printed twice by human authors, and it does not: on cross-language duplicates both arms score below the untrained base ($84.73$), and the recipe arm's lead over the verified arm shrinks from $45.33$ points on the easy tier to $9.74$ there. Second, who wins depends on who built the test (Table~\ref{tab:crosseval}, Appendix~\ref{app:results}). The verified arm wins on pairs that computer algebra built; the two arms trained on LLM rewrites win on pairs that an LLM built, whichever LLM and whichever prompt; and the untrained base wins on duplicates no generator wrote. A file that wins only on tests built the way it was built is not a better file; it is a matched one.

\paragraph{Two recipe-free references recover half to two thirds of the easy-tier gap; the remainder does not reach cross-language duplicates.}
The reference of Table~\ref{tab:controlled} is D4's LLM restatements, written under a prompt unrelated to the benchmark, paired with the verified arm's own negatives, so against the verified arm the only change is who wrote the positives, a computer algebra system or an LLM (Section~\ref{sec:mechanism}). That change alone recovers $29.85$ of the $45.33$ easy-tier points, matches the recipe arm on the hard tier and beats it at R@5. A second unrelated prompt with the same negatives recovers $22.15$, so against references built on verified negatives the part of the gap that needs the benchmark's own prompt is $15.47$ to $23.18$ points (Appendix~\ref{app:mechanism}). To check that the recovery is not simply a matter of the positives being paraphrased, we paraphrased them by machine translation out and back, with no LLM, and kept the same negatives: that recovers $1.76$ points. Up to the depth back-translation reaches, what recovers the gap is that an LLM wrote the text, not that the text was rewritten.

With the recipe arm's own near-misses attached to D4's restatements in place of the verified negatives (Section~\ref{sec:mechanism}), the recipe arm still leads on the easy tier, by $17.79$ points at the reference's negative count and by $24.59$ at its own volume; in the second comparison the two training sets differ only in which prompt wrote the positives (and by $2.7\%$ of rows), so $24.59$ is what the benchmark's own prompt buys with everything else held equal; the recipe-specific share is therefore $15.47$ to $24.59$ points across comparisons, larger at LLM negatives because counterexamples themselves lift the easy tier (Appendix~\ref{app:mechanism}). Both D4 cells beat the recipe arm on cross-language duplicates (Appendix~\ref{app:mechanism}). The recipe adds $15.47$ easy-tier and $4.62$ medium-tier points over the reference, and none of it reaches cross-language duplicates, where the recipe arm trails the reference by $11.77$ and the base by about $18$. As a benchmark-minus-real-data difference, the recipe arm's lead shrinks by $27.24$ points against the reference ($35.63$ against the second reference) and by $35.59$ against the verified arm; on same-language reprints the recipe arm and the reference are within noise (Table~\ref{tab:controlled}).

\paragraph{The gaming transfers across generators, backbones, and scale.}The effect is not tied to one LLM: Gemini-3-flash wrote the benchmark's pairs and Qwen3-32B-AWQ the recipe arm's, so what transfers is what the recipe writes, not one generator's fingerprint. Nor to one base model: on multilingual-e5-large~\citep{wang2024e5} the recipe arm leads by $+19.12$ easy-tier points but by only $+2.29$ on cross-language duplicates (seed $42$; BGE and late-interaction ColBERTv2: Appendix~\ref{app:results}). It holds at a larger size: with a $4$B model (same seed) the recipe arm doubles the verified arm's easy-tier score ($77.77$ against $39.51$), but on cross-language duplicates the order \emph{flips}, the recipe arm at $70.99$ against the verified arm's $83.72$, under an untrained base of $91.60$. Verified supervision largely keeps retention at that scale, and $39.51$ from verified pairs alone shows the tier is learnable.

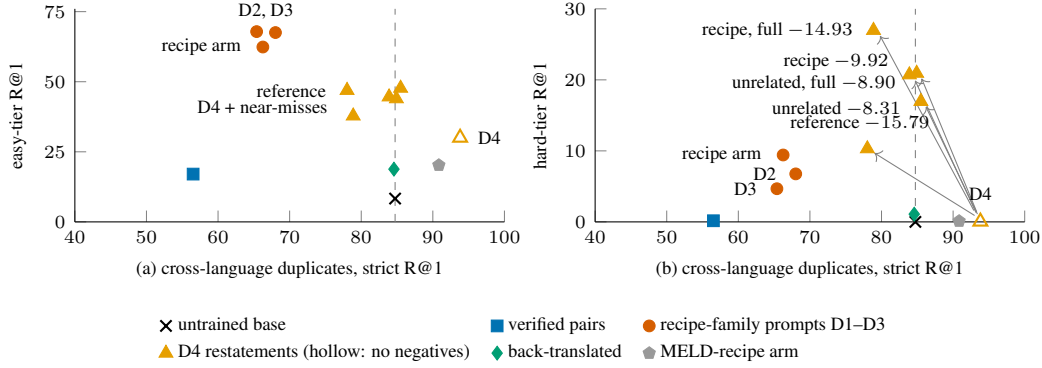
\begin{figure}[t]
\centering
\definecolor{cVer}{HTML}{0072B2}   
\definecolor{cRec}{HTML}{D55E00}   
\definecolor{cD4}{HTML}{E69F00}    
\definecolor{cBT}{HTML}{009E73}    
\definecolor{cMELD}{HTML}{999999}  
\begin{tikzpicture}
\begin{axis}[
  name=a,
  width=0.44\linewidth, height=4.4cm,
  xmin=40, xmax=100, ymin=0, ymax=76,
  xtick={40,50,60,70,80,90,100}, ytick={0,25,50,75},
  xlabel={(a) cross-language duplicates, strict R@1}, ylabel={easy-tier R@1},
  label style={font=\scriptsize}, tick label style={font=\scriptsize},
  axis y line*=left, axis x line*=bottom, clip=false,
]
\addplot[gray, dashed, forget plot] coordinates {(84.73,0) (84.73,76)};
\addplot[forget plot, only marks, mark=x, mark size=3pt, thick, black] coordinates {(84.73,8.32)};
\addplot[forget plot, only marks, mark=square*, mark size=2.2pt, cVer] coordinates {(56.52,17.05)};
\addplot[forget plot, only marks, mark=*, mark size=2.2pt, cRec] coordinates {(66.26,62.38) (68.02,67.54) (65.39,67.88)};
\addplot[forget plot, only marks, mark=triangle, mark size=3pt, thick, cD4] coordinates {(93.81,30.03)};
\addplot[forget plot, only marks, mark=triangle*, mark size=3pt, cD4] coordinates {(78.02,46.91) (83.89,44.59) (78.88,37.79) (85.50,47.69) (84.91,43.93)};
\addplot[forget plot, only marks, mark=diamond*, mark size=2.6pt, cBT] coordinates {(84.57,18.81)};
\addplot[forget plot, only marks, mark=pentagon*, mark size=2.4pt, cMELD] coordinates {(90.84,20.23)};
\node[font=\scriptsize, anchor=east] at (axis cs:64.5,62.38) {recipe arm};
\node[font=\scriptsize, anchor=south] at (axis cs:66.7,69.0) {D2, D3};
\node[font=\scriptsize, anchor=west] at (axis cs:95.0,30.03) {D4};
\node[font=\scriptsize, anchor=east] at (axis cs:76.4,46.91) {reference};
\node[font=\scriptsize, anchor=east] at (axis cs:76.6,41.2) {D4 + near-misses};
\end{axis}
\begin{axis}[
  name=b, at={(a.south east)}, anchor=south west, xshift=1.2cm,
  width=0.44\linewidth, height=4.4cm,
  xmin=40, xmax=100, ymin=0, ymax=30,
  xtick={40,50,60,70,80,90,100}, ytick={0,10,20,30},
  xlabel={(b) cross-language duplicates, strict R@1}, ylabel={hard-tier R@1},
  label style={font=\scriptsize}, tick label style={font=\scriptsize},
  axis y line*=left, axis x line*=bottom, clip=false,
  legend to name=invlegend, legend columns=3,
  legend style={font=\scriptsize, draw=none, /tikz/every even column/.append style={column sep=6pt}}, legend cell align=left,
]
\addplot[gray, dashed, forget plot] coordinates {(84.73,0) (84.73,30)};
\addplot[only marks, mark=x, mark size=3pt, thick, black] coordinates {(84.73,0.00)};
\addlegendentry{untrained base}
\addplot[only marks, mark=square*, mark size=2.2pt, cVer] coordinates {(56.52,0.16)};
\addlegendentry{verified pairs}
\addplot[only marks, mark=*, mark size=2.2pt, cRec] coordinates {(66.26,9.42) (68.02,6.76) (65.39,4.67)};
\addlegendentry{recipe-family prompts D1--D3}
\addplot[only marks, mark=triangle*, mark size=3pt, cD4] coordinates {(78.02,10.30) (83.89,20.71) (78.88,26.97) (85.50,16.96) (84.91,20.93)};
\addlegendentry{D4 restatements (hollow: no negatives)}
\addplot[forget plot, only marks, mark=triangle, mark size=3pt, thick, cD4] coordinates {(93.81,0.04)};
\addplot[only marks, mark=diamond*, mark size=2.6pt, cBT] coordinates {(84.57,1.09)};
\addlegendentry{back-translated}
\addplot[only marks, mark=pentagon*, mark size=2.4pt, cMELD] coordinates {(90.84,0.11)};
\addlegendentry{MELD-recipe arm}
\draw[->, thin, gray] (axis cs:93.0,0.9) -- (axis cs:79.0,9.7);
\draw[->, thin, gray] (axis cs:93.2,1.3) -- (axis cs:84.7,19.9);
\draw[->, thin, gray] (axis cs:92.9,1.6) -- (axis cs:79.8,26.2);
\draw[->, thin, gray] (axis cs:93.4,1.1) -- (axis cs:86.2,16.2);
\draw[->, thin, gray] (axis cs:93.3,1.4) -- (axis cs:85.6,20.2);
\node[font=\scriptsize, anchor=east] at (axis cs:64.5,9.42) {recipe arm};
\node[font=\scriptsize, anchor=east] at (axis cs:66.6,6.76) {D2};
\node[font=\scriptsize, anchor=east] at (axis cs:63.9,4.67) {D3};
\node[font=\scriptsize, anchor=south] at (axis cs:93.81,1.8) {D4};
\node[font=\scriptsize, anchor=south] at (axis cs:77.0,11.6) {reference $-15.79$};
\node[font=\scriptsize, anchor=east] at (axis cs:77.3,26.97) {recipe, full $-14.93$};
\node[font=\scriptsize, anchor=east] at (axis cs:82.3,22.5) {recipe $-9.92$};
\node[font=\scriptsize, anchor=east] at (axis cs:83.3,19.5) {unrelated, full $-8.90$};
\node[font=\scriptsize, anchor=east] at (axis cs:84.0,15.9) {unrelated $-8.31$};
\end{axis}
\node[anchor=north, yshift=-4pt] at (current bounding box.south) {\pgfplotslegendfromname{invlegend}};
\end{tikzpicture}
\caption{Benchmark score against real retrieval kept after training, one point per trained model. Horizontal axis: strict R@1 on cross-language duplicates; the dashed line and $\times$ mark the untrained base, so points left of it lost real retrieval by being trained. Colour gives the positives (legend); hollow means no negatives. (a)~Easy tier: every model trained on LLM rewrites scores above the verified arm. (b)~Hard tier: only deep LLM rewrites with negatives leave the floor. Arrows run from D4 alone to D4 with negatives attached; each label names the negatives (verified ones give the reference; ``recipe'' and ``unrelated'' are LLM near-misses from the recipe's prompt and from an unrelated one, at the reference's count or at the recipe arm's full volume) and the cross-language points lost. All models: Figure~\ref{fig:inversion-full}.}
\label{fig:inversion}
\end{figure}
\section{What the Recipe Adds, and the Fingerprint It Leaves}
\label{sec:mechanism}

\paragraph{Four rewriter prompts on one ladder, and a factorial over the pair.}
This section uses two designs. The first is a ladder of four prompts. \textbf{D1} is the benchmark's published Appendix-F prompt, near-verbatim, and its model is the recipe arm; \textbf{D2} is a close paraphrase of that prompt; \textbf{D3} asks for the same thing in a clearly different style; \textbf{D4} is a restatement prompt with no connection to the benchmark's. Everything else is held fixed: the same source problems, judge and training settings, and D1 to D3 share one output format, one positive and up to three near-misses, so the ladder changes the wording of the prompt while the shape of each training pair stays the same (Appendix~\ref{app:prompts}); D4 asks for restatements only and writes no negatives. The second design separates D4's prompt from its missing negatives: a grid crossing computer-algebra rewrites or D4 restatements with computer-algebra counterexamples or none, extended with LLM-written near-misses from the recipe's prompt and from an unrelated one and with back-translated positives (Table~\ref{tab:factorial}).

\paragraph{On the hard tier the pair structure scores, both halves are needed, and the recipe-specific part is an increment in the near-misses.}
Before running D2--D4 we predicted that hard-tier R@1 would fall step by step as the prompt moves away from the benchmark's. It did: $9.90 > 6.76 > 4.67 > 0.04$ from D1 to D4, at three seeds per rung and in every seed (Table~\ref{tab:dose-full}). The grid explains the last, large drop. Neither half of a training pair works alone: positives of any kind trained without negatives score $0.00$ to $0.05$. Negatives on shallow positives barely help either: near-copy and back-translated positives stay between $0.11$ and $4.18$ with negatives of any kind (Appendix~\ref{app:mechanism}). The positive has to be a deep LLM rewrite (Section~\ref{sec:limitations}). Give D4's restatements the verified negatives and the result is the reference, at $10.30$ over eight seeds, level with the recipe arm, from a file with no benchmark wording in it; a second recipe-free prompt's restatements with the same negatives reach only $3.97$, so the level a recipe-free file reaches depends on which prompt wrote the restatements. Now change only the negatives, keeping D4's positives (three seeds each; matched to the reference's count, $0.08$ fewer per row): near-misses written under a prompt unrelated to the recipe lift the score to $16.96$ and near-misses written under the benchmark's own prompt to $20.71$; with as many negatives as the recipe arm has, $20.93$ and $26.97$, the last nearly three times the recipe arm (Table~\ref{tab:factorial}). So the tier rewards the recipe's \emph{pair structure}, a deep rewrite against an LLM-written near-miss. Near-misses from either LLM prompt carry most of the step up from verified negatives; the recipe's own add $3.75$ to $6.03$ on top. The template's own positives are not required, and restatements from an unrelated prompt outscore them at either kind of negative: $26.97$ against $9.42$ with the recipe's near-misses, $10.30$ against $2.09$ with verified negatives. One tempting account, that a positive and its near-misses written in one call share phrasing, is ruled out: regenerating the recipe's positives and near-misses in separate calls leaves the tier where it was ($9.79$). Two other accounts are untested (Appendix~\ref{app:mechanism}).

\paragraph{On the easy tier no gradient appears: the genre scores.}We predicted the same step-by-step fall for the easy tier. It does not happen. The only rung that drops is D4, the one with no negatives, and once D4 gets the verified negatives it becomes the reference of Table~\ref{tab:controlled}. D2 and D3 cannot be told apart, their difference flipping sign across seeds, and both score five to six points \emph{above} the recipe arm (Table~\ref{tab:dose-full}), the opposite of the registered direction (Appendix~\ref{app:setup}). The easy tier does not care which LLM prompt wrote the training pairs: any prompt that produces LLM-style rewrites scores, so the tier is gamed at the \emph{genre} level.

\paragraph{The real-data prediction failed: D4 tops cross-language retrieval and every negative that lifts it costs points.}
We had registered that all four rungs would sit just below the base on cross-language duplicates, in a narrow band with no trend; that failed: the rungs spread almost three times wider than the bound and D4 landed above the base (Appendix~\ref{app:mechanism}). D4 sits at the top on cross-language duplicates, $93.81$ against the base's $84.73$, within noise of back-translated positives trained alone ($94.57$; Appendix~\ref{app:mechanism}), at $+0.04$ hard R@1. Every way of giving it negatives buys hard-tier score and pays in cross-language score: the verified negatives that lift it to $10.30$ cost it $15.79$ of those points, the recipe's own near-misses at the reference's count lift it to $20.71$ and cost $9.92$, and near-misses from an unrelated prompt lift it to $16.96$ and cost $8.31$, leaving that model at the base (Figure~\ref{fig:inversion}b). The recipe arm itself, which leads the reference on the easy tier, trails it by $11.77$ on cross-language duplicates (Section~\ref{sec:gaming}; Figure~\ref{fig:inversion}).

\paragraph{An $n$-gram probe finds the LLM fingerprint, and recipe-prompt models track it.}
A plain $n$-gram classifier, trained on our rewrites against their own source problems and tested on sources it never saw, tells LLM rewrites from human-written problems at $0.880$ AUC, detects the benchmark's Gemini-3-flash golds at $0.804$ without having seen a token from that vendor, and separates the benchmark's golds from its own near-misses at $0.750$ with no retriever involved, so a style difference between right and wrong answers is built into its pairs; the recipe-free D4 rewrites are detected at $0.676$ and D1--D3 somewhat more strongly, and that climb is what the recipe adds (Table~\ref{tab:panel}). Correlating each model's similarities with that style score across documents, query by query, shows neither the base nor the verified arm tracking style; the recipe arm does ($\rho = +0.171$, above the verified arm on $94.2\%$ of queries), D2 and D3 do too, D4 does not, and the reference and the D4 cells trained with the recipe's near-misses show only a trace ($+0.025$ to $+0.050$). The habit comes from the recipe's rewrite prompts as a family, not the template alone, and the cells that buy the most hard-tier score carry little of it: chasing style is not what buys the tier (Table~\ref{tab:style}). The $1{,}401$ hard-tier queries the recipe arm alone wins skew toward golds sharing the query's wording, not toward LLM-styled golds (Section~\ref{sec:limitations}).

\section{What Transfers to Real Data, and Two External Benchmarks}
\label{sec:ood}

\paragraph{Most of the benchmark gap reaches neither duplicate probe.}
On cross-language duplicates both arms score below the untrained base ($84.73$), the recipe arm less: over eight seeds it keeps a $9.74$ $[3.96, 15.71]$ lead over the verified arm, positive in every seed (Table~\ref{tab:ood}). But on the easy tier the same two arms are $45.33$ points apart, so the lead shrinks by $35.59$ $[29.51, 41.47]$ points from benchmark to real data. That shrinkage is our central statistic, and $27.24$ $[21.40, 32.93]$ of it is recipe-specific: the reference, with no benchmark wording, loses far less of its lead (Table~\ref{tab:controlled}; Figure~\ref{fig:inversion}). It is a rank-$1$ effect: at R@5 the recipe arm's leads over the verified arm on the easy tier and on cross-language duplicates are equal ($7.33$ against $7.48$), and against the reference both are within a point of zero, so what recipe-matching buys is the top rank (Appendix~\ref{app:results}). Same-language reprints say the same thing more sharply: there the recipe arm scores $5.00$ points below the verified arm ($[-13.42, 3.35]$, no lead at all) and the lead shrinks by $50.33$ $[41.85, 58.75]$ points, the reading we registered. On same-language data the recipe arm and the reference are within noise of each other and neither holds a measurable lead over the verified arm, so the recipe-specific share is visible only in the cross-language comparison. The medium tier behaves the \emph{opposite} way against the verified arm: the arms' gap there is smaller than on cross-language duplicates in seven of eight seeds. Against the reference the recipe arm leads by $4.62$ on the tier and trails by $11.77$ on duplicates, a recipe-specific shrinkage of about $16$ points by subtraction, so the tier carries a smaller version of the easy tier's effect, and the inflation claim rests on the easy tier, where it is largest (Appendix~\ref{app:results}).

Off the benchmark's genre, training hurts: on BRIGHT's math splits and on ImpliRet's $9{,}000$ queries~\citep{taghavi2025impliret} both arms score below the untrained base, and on ImpliRet the recipe arm is the worst model we measure. Eleven human-curated MIRB tasks~\citep{ju2025mirb} agree: the base leads both arms on ten (Appendix~\ref{app:results}). We had registered that the arms would not differ on MELD or BRIGHT; it failed on both, in the recipe arm's favour: on MELD the recipe arm leads the verified arm by $+9.45$ R@1, and MELD's pairs are LLM rewrites of mathematics, so the recipe arm's advantage on that genre appears there too; on BRIGHT's two splits it leads by $8.36$ $[3.42, 13.99]$ and $0.75$ $[0.16, 1.52]$ nDCG points while both arms sit below the base (Table~\ref{tab:external}).

\paragraph{SABER-Math: the document attack fails; training on its summaries buys a small gain that survives a matched budget.}
SABER-Math~\citep{georgiev2026saber} is built differently, and its pipeline is public, so we could copy the stage we attack. Its documents are organic, human-written problems, at most translated or cleaned by its LLM; the one text it writes is a short \emph{summary} of each problem's solution idea. Summary overlap is one of SABER's two candidate-finding signals, the other being topic overlap; the pairs then go to an LLM-judged tournament, and the summaries themselves are never scored. Our registered attack, aimed at the documents, fails: nDCG@10 \emph{drops} where we had required a gain of $+0.02$, and the correlation between the model's similarities and SABER's summary-overlap rule rises only $1.14\times$ where we had required $2\times$. We then attacked each surface alone at the same row budget, three seeds each: training on summaries alone gains $+0.0343 \pm 0.0018$ nDCG@10 over the base in every seed, as registered, while training on documents alone loses. Two things favoured that run for reasons outside the recipe, $1.8\times$ the steps and twice the distinct positives; at the document arm's $57$ steps and $2{,}368$ positives it still gains $+0.0255 \pm 0.0041$ (three seeds). This is a score that moved, not proof that the model matched SABER's rule, and we attacked only the discovery stage, not the tournament that scores the pairs (Appendix~\ref{app:results}).

\paragraph{MELD: a reconstructed prompt moves its score at no cost to retention.}
MELD~\citep{ye2026meld} is $270$ pairs of mathematical statements restated across subfield dialects, written by Claude Opus 4.7, edited by hand, then checked by GPT-5.5. It publishes the prompts behind its \emph{training} pipeline but not the one behind its evaluation set. We rebuilt that procedure from its description and ran it through Qwen3-32B-AWQ, another vendor's model, each arm on $3{,}150$ matched rows at three seeds, gated item by item against MELD's evaluation texts; four readings were registered beforehand. The attacked arm beats the base in every seed, both scored with the same query prompt, by $+26.97$ R@1 on average. At seed $42$ it is within noise of MELD's best published model in our harness ($+2.22$ $[-2.96, 7.22]$ against MathLeap-Octen-8B; both read higher here than in MELD's own table, Appendix~\ref{app:setup}) and leads the MathNet recipe arm by $+17.78$ on MELD, while inheriting nothing on MathNet's hard tier ($0.11$ R@1): what it learned is MELD's, not general. An arm trained with none of MELD's subfield names still captures most of the gain, so the effect is genre-level, not vocabulary. A third arm generated under a rewritten prompt, procedure matched, keeps $80\%$ of the gain, so most of the effect is the cross-dialect pairing and the rest the wording (Appendix~\ref{app:results}). One registered reading landed the other way: we predicted a fall below the base on cross-language duplicates; MELD's arm reaches $90.67$, above the base in every seed, and leaves same-language retention at the base ($-1.60$ $[-6.08, 2.46]$; Appendix~\ref{app:setup}). Unlike on MathNet, most of what it learns is the task.

Only on MathNet-Retrieve can the inversion be pinned to a single edit of the training file, one that raises benchmark score and lowers real retrieval at once. MELD's recipe raises the score while the model keeps its real retrieval, so its gain is mostly the task. SABER's gain comes from summaries it never scores, is small but holds at a matched budget, and its arms lose cross-language retrieval.
\section{Limitations}
\label{sec:limitations}

\textbf{Each recipe surface is attacked on one benchmark, and SABER's evidence is thin.} We attacked each kind of LLM-written surface, documents, summaries and evaluation pairs, on one benchmark each, so we cannot say how general each result is. On SABER the summary arm's nDCG gain holds in every seed, matched budget or not, but the correlation with SABER's construction rule barely rises, and the attack never touches the tournament that produces the scores, so it does not show that the score can be bought by matching the rule; each pure-channel arm has three seeds (one seed for real duplicates), and the mostly-document $6{,}145$-row attack and SABER's controlled arms are single runs (Appendix~\ref{app:results}).

\textbf{Two MELD caveats.} MELD's evaluation-set prompt is unpublished, so what we attacked is our reconstruction of it; the prompt control puts most of the gain on the cross-dialect pairing rather than on exact wording, but it is still a reconstruction. And MELD's pairs are written by an LLM before humans review them, so the MathNet recipe arm's $+9.45$ lead over the verified arm on MELD (Table~\ref{tab:external}) is a model trained on LLM rewrites scoring well on other LLM rewrites; whether that lead reaches anything a human wrote is untested.

\textbf{Residual contamination.} The headline arms trained under the leaky exact-text filter, with about $33\%$ of each arm's anchors a benchmark query's own statement; the three corrected-filter retrains confirm rather than replace them, and one used different source lists (Appendix~\ref{app:contamination}).

\textbf{The hard-tier evidence is automated; the probes have limits.} No human has read the $1{,}401$ hard-tier queries the recipe arm alone wins at rank~$1$. Our reading that those wins come from shared wording rests on a word-overlap classifier that separates the wins from the rest (AUC $0.639$) while the gold's LLM-style score does not ($0.511$), and shared judge errors between the benchmark's golds and the recipe arm's positives are not ruled out; an audit sheet (\texttt{data/hard\_audit}) lets both readings be checked. The duplicate probes carry no near-misses, so we built one from the verified arm's machinery on $1{,}107$ held-out benchmark queries: on it the recipe arm leads the reference by $6.98$ R@1, so recipe-matching does buy some near-miss discrimination, but the cells with the highest hard-tier scores trail the reference on it, so a higher tier score does not mean better discrimination (Appendix~\ref{app:setup}). The cross-language probe mixes two effects, equivalence never learned and cross-lingual alignment forgotten by a monolingual fine-tune; the same-language slice has no alignment to forget but resolves only $12$- to $14$-point differences at $80\%$ power. Beyond the two arms and the reference every cell has three seeds, the ladder cannot separate the positives' wording from the near-misses' quality within D1--D3, and the back-translated control separates paraphrase depth from LLM authorship only down to its own depth (Appendix~\ref{app:setup}).

\section{Conclusion}
\label{sec:conclusion}
On MathNet-Retrieve the two tiers reward two different things. The easy tier rewards the genre of LLM rewrites, whichever prompt wrote them; the hard tier rewards the recipe's pair structure, a deep rewrite set against a minimal-edit near-miss. Half to two thirds of the recipe arm's easy-tier lead over the verified arm is bought by any LLM-written training pairs, and back-translation, as deep as a non-LLM paraphrase reaches, attributes this to the authorship rather than the depth of the paraphrase; the rest needs the benchmark's own prompt and vanishes on real duplicates that no generator wrote. MELD also moves when trained on pairs built its way, SABER-Math only through a small gain robust to a matched budget, and only on MathNet-Retrieve can one edit to a training file be shown to raise the score and lower real retrieval at once. A benchmark should not make score and retention point in opposite directions.

Four practices follow for builders. (i)~Assume the recipe will get out or be rebuilt: MELD's score moved under a reconstruction of a prompt it never published. (ii)~Prefer test material that people wrote to material an LLM wrote. (iii)~Keep a subset no generator touched and score models on it too; a model that does much better on the LLM-written part is showing the gaming signal. (iv)~Publish a recipe-sensitivity number, how much a model gains by matching the recipe, as contamination tests publish how much of a score does not carry over to reworded or synthetic copies~\citep{dekoninck2024constat}. Regenerating the golds is a weaker lever: with the benchmark's own recipe it removes under a quarter of the gap, and drawing golds from several prompts under two fifths (Appendix~\ref{sec:forgetting}).

\section*{Ethics and Artefact Release}
This paper reports attacks on three benchmarks. We release no attack artefact that is not already derivable from the benchmarks' own construction descriptions. There is no non-public exploit to disclose: two of the three prompts are public, and the third we reconstructed from a published procedure paragraph, so we treat this as a construct-validity finding to publish rather than a vulnerability to embargo. For MathNet-Retrieve the gaming vector follows from its published prompt and the pair structure it writes, and the point of the paper is that a well-meaning practitioner can reproduce it by accident; for MELD, whose evaluation-set prompt is not published, a reconstruction of the procedure paragraph sufficed, which is the more uncomfortable finding. The published rewriter prompt is the entire content of our recipe arm, and so is the attack vector for the headline experiment rather than background context. The analysis artefacts needed to re-derive every claim here, including those unfavourable to us, are released, together with the commands to regenerate the large intermediates (per-query rank dumps, attack pair files) that the repository does not carry; and the two mined duplicate evaluations, cross-language and same-language, are released as remediation: generator-free targets for the ability the benchmark intends to measure. The three trained models we release are the recipe-free rung (D4), the $4$B verified arm and the mixed-supervision model V2.

\section*{Reproducibility Statement}
Every number in this paper is regenerable from the accompanying repository, \repourl. \texttt{reproducibility.json}
records, for each result, the exact command, the SLURM job id, the input files and the output
artefacts; the evaluation harnesses are calibrated against externally published rows wherever one
exists, and every calibration check that fails (the MathNet-Retrieve $4$B row, MELD and ImpliRet) is
reported in Appendix~\ref{app:setup}. Pre-registered predictions are stored in the headers of the job
scripts that produced them, so their wording can be checked against the code that ran. Training data,
per-query rank dumps and attack pair files that the repository does not carry are listed in its README
with the commands that rebuild them.

\section*{AI Use Statement}
Generative AI is both a tool and the object of study in this work, so we list its uses by
category. \emph{Synthetic data.} Every LLM-written training
set in the paper (the recipe arm, the rungs D1 to D5, the LLM-written near-misses, the regenerated
golds of the defence experiment and the MELD and SABER attack sets) was generated by Qwen3-32B-AWQ
and filtered by LLM judges (Qwen3-32B-AWQ, and Mistral-Small-24B-Instruct-2501 for the second-judge
control), as Sections~\ref{sec:setup} and~\ref{sec:mechanism} and Appendices~\ref{app:mechanism}
and~\ref{app:prompts} describe. That generation is the experimental manipulation rather than a
convenience; the verified arm, the two duplicate evaluations and the near-miss probe contain no
LLM-written text, and the back-translation controls use the NLLB-200 machine-translation model.
\emph{Implementation.} An AI coding assistant was used to write and debug the experiment scripts,
evaluation harnesses and analysis code. All numbers in the paper were re-derived from the result
files in the accompanying repository rather than accepted from generated text. \emph{Feedback on
methodology.} The same assistant ran six adversarial review passes over the completed project, and
their findings drove a large fraction of the corrections and controls reported here. The research
questions, the experimental designs, the pre-registered predictions and every interpretive claim are
the authors'. \emph{Writing.} The assistant was used to draft and revise this manuscript; errors it
introduced and the passes caught were corrected. We did not use generative AI to formulate or prove
mathematical claims, to propose or refine the hypotheses, or to interpret the results, and the
remaining categories (theoretical or conceptual frameworks, translation of the
manuscript, qualitative or thematic analysis) do not apply. The authors have reviewed all
AI-assisted work and take responsibility for the final content, including all text, claims and
artefacts.

\section*{Acknowledgments}
 We thank Prof Sultan Albarakati, KAUST, for his support in making this paper possible. For computer time, this research used Ibex managed by the Supercomputing Core Laboratory at King Abdullah University of Science \& Technology (KAUST) in Thuwal, Saudi Arabia.

\bibliographystyle{unsrtnat}
\bibliography{references}

\appendix
\section{Data construction, matched-budget asymmetries, and harness calibration}
\label{app:setup}

\begin{table}[htbp]
\centering
\caption{Training settings shared by both arms. Every row is identical for the verified arm and the recipe arm; only the training file differs. Two other experiments reuse these settings with stated deviations, applied identically to both arms: the SABER-Math attack of Section~\ref{sec:ood} caps training sequences at $1{,}024$ tokens where the controlled arms use the backbone's default, and the $4$B replication uses rank-$16$ LoRA adapters instead of full fine-tuning, a $1 \times 10^{-4}$ learning rate instead of $2 \times 10^{-5}$, and gradient checkpointing.}
\label{tab:hparams}
\small
\begin{tabular}{l p{9.4cm}}
\toprule
Base model & Qwen3-Embedding-0.6B, bf16, all weights trained \\
Loss & the in-batch contrastive loss of Eq.~\ref{eq:loss}, computed with gradient caching~\citep{gao2021gradcache} \\
Batch size / mini-batch size & $256$ / $16$ (the cache makes the gradient identical to a plain batch of $256$) \\
Learning rate & $2 \times 10^{-5}$ \\
Epochs & $3$ \\
Seeds & $42$--$49$ (eight per arm) \\
Training rows per arm & $6{,}145$, with all rows of one source problem kept together \\
Dev split & held out by source problem, so no problem is in both training and dev \\
\bottomrule
\end{tabular}
\end{table}

\paragraph{Task, metric, and training objective.}
A retriever is an encoder $f_\theta$ that maps a query $q$ and a document $d$ to vectors and ranks the corpus $\mathcal{C}$ by cosine similarity, $s_\theta(q,d) = \cos(f_\theta(q), f_\theta(d))$. Each query in an evaluation set $\mathcal{E}$ has one gold document $d_q^{+} \in \mathcal{C}$, and $\mathrm{R}@k(\theta;\mathcal{E})$ is the share of queries whose gold lands in the top $k$. A training row $r = (a_r, p_r, N_r)$ holds an anchor problem, one positive $p_r \sim a_r$ that is equivalent to it, and a set of hard negatives $n \not\sim a_r$ that are not. Each negative becomes its own training example $(a_r,p_r,\{n\})$; a row with no negatives becomes the single example $(a_r,p_r,\emptyset)$. The examples of all rows in a training set $\mathcal{D}$ form $E(\mathcal{D})$. Every model in the paper is trained with the same in-batch contrastive loss: for each example $i$ in a batch $B$, the anchor's positive must beat every other candidate in the batch, its own negatives and the positives and negatives of every other example, the set $\mathcal{K}(i,B) = \{p_i\} \cup N_i \cup \bigcup_{j \in B \setminus \{i\}} (\{p_j\} \cup N_j)$:
\begin{equation}
\label{eq:loss}
\mathcal{L}(\theta;\mathcal{D}) = \frac{1}{|E(\mathcal{D})|}
  \sum_{B \in \mathcal{P}} \sum_{i \in B} -\log
  \frac{\exp(\gamma\, s_\theta(a_i,p_i))}
       {\sum_{d \in \mathcal{K}(i,B)} \exp(\gamma\, s_\theta(a_i,d))},
\end{equation}
with scale $\gamma = 20$ and $\mathcal{P}$ the partition of the examples into batches by a sampler that never puts two examples sharing any text in one batch (sentence-transformers' \texttt{NO\_DUPLICATES}). Since every row of a source problem uses that problem's text as its anchor, the rows of one source problem never meet in a batch either, so a source's equivalent positives cannot act as in-batch negatives for one another; this holds for the verified arm's up to three rows per source as much as for the recipe arm's one. One consequence matters later: a row with three negatives contributes three examples and a row with none contributes one, so rows weigh $\max(|N_r|,1)$, and a training file with more negatives per row gets more examples and more optimiser steps. The signal-matched control of Table~\ref{tab:ablations} removes that imbalance. The arms differ in one thing only, where their rows come from: $\mathcal{D}_{\mathrm{CAS}}$ holds pairs a computer algebra system certifies, and $\mathcal{D}_{\pi}$ holds pairs an LLM writes under a rewriter prompt $\pi$ and an LLM judge accepts; the trained models are $\theta_{\mathrm{CAS}}$ and $\theta_{\pi}$. The benchmark's own golds were produced the same way, under its published prompt $\pi^{\star}$. Writing $\Delta_{\mathcal{E}}(\pi) = \mathrm{R}@1(\theta_\pi;\mathcal{E}) - \mathrm{R}@1(\theta_{\mathrm{CAS}};\mathcal{E})$ for the lead of the LLM-trained model over the verified one on a set $\mathcal{E}$, the \emph{recipe-matching gain} $\Delta_{\mathrm{bench}}(\pi)$ is that lead on the benchmark tier under test. The paper's central claim is that the lead is larger on the benchmark than on real duplicates: $\Delta_{\mathrm{bench}}(\pi^{\star}) - \Delta_{\mathrm{real}}(\pi^{\star}) > 0$, where $\Delta_{\mathrm{real}}$ is the same lead on real duplicates, cross-language ones unless stated (Section~\ref{sec:ood}). Putting the reference $\theta_{\mathrm{ref}}$ in place of $\theta_{\mathrm{CAS}}$ in both terms isolates the share of each that is specific to the benchmark's own prompt (Section~\ref{sec:gaming}).

\paragraph{The verified arm (CAS), in detail.}
The verified arm's pairs are built from the public $27{,}817$-problem corpus with no LLM involved. That corpus is MathNet's own problem collection, the collection the benchmark's queries were drawn from: the \texttt{ShadenA/MathNet} dataset on Hugging Face (configuration \texttt{all}), problem and solution text with metadata and no images, $27{,}817$ records as downloaded on 2026-07-29 by \texttt{scripts/download\_corpus.py}; the same file serves as the search corpus of both duplicate evaluations below. A problem qualifies when a first pass finds at least one math span that parses as an equation or inequality and its id is not on the exclusion list of anchor-matched corpus problems as it stood before the correction of Appendix~\ref{app:contamination}: $9{,}012$ problems qualify, and a further $4{,}666$ parse but sit on that list. We attempted all $9{,}012$ rather than sampling, so the arm's sources follow the corpus's own mix among parseable problems: $3{,}289$ geometry, $3{,}210$ algebra, $1{,}340$ number theory, $1{,}009$ discrete mathematics and $164$ from the remaining domains. A \LaTeX{} normaliser first cleans each source so that SymPy can parse more of it, lifting its unique-expression coverage from $86.61\%$ to $92.29\%$. Under the full parse $1{,}230$ of the $9{,}012$ yield no usable relation, $61$ exceed the $30$-second limit per problem and $44$ crash the parser, leaving $7{,}677$ processed. From each parsed equation or inequality we make positives in three ways: renaming variables, done by substitution inside the source's own math spans and checked by parsing the result back and confirming the same structure; reformulating, by moving terms, expanding or scaling, checked by confirming that the difference between the original and the rewrite is zero at ${\geq}20$ random complex points; and mirroring, which swaps the two sides of an inequality and reverses its direction, checked by the same numeric test with the sign reversed. Negatives are minimal edits, a changed constant, exponent or operator, or a flipped inequality, and each is kept only when a numeric counterexample proves it is not equivalent. Reformulated and mirrored positives and all negatives are printed back into the problem by SymPy's \LaTeX{} printer, so a printing style shared by positives and negatives cannot tell them apart, and a printed relation is used only if it re-parses to exactly the relation it came from, since the printer drops information for some constructs. This yields $14{,}361$ verified positives and $9{,}408$ distinct counterexampled negatives from $7{,}397$ problems, $5{,}175$ of them with at least one negative; because one negative can be attached to several training records, there are $22{,}002$ negative attachments in all. A manual audit of $20$ sampled records ($10$ positives, $10$ negatives) found $20/20$ correct.

\paragraph{The recipe arm, in detail.}
The recipe arm's pairs come from the benchmark's own Appendix-F rewriter prompt, adapted near-verbatim with only the output format changed so that the LLM returns our fixed JSON layout. We ran it through Qwen3-32B-AWQ, on purpose a different generator from the benchmark's Gemini-3-flash, and kept or dropped each pair with a single LLM judge and no symbolic check. The benchmark itself is stricter, keeping a pair only when Gemini-3-flash and GPT-5 both judge it equivalent, so our filter admits more than the benchmark's does, and whether the pairs it admits are worse or merely different we cannot tell. The judge accepts $95.8\%$ of the positives the generator writes ($6{,}145$ of $6{,}414$) and $89.5\%$ of all generated flat rows, positives and near-misses together ($22{,}965$ of $25{,}656$), which grouped by source problem gives $6{,}145$ rows carrying $16{,}169$ negatives. Both arms draw on the same $7{,}089$ source problems, the verified arm's $7{,}397$ minus $308$ that belong to the real-duplicate sets, and both are capped at $6{,}145$ rows, the recipe arm's size after judging.

\paragraph{What the matched budget does and does not match.}
Both arms have $6{,}145$ rows, but they are not equal in three ways; the main text quotes only the headline $1.7\times$ figure. (i)~Negatives per row. The verified arm's file attaches $22{,}002$ negatives over $14{,}361$ records, $1.53$ per record ($1.52$ in the $6{,}145$-row draw that trains); the recipe arm's file attaches $16{,}169$ over $6{,}145$ rows, $2.63$ per row, a $1.7\times$ asymmetry in the recipe arm's favour. (ii)~Training volume. Because each negative becomes its own training example, the recipe arm's $6{,}145$ rows hold $16{,}248$ examples ($16{,}169$ negatives plus the $79$ rows without one); the $5\%$ dev split, drawn by source problem, holds out $307$ rows ($817$ examples) for checkpoint selection, so $5{,}838$ rows and $15{,}431$ examples train, over $177$ optimiser steps, against $9{,}921$ examples and $114$ steps for the verified arm (which holds out $308$ rows the same way), and they cover $6{,}145$ distinct source problems against $4{,}621$. Those step counts are the full training length; the checkpoint actually selected at seed $42$ had $100$ updates for the recipe arm and $114$ for the verified arm, so the raw step gap overstates the advantage conceded. (iii)~Seeds. Every recipe-arm seed trains on the identical $6{,}145$ rows, whereas each verified-arm seed draws its own $6{,}145$ from the $13{,}747$ rows that pass the gate (seeds $42$ and $43$ share $44\%$ of their rows, $4{,}621$ vs.\ $4{,}584$ sources). The verified arm's spread across seeds therefore includes variation from the data draw as well as from training, which widens the gap's interval rather than narrowing it.

\paragraph{Training-signal volume does not predict the score.}The prompt ladder gives one more check that the arm gap is not about how much training a model gets, and it works from both sides. From below: D4 trains on $5{,}838$ examples, fewer than either controlled arm, yet posts $30.03 \pm 1.45$ easy R@1 against the verified arm's $17.05 \pm 0.79$. From above: the uncapped verified arm has more of everything than the recipe arm, $13{,}747$ rows, $7{,}089$ source problems and, after its $5\%$ development split, $22{,}346$ contrastive examples, yet scores $15.92$, below even the capped verified arm. More training data of the verified kind does not raise the score, and less data of the LLM-written kind does not lower it.

\paragraph{The signal-matched verified arm: construction and residuals.}The second verified arm, built to match the recipe arm's training volume, was pre-registered in \texttt{scripts/e2\_negmatched.slurm} (verdict \texttt{scripts/e2\_verdict.py}; construction certificate \texttt{results/e2\_negmatched\_construction.json}). It has one row for each of the recipe arm's exact $6{,}145$ source problems, the same $16{,}169$ negative attachments ($2.63$ per row), and the recipe arm's exact dev split at every seed. To reach that many negatives without pulling in unrelated problems, each source's own counterexampled negatives were reused where needed. Totals and means match exactly; the per-row distribution cannot, for two reasons: $1{,}797$ of the source problems have no counterexampled negative at all, against $79$ zero-negative rows in the recipe arm, and a reused negative appears more than once, so under the batch sampler, which never places two examples of one row in the same batch, those rows are seen more often, which acts as oversampling.

\begin{table}[htbp]
\centering
\caption{Does the easy-tier gap survive when a possible explanation for it is removed? Each row removes one and reports the recipe arm's lead over the verified arm on the easy tier (first row: the headline comparison of Table~\ref{tab:controlled}). Rows marked ``intact'' change the verified arm only, so they report its score rather than a gap. No $0.6$B control brings the gap below $41$ points; the other-backbone and $4$B rows shrink it but keep its sign (Appendix~\ref{app:results}). The last two rows are the two recipe-arm variants of Appendix~\ref{app:mechanism}, each against the eight-seed verified arm.}
\label{tab:ablations}
\footnotesize
\setlength{\tabcolsep}{4pt}
\begin{tabular}{@{}p{7.3cm}p{3.1cm}>{\centering\arraybackslash}p{3.0cm}@{}}
\toprule
Control & Explanation it removes & Easy-tier gap (R@1) \\
\midrule
Matched budget, eight seeds (Table~\ref{tab:controlled}) & seed noise & $45.33 \pm 0.93$ \\
Queries the original gate kept out of both training files ($8{,}663$) & leaked test items & $46.37 \pm 0.85$ \\
No copy of the query in either arm's rows ($12{,}760$, strictest) & leaked test items & $44.73 \pm 0.86$ \\
Retrain under the corrected filter, $33\%$ fewer rows & near-verbatim copies of queries & $43.11 \pm 0.90$ \\
Retrain under the corrected filter, removed third regenerated from fresh sources & near-verbatim copies of queries & $44.45 \pm 0.87$ \\
Retrain under the corrected filter, removed third regenerated from the same sources & near-verbatim copies of queries & $44.38 \pm 1.46$ \\
Verified arm given the recipe arm's negative volume & more contrastive signal per row & intact ($16.57$ vs $17.05$, verified arm) \\
Uncapped verified arm ($2.2\times$ rows, more sources) & more training data & intact ($15.92$, verified arm) \\
e5-large and BGE-large base models (seed $42$) & the base model & $+19.12$ / $+15.91$ \\
$4$B base model, rank-$16$ LoRA (seed $42$) & model size & $77.77$ vs $39.51$ (base $11.95$) \\
Recipe arm kept to rows a second vendor's judge also admits ($5{,}398$; three seeds) & the single self-judge & $47.66$ ($64.71$ vs $17.05$) \\
Recipe arm with positives and near-misses written in separate calls (three seeds) & shared phrasing within a pair & $41.42$ ($58.47$ vs $17.05$) \\
\bottomrule
\end{tabular}
\end{table}

\paragraph{Signal matching leaves the verified arm's easy-tier score where it was.}Over seeds $42$--$44$ the signal-matched verified arm scores $16.57 \pm 1.07$ on the easy tier, $0.30 \pm 0.12$ on the hard tier and $50.13 \pm 4.09$ strict R@1 on cross-language duplicates. The easy-tier score sits inside the eight-seed verified arm's band, the hard-tier score a little above it ($0.16 \pm 0.06$) and the duplicate score a little below. Of the two outcomes registered in advance, the confirming one fired: giving the verified arm the recipe arm's negative volume and example count does not raise its easy-tier score, so the imbalance in training signal is not a first-order driver of the $45.33$-point gap.

\paragraph{Paraphrase depth separates the arms, does not order the LLM rungs, and does not carry the reference's gain.}
The verified arm's positives keep the source problem's wording while the LLM positives are free paraphrases, so the arm gap could in principle be about how much the text was changed rather than who changed it. We therefore measured paraphrase depth directly, as the overlap of three-character chunks (character-$3$-gram Jaccard similarity) between each row's source and its positive, where $1$ means identical text. The verified file's positives are close to their sources, $0.865$ over its $14{,}361$ rows (median $0.899$); the LLM files are far, $0.463$ for D1, $0.448$ for D2, $0.397$ for D3 and $0.421$ for D4 (whitespace-token figures in \texttt{results/paraphrase\_depth.json}). Depth thus separates the verified file from every LLM file, but within the LLM family it orders nothing on the easy tier: the four files sit in the narrow band $0.397$--$0.463$ while their scores run from $30.03$ to $67.88$, and D4's positives are slightly farther from their sources than D1's yet score less than half as much. On the hard tier D1--D3 do line up with depth, the closer paraphrase scoring higher, but those prompts write both the positive and the near-misses, so the factorial of Table~\ref{tab:factorial} cannot tell that reading from a difference in negative quality; D4 sits at the floor there because it has no negatives. The direct test is a paraphrase no LLM wrote: positives back-translated through the NLLB-200 translation model, at depth $0.783$, with the reference's negatives. That model scores $18.81$ easy R@1 against the reference's $46.91$ and the verified arm's $17.05$ (Appendix~\ref{app:mechanism}), so of the reference's lead over the verified arm (Table~\ref{tab:controlled}) depth accounts for $1.76$ $[0.62, 2.82]$ points and LLM authorship for the rest, down to that depth. We could not build a deeper paraphrase without an LLM, and the mined duplicate graph cannot supply one at the matched budget: its $661$ cross-language and $242$ monolingual clusters are the two retention probes themselves and hold far fewer pairs than the $6{,}145$-row budget, so any split of them would train on part of the probe.

\paragraph{Harness calibration: the checks that pass.}Before trusting our evaluation code we checked that it reproduces numbers the benchmark authors published. Our sentence-transformers~\citep{reimers2019sentencebert} reimplementation gives the same all-mpnet-base-v2 easy-tier result as their paper, $6.75/82.47/91.07$ R@1/5/10, matching the per-tier Table~4 row of their updated revision to the second decimal. It also reproduces their finding that near-misses almost always outrank the gold at the top of the list: on the hard tier only $11$ of $14{,}993$ Qwen3-Embedding-4B queries rank the gold above every near-miss. For the three large baselines, RaDeR, ReasonIR and MathLeap, we checked our encoding settings against each model's reference implementation.

\paragraph{The one MathNet-Retrieve calibration check that fails.}
For Qwen3-Embedding-4B the benchmark publishes $14.76$ easy R@1 and $36.87$ hard R@10; our best row is $11.96$ and $28.58$. Changing the instruction string and the input length did not close the gap, and truncation cannot be the cause, since exactly one of the $117{,}088$ corpus documents exceeds ${\sim}1{,}024$ tokens. We later ran a fourteen-setting grid (\texttt{scripts/calib4b.slurm}: bfloat16 and float16; no instruction, the model's own query prompt, or a math-retrieval instruction; $2{,}048$ or $8{,}192$ tokens; plus mean pooling and float32 at the model's own prompt; \texttt{results/calib4b\_summary.json}). It narrows the gap but does not close it: every setting with a math instruction scores $13.15$ to $13.19$ easy R@1, the closest being bfloat16 at $8{,}192$ tokens ($13.19$, still $1.57$ below $14.76$); the model's own query prompt gives $11.89$ to $11.95$, no instruction $11.10$ to $11.18$, float32 exactly matches bfloat16, and mean pooling collapses to $4.97$. We had registered that a setting would replace the one in the baseline tables only if it came within $0.5$ R@1 of $14.76$; none does, so the tables keep their existing settings and the discrepancy stays unresolved. As a result, three numbers for the untrained $4$B baseline appear in this paper, and each place names the one it uses: prompted at $2{,}048$ tokens ($11.96$, the best row, quoted in Section~\ref{sec:setup}); prompted at $1{,}024$ tokens ($11.95$, the base for the $4$B replication of Section~\ref{sec:gaming} and Appendix~\ref{app:results}, whose evaluation runs at $1{,}024$ tokens); and unprompted at $2{,}048$ tokens ($11.15$, Table~\ref{tab:leaderboard-full}).

\paragraph{MathNet-Retrieve protocol: full corpus, native baseline conventions.}Every MathNet-Retrieve number ranks the full $117{,}088$-document corpus for all $15{,}000$ queries of a tier; nothing is subsampled. Our own models encode in bfloat16 with inputs cut at $1{,}024$ tokens. V2 is evaluated with the instruction prompt it was trained with. Baselines are run the way their authors run them, with one documented exception: the MathLeap models ship with inputs cut at $128$ tokens, too short for these problems, so we evaluate them at $2{,}048$ tokens.

\paragraph{SABER: the strongest calibration case.}For the three external benchmarks, SABER-Math~\citep{georgiev2026saber}, MELD~\citep{ye2026meld} and the BRIGHT theorem splits~\citep{su2025bright}, we state separately how closely our evaluation matches the official one. SABER matches most tightly: our nDCG computation gives bit-identical output to SABER's own metric code on $500$ random cases, and our scores for three reference models, e5-large, bge-m3 and the untrained base, reproduce the published numbers to three decimals ($0.4879$ vs.\ $0.488$; $0.5114$ vs.\ $0.511$; $0.5747$ vs.\ $0.575$).

\paragraph{BRIGHT: comparable across our models, not to the leaderboard.}On BRIGHT we follow the official scoring procedure, dropping the excluded ids, keeping the top $1{,}000$ results and reporting nDCG@10, with one deviation that applies to every model we run: inputs are cut at $1{,}024$ tokens, where the official setup lets each model use its own maximum, up to $8{,}192$. Our BRIGHT numbers can therefore be compared with one another but not with the public leaderboard, and every output file records this with the flag \texttt{comparable\_to\_official=false}.

\paragraph{MELD: reimplemented from its paper, and it does not calibrate.}
MELD has no official evaluation code, so we reimplemented the protocol its paper describes: retrieval over the pairs alone, plus an AUC for how well a model separates true pairs from non-pairs, with the AUC code checked against scikit-learn to $10^{-12}$. That reimplementation does not reproduce MELD's published numbers. For the two models MELD publishes that we also run, we score higher than they do: Qwen3-Embedding-4B $18.89$ against their $13.7$ ($63.15$ vs.\ $58.5$ at R@10, mean gold rank $13.02$ vs.\ $21.4$) and MathLeap-Octen-8B $34.44$ against their $28.9$. MELD does not state the prompt it evaluates with, so we cannot tell how much of that gap is our harness and how much is the prompt. As on BRIGHT, inputs are cut at $1{,}024$ tokens. Our MELD numbers are therefore comparable among our own models only, and we make no claim against MELD's published table.

\paragraph{ImpliRet: the calibration check fails.}On ImpliRet~\citep{taghavi2025impliret} we follow the official procedure: six subsets, each a category paired with a style, each of $1{,}500$ queries, every query ranked against its own subset's $1{,}500$ documents, scored by nDCG@10 per subset. Its one calibration check fails: our ReasonIR-8B row is $19.45$ where the benchmark's authors report $13.64$, further apart than the tolerance of $2.0$ we fixed in code before running. The likely cause is a settings difference, since the official ReasonIR run encodes queries with no instruction while we use the one ReasonIR's authors recommend. As on BRIGHT and MELD, our ImpliRet numbers are comparable among our own models only, and we make no claim against the published table.

\paragraph{Real-duplicate evaluation set, in detail.}
To measure real rather than generated equivalence, we search the public corpus for the same competition problem printed in several national olympiad booklets in different languages. Three signals find such pairs: rare formulas shared as $4$-token chunks ($519$ of the $754$ pairs), exact and near-exact text matches, and matches corroborated by a shared answer. This yields $754$ verified pairs. Hand-reading an unbiased random sample of $20$ accepted pairs found $17$ correct; the miner's last two fixes then removed two of the three false positives, leaving $17$ of the $18$ surviving sampled pairs correct, so we quote precision as $85$--$90\%$ (the mining thresholds were tuned separately on about $80$ hand-labelled candidate pairs over three rounds). Chaining the pairs (if A matches B and B matches C, all three are one problem) gives $661$ clusters over $1{,}377$ problems; their $393$ non-English members, in $14$ languages, become the queries, searched against the full $27{,}817$-document corpus, with $457$ correct answers in all. A hit counts only if the retrieved gold is the same problem in a different language from the query, and the query's own corpus entry is always removed from the ranking so that a model cannot score by finding itself.

\paragraph{Three caveats.}First, the mining favours problems that share formulas, and shared rare formulas are exactly the signal a formula-matching embedder exploits; a strong multilingual embedder already scores $91.60$ strict R@1 on the set with no training (Qwen3-Embedding-4B). Scores above the set's $85$--$90\%$ hand-read precision, such as D4's $93.81$, mean that the mislabelled pairs are retrieved too, presumably through the shared formulas that mined them, so part of any score on this set is formula matching; the slices below bound that part for the two arms. Second, the set has no constructed near-miss distractors, so it cannot test whether a model puts a true equivalent above a minimal edit at rank~$1$, which is what the hard tier targets. Third, no prompt generated it, so there is no recipe a model could match. For these reasons we use it to measure whether training has destroyed real retrieval, not as a headline benchmark, and the inversion of Appendix~\ref{sec:forgetting} is a statement about benchmark score against retained ability.

\paragraph{Both controlled arms trail the base on every slice of the real-duplicate set.}
Since the mining found these duplicates by matching formulas, the trained arms' losses on the set could be an artefact of an evaluation that favours formula matchers. To check, we split the $393$ queries three ways. By how the pair was found: answer-corroborated if any pair supporting the query's cluster shared an answer, text-mined if any was a text match, formula-only otherwise. By how much rare-formula content the query and its gold share, weighted by inverse document frequency and cut at its quartiles; the miner's own containment ratio could not be used here, since it saturates at $1.0$ for $316$ of the $393$ queries, so the cut uses the absolute shared mass. And by how formula-heavy the query text is, cut at its median. Inside every slice we recompute each arm's gap to the untrained base with a nested bootstrap over queries and seeds ($B = 2{,}000$, queries resampled individually; Table~\ref{tab:xling-slices}, \texttt{results/xling\_slice\_analysis.json}).

Both controlled arms fall below the base on every slice, including the $25$ answer-corroborated queries (all but one of which the formula miner never matched) and the bottom quartile of shared rare-formula mass; the interval excludes zero on every slice for the verified arm, and on every slice but the $15$ text-mined queries for the recipe arm. If the losses were an artefact of formula matching, the recipe arm would lose less where there are fewer formulas to match. It loses more: $-20.45$ on the bottom quartile of shared formula mass against $-12.12$ on the top, $-20.75$ on formula-light against $-16.20$ on formula-heavy queries, and about the same ($-18.51$ against $-18.47$) on the saturated containment split. D4, with no recipe wording and no negatives, falls below the base on no slice and exceeds it by most on the bottom quartile ($+15.15$ $[6.06, 23.23]$), the reverse of what an evaluation that merely rewarded formula matching would show. One slice does reorder the two arms: on the $25$ answer-corroborated queries the verified arm scores $60.00$ against the recipe arm's $52.50$, with wide and heavily overlapping intervals, so the recipe arm's cross-lingual lead of Section~\ref{sec:ood} holds for the full set, not for every slice of it.

\begin{table}[htbp]
\centering
\caption{Cross-language retention by query slice. Each row is a subset of the $393$ cross-language queries; Base is the untrained model's strict R@1 on that subset, and the three gap columns give how far the verified arm, the recipe arm and D4 sit from it, with $95\%$ nested bootstrap intervals over queries and training seeds ($B = 2{,}000$; the arms eight seeds, D4 three). Rows two to four split the queries by how the pair was found, the two quartile rows keep the queries with the least and the most shared rare-formula content, and the last two split by how formula-heavy the query text is. On no slice does either arm reach the base; the recipe arm loses more where formulas are scarce (bottom quartile, formula-light) than where they are plentiful, the reverse of a formula-matching artefact; and D4 falls below the base on no slice.}
\label{tab:xling-slices}
\footnotesize
\setlength{\tabcolsep}{3pt}
\resizebox{\linewidth}{!}{%
\begin{tabular}{lrcccc}
\toprule
Query slice & $n$ & Base & Verified $-$ base & Recipe $-$ base & D4 $-$ base \\
\midrule
All queries & 393 & 84.73 & $-28.21$ $[-33.21, -23.15]$ & $-18.48$ $[-23.47, -13.68]$ & $+9.08$ $[5.85, 12.55]$ \\
Pair found through a shared answer & 25 & 88.00 & $-28.00$ $[-45.00, -12.50]$ & $-35.50$ $[-55.50, -16.00]$ & $0.00$ $[-12.00, 12.00]$ \\
Pair found through matching text & 15 & 86.67 & $-18.33$ $[-37.50, -3.33]$ & $-8.33$ $[-23.33, 0.00]$ & $0.00$ $[0.00, 0.00]$ \\
Pair found through formulas only & 353 & 84.42 & $-28.65$ $[-33.50, -23.16]$ & $-17.71$ $[-22.59, -12.96]$ & $+10.10$ $[6.42, 13.69]$ \\
Least shared formula content (bottom quartile) & 99 & 76.77 & $-29.42$ $[-38.89, -20.20]$ & $-20.45$ $[-31.19, -9.47]$ & $+15.15$ $[6.06, 23.23]$ \\
Most shared formula content (top quartile) & 99 & 93.94 & $-23.36$ $[-31.69, -15.40]$ & $-12.12$ $[-19.44, -5.18]$ & $+3.03$ $[-1.01, 7.41]$ \\
Formula-light query text (at or below median) & 197 & 78.68 & $-28.11$ $[-35.34, -21.45]$ & $-20.75$ $[-27.54, -13.90]$ & $+12.52$ $[7.28, 17.77]$ \\
Formula-heavy query text (above median) & 196 & 90.82 & $-28.32$ $[-34.95, -22.07]$ & $-16.20$ $[-22.58, -9.82]$ & $+5.61$ $[1.87, 9.35]$ \\
\bottomrule
\end{tabular}}
\end{table}

\paragraph{A same-language duplicate set addresses the cross-lingual confound.}
The cross-lingual set could in principle punish any model fine-tuned on one language, recipe-like or not. The mined duplicate graph also holds $242$ clusters whose members share a language, which the cross-lingual set skipped, and we built a second set from them (\texttt{scripts/build\_samelang\_eval.py}, \texttt{data/samelang\_eval}): every member of such a cluster whose language is known queries the same $27{,}817$-problem corpus, with its own entry masked out, for its same-language reprints ($498$ queries). The clusters come from the same miner whose hand-read cross-lingual sample put precision at $85$--$90\%$; the same-language slices were not separately hand-checked, and like the cross-lingual set they carry no constructed near-misses. The set is split by two flags. First, $143$ clusters were found by the exact-text rule, so their reprints are nearly byte-identical and any model should retrieve them; we report them only as a ceiling check. Second, contamination: the trainer's evaluation gate removed only the $779$ ids that are queries or golds of the cross-lingual set, so members of same-language clusters could have been trained on. Of the $99$ non-exact clusters, $61$ have no member in any of our training pair files, and their $125$ queries ($115$ English) form the primary slice; the $38$ leaked clusters ($82$ queries) are reported separately. The reading was registered before any number existed (\texttt{scripts/samelang\_eval.slurm}, verdict in \texttt{scripts/samelang\_verdict.py}). The test statistic is the same-language difference-in-differences, the easy-tier arm gap minus the primary-slice arm gap: it passes at or above $22.67$ with an interval excluding zero, is partial between zero and that bar or whenever the interval includes zero, and refutes the easy-tier reading at or below zero. Every model of ours with a cross-lingual result was scored under the encoding convention recorded in that result (Table~\ref{tab:samelang}).

\begin{table}[htbp]
\centering
\caption{Same-language reprints: R@1 for retrieving a reprint of the query in the same language, for every model we trained. The four score columns are the four slices: the primary slice ($125$ queries from reworded clusters that no training file touches), its English queries, the leaked slice (queries from reworded clusters that some training file does touch), and the exact reprints (near-identical copies that any model should retrieve, a ceiling check). The last column is each model's gap to the untrained base on the primary slice, with a $95\%$ nested bootstrap interval over clusters and seeds. Every model trained with negatives of any kind sits well below the base; only the models trained without negatives, the back-translated model, V2 and the MELD arm stay within noise of it. Seeds: the two arms and the reference eight; the ladder rungs, CAS positives without negatives, the other factorial cells, V2 and the MELD arm three; the rest single runs.}
\label{tab:samelang}
\scriptsize
\setlength{\tabcolsep}{2pt}
\begin{tabular}{@{}p{3.7cm}cccc>{\centering\arraybackslash}p{3.1cm}@{}}
\toprule
Model & Primary ($125$) & English ($115$) & Leaked ($82$) & Exact reprints ($291$) & Primary gap to base \\
\midrule
Base $0.6$B (untrained) & 62.40 & 59.13 & 50.00 & 100.00 & \\
Base $+$ V2 instruction & 62.40 & 59.13 & 52.44 & 100.00 & $0.00$ $[-3.23, 3.17]$ \\
BM25 (lexical baseline) & 54.40 & 51.30 & 32.93 & 100.00 & $-8.00$ $[-17.07, 0.79]$ \\
Verified arm ($8$ seeds) & 51.30 & 47.07 & 34.91 & 99.40 & $-11.10$ $[-18.90, -3.89]$ \\
Recipe arm ($8$ seeds) & 46.30 & 41.63 & 35.82 & 85.61 & $-16.10$ $[-26.37, -6.10]$ \\
D4 (recipe-free, no negatives) & 61.07 & 57.68 & 44.72 & 100.00 & $-1.33$ $[-7.20, 3.73]$ \\
CAS positives, no negatives & 58.13 & 54.49 & 33.33 & 100.00 & $-4.27$ $[-11.38, 1.63]$ \\
Reference (D4 $+$ CAS negatives, $8$ seeds) & 39.20 & 40.33 & 21.80 & 63.49 & $-23.20$ $[-33.30, -12.80]$ \\
D4 $+$ LLM negatives (recipe prompt), count-matched & 47.73 & 43.77 & 35.77 & 90.15 & $-14.67$ $[-24.80, -5.25]$ \\
D4 $+$ LLM negatives (recipe prompt), full & 39.47 & 35.07 & 30.89 & 78.58 & $-22.93$ $[-34.42, -11.29]$ \\
D4 $+$ LLM negatives (unrelated prompt), count-matched & 50.13 & 45.80 & 36.59 & 94.96 & $-12.27$ $[-21.07, -4.03]$ \\
D4 $+$ LLM negatives (unrelated prompt), full & 45.33 & 40.58 & 32.11 & 88.77 & $-17.07$ $[-27.73, -7.14]$ \\
CAS positives $+$ recipe LLM negatives & 34.40 & 30.43 & 24.39 & 91.75 & $-28.00$ $[-37.37, -18.93]$ \\
Back-translated positives $+$ CAS negatives & 57.87 & 54.20 & 46.75 & 100.00 & $-4.53$ $[-14.09, 5.07]$ \\
Three-hop back-translated positives $+$ CAS negatives & 46.67 & 42.61 & 43.50 & 91.07 & $-15.73$ $[-26.34, -5.42]$ \\
D5 (second recipe-free prompt, no negatives) & 64.53 & 61.45 & 47.56 & 100.00 & $+2.13$ $[-2.98, 6.78]$ \\
D5 $+$ CAS negatives (second reference) & 41.33 & 42.61 & 24.80 & 68.96 & $-21.07$ $[-31.47, -10.84]$ \\
Recipe arm, positives and near-misses in separate calls & 36.53 & 32.46 & 30.89 & 62.43 & $-25.87$ $[-37.87, -13.87]$ \\
Recipe arm, both judges ($5{,}398$ rows) & 46.13 & 41.45 & 39.02 & 89.92 & $-16.27$ $[-26.40, -6.04]$ \\
CAS positives $+$ unrelated-prompt near-misses & 32.00 & 27.83 & 21.95 & 87.74 & $-30.40$ $[-40.21, -21.51]$ \\
Back-translated positives, no negatives & 63.47 & 60.29 & 49.19 & 100.00 & $+1.07$ $[-5.60, 7.38]$ \\
Back-translated positives $+$ unrelated-prompt near-misses & 47.47 & 44.64 & 34.15 & 94.85 & $-14.93$ $[-24.07, -5.69]$ \\
Back-translated positives $+$ recipe-prompt near-misses & 46.93 & 44.06 & 38.62 & 98.17 & $-15.47$ $[-24.39, -6.99]$ \\
D1 positives, no negatives & 64.53 & 61.45 & 49.19 & 100.00 & $+2.13$ $[-0.81, 5.65]$ \\
D1 positives $+$ CAS negatives & 38.13 & 39.71 & 20.73 & 62.66 & $-24.27$ $[-34.91, -13.55]$ \\
D1 positives $+$ unrelated-prompt near-misses & 40.53 & 35.36 & 29.67 & 71.71 & $-21.87$ $[-32.55, -11.92]$ \\
V2 (mixed data $+$ instruction) & 61.87 & 58.55 & 42.68 & 100.00 & $-0.53$ $[-4.57, 3.20]$ \\
MELD-recipe arm & 60.80 & 57.39 & 42.28 & 100.00 & $-1.60$ $[-6.08, 2.46]$ \\
\bottomrule
\end{tabular}
\end{table}

\paragraph{The easy-tier lead does not reach same-language duplicates.}
On the primary slice the recipe arm scores $46.30$ against the verified arm's $51.30$, a gap of $-5.00$ $[-13.42, 3.35]$ whose interval includes zero, so there is no measurable difference between the arms. Subtracting it from the easy-tier gap gives a same-language difference-in-differences of $50.33$ $[41.85, 58.75]$, far above the registered bar, so the reading passes; on the English queries alone the gap is $-5.44$ $[-14.47, 3.48]$. This slice is harder for the untrained base than the cross-lingual one, $62.40$ against $84.73$, because these are reprints reworded within one language rather than translations. Several models stay at the base here: D4 ($-1.33$), CAS positives without negatives ($-4.27$), V2 and the MELD arm. That settles one earlier puzzle: on cross-language duplicates D4 and CAS positives alone were far apart ($93.81$ against $52.08$), and on same-language data they are not, so that cross-lingual gap was the CAS model forgetting how to work across languages, not a difference in recognising equivalence. Attaching negatives to D4's restatements always costs same-language retention: D4's $61.07$ falls to $39.20$ with the verified arm's counterexamples, to $47.73$ with the recipe's near-misses and to $50.13$ with an unrelated prompt's. Yet the same counterexamples attached to back-translated positives leave that model at the base ($57.87$, $-4.53$ $[-14.09, 5.07]$), so the cost depends on the positives as well as the negatives. The later cells fit the pattern: D5 alone sits at the base ($64.53$, $+2.13$ $[-2.98, 6.78]$), D5 with counterexamples next to the D4 reference ($41.33$ against $39.20$), the three-hop back-translated model at $46.67$, the two-judge recipe arm at the recipe arm's level ($46.13$), and the separate-call recipe arm lowest of the recipe variants ($36.53$). Finally, the reference trails even the verified arm here, so its real-data advantage over the recipe arm is a cross-lingual result: on this slice the two are within noise of each other ($+7.10$ $[-2.34, 17.04]$ in the recipe arm's favour), and the difference-in-differences between them falls from $27.24$ cross-lingually to $8.37$ $[-1.14, 18.16]$.

\paragraph{The exact-text ceiling check: LLM-written near-misses cost a share of near-identical reprints.}
The $291$ exact-text queries each have a near-identical copy in the corpus, so any working retriever should find every one, and the untrained base does. The check shows which training data breaks even that. Every model trained with LLM-written near-misses misses some, whatever its positives: the recipe arm $85.61$, the D4 cells with the recipe's near-misses $90.15$ and $78.58$, the cells with an unrelated prompt's near-misses $94.96$ and $88.77$, and CAS positives with the recipe's near-misses $91.75$. Computer-algebra counterexamples do the same damage only beside D4's restatements, where the reference retrieves $63.49$; beside near-copy or back-translated positives they cost little or nothing (the verified arm $99.40$, back-translated positives $100.00$). The models with no negatives (D4 and CAS positives alone), V2, whose counterexamples are diluted by problem-to-solution replay, and the MELD arm retrieve all $291$.

\paragraph{What the same-language slice can and cannot detect.}
With $125$ queries in $61$ clusters the primary slice is a blunt instrument. From the width of its arm-gap intervals, $[-13.42, 3.35]$ against the verified arm and $[-2.34, 17.04]$ against the reference, the standard errors are about $4.3$ and $4.9$ points, so the smallest true arm difference the slice would reliably detect, at $80\%$ power, is about $12$ to $14$ points ($8.4$ to $9.7$ at $50\%$). The $7.10$-point recipe-minus-reference gap is below all of those, so ``no measurable difference'' here means too small for this test to see, not zero. The $82$-query leaked slice, whose clusters some training file touched, does not change the reading: recipe minus reference is $+14.0$ there, the same sign as the primary slice's $+7.10$, and recipe minus verified $+0.9$ (Table~\ref{tab:samelang}). No model we trained beats the untrained base on the primary slice beyond noise. The closest is SABER's summary arm at $64.80$, $2.4$ ahead with an interval touching zero ($[0.00, 5.51]$); the mixed-data models, the soups (the $0.7$ soup at $-7.20$ $[-15.08, 0.00]$, touching zero), the MELD arms and the back-translated control ($57.87$, $-4.53$ $[-14.09, 5.07]$) all sit within noise of it.

\paragraph{A generator-free near-miss probe: recipe-matching buys some near-miss discrimination, and the hard tier's score does not order it.}
Neither duplicate probe has constructed near-misses, so we built a near-miss test that no generator wrote, with the same machinery that makes the verified arm's pairs (\texttt{scripts/build\_cas\_probe.py}). Starting from benchmark query problems none of whose corpus copies is a source of any training file or a member of either duplicate cluster set ($4{,}759$ eligible, $2{,}000$ attempted), each item gets a computer-algebra-verified positive and up to three minimal-edit negatives, each refuted by a numeric counterexample, with no LLM anywhere: $1{,}107$ items ($901$ rename, $206$ reformulate; $1.885$ negatives per item; $633$ English-tagged, $78$ tagged with another language or a bilingual pair (Spanish, Russian, Chinese, Mongolian, Arabic and German among them) and $396$ untagged). The held-out check scanned the pair files of the arms and factorial cells; one of the $1{,}107$ problems also appears as a replay row in the mixed training set of V1 and V2, so those two models may have seen it. Every model we trained is scored two ways under its recorded encoding convention (\texttt{scripts/casprobe\_eval\_driver.py}): ranking the full public corpus plus every probe document, which is the hard tier's shape without a generator, and ranking the probe documents alone, a pure near-miss test (Table~\ref{tab:casprobe}). Five models are in-family, because the probe uses their own transformation families: the verified arm, V2 and the three cells trained on CAS positives. Their scores are disclosed rather than read. Before reading any comparison we registered (R14-a to R14-d) a noise band from the verified arm's eight seeds on the probe, $\sigma_{\mathrm{probe}} = \sqrt{2}$ times their sd $= 0.46$, band $0.91$: the recipe arm within the band of the reference would be consistent with ``buyable, not yet shown to be measured'', above it ``a measured equivalence gain; the hard-tier claim narrows''. Outcomes: the recipe arm scores $60.49 \pm 2.50$ R@1 on the full corpus and $68.38 \pm 2.40$ on the near-miss ranking, the reference $53.51 \pm 3.44$ and $65.42 \pm 2.22$ (eight seeds each), the base $10.84$ and $10.93$. The other two rungs of the ladder, scored by the same driver over three seeds, land above the recipe arm: D2 at $62.54 \pm 1.23$ and $69.35 \pm 1.99$, D3 at $68.50 \pm 1.38$ and $72.30 \pm 2.21$, so what the recipe family buys here is not specific to the benchmark's template, and a prompt in a different style buys more of it. The recipe arm leads the reference by $6.98$ and $2.96$ and the base by $49.65$ and $57.45$, above the band on every registered contrast: recipe-matching does buy generator-free near-miss discrimination beyond what the recipe-free reference learns, and we concede it in Appendix~\ref{sec:forgetting} and Section~\ref{sec:limitations}. The probe also shows that hard-tier score does not track that discrimination. The two cells with the highest hard-tier scores, D4's restatements with the recipe's near-misses at full and at matched volume ($26.97$ and $20.71$ hard R@1), score $30.50$ and $38.06$ on the probe, below D4 alone ($42.31$ at $0.04$ hard R@1) and the reference ($53.51$ at $10.30$), while back-translated positives with verified negatives ($1.09$ hard R@1) reach $74.79$. Among D4's restatements with negatives, more hard-tier score comes with less generator-free discrimination (from $53.51$ at $10.30$ to $30.50$ at $26.97$), and across the factorial the score does not order it (Spearman's $\rho$ between hard-tier R@1 and probe R@1 over the $22$ out-of-family cells of Table~\ref{tab:casprobe} is $+0.06$ on the full corpus and $+0.19$ on the near-miss ranking), so a high hard-tier score has not been shown to measure the discrimination the tier was built to test. An exploratory link to the sole wins (R14-e): $89$ probe queries are among the $1{,}401$ hard-tier queries only the recipe arm gets right; on them its probe R@1 is $56.18$ against $60.87$ on the other $1{,}018$ (eight-seed means; the reference $49.30$ against $53.88$), so the queries it alone wins on the hard tier are not ones where it discriminates better without a generator. Two limits: the probe's edits are the verified arm's families, rename and reformulation, shallower than the benchmark's deep disguises, so it measures near-miss discrimination at that depth and not at the hard tier's; and every cell but the two arms and the reference carries three seeds.

\begin{table}[htbp]
\centering
\caption{The generator-free near-miss probe: $1{,}107$ held-out benchmark queries, each with a computer-algebra-verified positive and counterexampled minimal-edit negatives. Each model's benchmark hard-tier R@1 is set beside its probe R@1 scored two ways: against the full public corpus plus the probe documents (the hard tier's shape without a generator) and against the probe documents alone (a pure near-miss test). Means $\pm$ sample std over seeds where more than one: the arms and the reference eight, the other cells three, the base one. Rows marked $\dagger$ are in-family, since the probe uses their own transformation families (the verified arm, V2 and the three CAS-positive cells); their scores are shown but not read as evidence. Among the other rows, the two with the highest hard-tier scores have the lowest probe scores of D4's cells, while back-translated positives, near the hard-tier floor, score high on the probe.}
\label{tab:casprobe}
\footnotesize
\setlength{\tabcolsep}{3pt}
\begin{tabular}{@{}p{5.6cm}>{\centering\arraybackslash}p{1.9cm}>{\centering\arraybackslash}p{2.6cm}>{\centering\arraybackslash}p{2.6cm}@{}}
\toprule
Model & Benchmark hard-tier R@1 & Probe R@1, full corpus & Probe R@1, near-misses only \\
\midrule
Base $0.6$B (untrained) & 0.00 & 10.84 & 10.93 \\
BM25 (lexical baseline) & 0.01 & 72.27 & 73.8 \\
Verified arm ($8$ seeds)$^\dagger$ & 0.16 & 97.42 $\pm$ 0.32 & 97.53 $\pm$ 0.33 \\
Recipe arm ($8$ seeds) & 9.42 & 60.49 $\pm$ 2.50 & 68.38 $\pm$ 2.40 \\
D2 arm (paraphrase of the benchmark's prompt) & 6.76 & 62.54 $\pm$ 1.23 & 69.35 $\pm$ 1.99 \\
D3 arm (different-style prompt) & 4.67 & 68.50 $\pm$ 1.38 & 72.30 $\pm$ 2.21 \\
D4 (recipe-free, no negatives) & 0.04 & 42.31 $\pm$ 2.64 & 42.40 $\pm$ 2.72 \\
CAS positives, no negatives$^\dagger$ & 0.00 & 45.26 $\pm$ 1.10 & 45.26 $\pm$ 1.10 \\
Reference (D4 $+$ CAS negatives, $8$ seeds) & 10.30 & 53.51 $\pm$ 3.44 & 65.42 $\pm$ 2.22 \\
D4 $+$ LLM negatives (recipe prompt), count-matched & 20.71 & 38.06 $\pm$ 1.77 & 44.90 $\pm$ 2.04 \\
D4 $+$ LLM negatives (recipe prompt), full & 26.97 & 30.50 $\pm$ 2.37 & 40.83 $\pm$ 1.59 \\
D4 $+$ LLM negatives (unrelated prompt), count-matched & 16.96 & 39.54 $\pm$ 2.63 & 45.05 $\pm$ 1.86 \\
D4 $+$ LLM negatives (unrelated prompt), full & 20.93 & 39.21 $\pm$ 0.24 & 47.88 $\pm$ 0.24 \\
CAS positives $+$ recipe LLM negatives$^\dagger$ & 0.11 & 85.91 $\pm$ 0.86 & 85.97 $\pm$ 0.91 \\
Back-translated positives $+$ CAS negatives & 1.09 & 74.79 $\pm$ 1.34 & 80.34 $\pm$ 0.77 \\
Three-hop back-translated positives $+$ CAS negatives & 4.47 & 52.06 $\pm$ 1.65 & 62.51 $\pm$ 1.57 \\
D5 (second recipe-free prompt, no negatives) & 0.01 & 24.60 $\pm$ 1.23 & 24.93 $\pm$ 1.18 \\
D5 $+$ CAS negatives (second reference) & 3.97 & 55.07 $\pm$ 2.53 & 64.77 $\pm$ 1.66 \\
Recipe arm, positives and near-misses in separate calls & 9.79 & 39.26 $\pm$ 3.81 & 52.42 $\pm$ 3.75 \\
Recipe arm, both judges ($5{,}398$ rows) & 7.58 & 58.96 $\pm$ 2.15 & 66.49 $\pm$ 2.22 \\
CAS positives $+$ unrelated-prompt near-misses$^\dagger$ & 0.23 & 86.45 $\pm$ 1.34 & 86.45 $\pm$ 1.34 \\
Back-translated positives, no negatives & 0.05 & 18.34 $\pm$ 0.27 & 18.61 $\pm$ 0.27 \\
Back-translated positives $+$ unrelated-prompt near-misses & 1.88 & 35.65 $\pm$ 2.81 & 39.39 $\pm$ 2.01 \\
Back-translated positives $+$ recipe-prompt near-misses & 4.18 & 42.22 $\pm$ 1.52 & 45.32 $\pm$ 1.13 \\
D1 positives, no negatives & 0.00 & 49.32 $\pm$ 2.22 & 49.32 $\pm$ 2.22 \\
D1 positives $+$ CAS negatives & 2.09 & 65.22 $\pm$ 0.54 & 74.16 $\pm$ 0.77 \\
D1 positives $+$ unrelated-prompt near-misses & 3.45 & 53.15 $\pm$ 1.77 & 62.90 $\pm$ 1.79 \\
V2 (mixed data $+$ instruction)$^\dagger$ & 0.01 & 85.42 $\pm$ 1.47 & 85.42 $\pm$ 1.47 \\
MELD-recipe arm & 0.11 & 24.63 $\pm$ 0.45 & 24.66 $\pm$ 0.50 \\
\bottomrule
\end{tabular}
\end{table}

\paragraph{Statistical reporting.}Unless noted otherwise, a gap between two arms is reported as a point estimate with a $95\%$ confidence interval from a nested bootstrap of $B = 5{,}000$ replicates, each of which redraws the queries with replacement and also redraws the eight training seeds, so the interval covers both query-sampling noise and training noise. Summaries across seeds are quoted as mean $\pm$ sample standard deviation. When two models exist as single runs, with no seeds to resample, we use a paired bootstrap over queries. Whenever we subtract a real-duplicate gap from a benchmark gap, the subtraction is done inside each bootstrap replicate, so the difference-in-differences (DiD) gets its own interval.

\paragraph{The cluster bootstrap governs the main text's real-duplicate intervals.}
Several duplicate queries can come from one cluster, the same problem in several languages, so they are not independent. Every cross-language real-duplicate interval in the main text therefore resamples clusters rather than queries: it draws clusters with replacement from the $370$ clusters behind the $393$ queries, then redraws queries within each drawn cluster. The same-language intervals of Tables~\ref{tab:controlled} and~\ref{tab:samelang} resample the $61$ primary-slice clusters the same way. For the eight-seed arm gap ($9.74$ $[3.96, 15.71]$) and difference-in-differences ($35.59$ $[29.51, 41.47]$), the cluster resampling is nested inside the seed resampling. Two exceptions: the per-slice intervals of Table~\ref{tab:xling-slices} resample queries individually, and the retrain rows of Table~\ref{tab:cleangate} are eight-seed means with no query resampling.

\paragraph{Families of tests, and what is corrected.}
We apply no correction for multiple testing, and the results should be read with that in mind. The tests fall into two groups. The confirmatory group is the predictions written down before the data existed, each with its threshold and its falsifier: the prompt ladder, the SABER attack, the MELD attack, the same-language readout and the factorial's reading map. These are reported as pass or fail against the pre-set bar. Everything else is exploratory: the eighteen rung${\times}$tier verdicts, the sixty-six evaluation bootstraps, the fifteen MELD paired intervals, the five contamination slices, the real-duplicate query slices and the hit-accounting tests. Their $p$-values are uncorrected and describe the data rather than establish significance. Two cautions: the easy-tier D1$-$D3 comparison (Appendix~\ref{app:mechanism}) skips a rung and was not part of the registered ladder, so it is post hoc; and the real-duplicate column of Table~\ref{tab:dose-full} is reported for completeness, not as a ranking of the rungs.

\paragraph{Which estimator each $\pm$ carries.}
Not every $\pm$ measures the same thing. When a training file holds exactly the $6{,}145$-row budget, every seed trains on the same rows and the spread across seeds is pure training randomness; when the file is larger, each seed draws a different $6{,}145$ rows and the spread also carries variation from the draw. The recipe arm is the first kind (its file holds exactly $6{,}145$ rows) and the verified arm the second (each seed draws $6{,}145$ of $13{,}747$ gated rows). D1 and D3 train their whole files, so their three-seed spreads are pure training randomness, whereas D2, D4 and the eight-seed reference draw $6{,}145$ of $6{,}553$, $6{,}768$ and $6{,}768$ rows per seed and carry a small data-draw component. Among the factorial cells, the two with the recipe's near-misses ($5{,}978$ rows) and the CAS-positive cell with the recipe's near-misses ($6{,}145$) train their whole files, while the two unrelated-prompt cells ($6{,}393$ rows) and the back-translated cell ($6{,}576$) draw per seed and carry the same small component. Setting a three-seed rung's spread beside the eight-seed verified arm's therefore compares unlike quantities; where we do so, as with D4's $30.03 \pm 1.45$ against the verified arm's $17.05 \pm 0.79$, the difference between the means is far larger than either spread, so the mismatch does not affect the reading.

\paragraph{Evaluation-set, scale and supervision caveats.}
Further caveats. The real-duplicate set shares $220$ source problems with MathNet-Retrieve; removing them changes nothing. Strong prompted models already score above $90$ on that set, so differences near the top of the range are compressed. The $4$B replication changes three training settings at once, LoRA adapters, a $5\times$ learning rate and gradient checkpointing, identically for both arms, and neither the V2 mixing fix nor the prompt ladder was rerun at $4$B. The recipe arm's generator and judge are the same model, so a stricter judge could change that arm's pair quality, though not the mechanism. Each arm selects its checkpoint on a dev split drawn from its own training file, so selection is in-distribution for both arms and, for the recipe arm, is itself a recipe-matching criterion; we did not re-select on a neutral dev set. The computer-algebra negatives treat every variable as a free real or complex number, so a problem that constrains its variables, say to positive integers, can be formally non-equivalent to its edit yet be the same problem to a competitor. That does not drive the verified arm's cross-language loss: the cell that swaps its counterexamples for the recipe's near-misses collapses further (Appendix~\ref{app:mechanism}).

\paragraph{Seed counts, and how far the ladder and the grid reach.}
The ladder's rungs have three seeds each (D1 eight), and the easy-tier D2$-$D3 difference flips sign across seeds, so it should not be read. Inside D1--D3 one prompt writes both the positive and the near-misses; the grid separates them for D1 alone, whose positives were also trained with verified negatives and with an unrelated prompt's near-misses (Appendix~\ref{app:mechanism}), not for D2 or D3; its cells beyond the controlled arms have three seeds (the reference eight), its unrelated near-misses come from one prompt, and its back-translated control is a shallower paraphrase than a free restatement (character-$3$-gram overlap with the source $0.783$, or $0.640$ after three hops that add noise rather than restatement, against the verified arm's $0.865$ and D4's $0.421$; Appendix~\ref{app:mechanism}). Paraphrase depth and LLM authorship are therefore separated only down to that depth.

\paragraph{The difference-in-differences holds on all three scales; the share it implies moves with them.}
A percentage point is not worth the same everywhere: it is worth less where easy-tier accuracy sits ($17 \to 62\%$) and more where real-duplicate accuracy sits ($57 \to 66\%$). We therefore recompute the central contrast on three scales, seed-paired across the eight seeds with a paired bootstrap over seeds ($B = 10{,}000$; \texttt{scripts/did\_scale\_robustness.py}, \texttt{results/did\_scale\_robustness.json}): percentage points, log-odds, and error reduction, the fraction of remaining errors a model removes. The difference-in-differences is positive in every seed on every scale and every interval excludes zero: $35.59$ $[32.57, 38.32]$ in points (narrower than Table~\ref{tab:controlled}'s interval because this version does not resample queries), $1.676$ $[1.562, 1.785]$ in log-odds and $0.328$ $[0.272, 0.384]$ as error reduction. What does depend on the scale is the share of the benchmark gap that fails to reach real data: $78.5\%$ on points, $80.3\%$ in log-odds and $60.0\%$ as error reduction. The main text therefore reports the point-scale difference-in-differences ($35.59$), and this paragraph the range.

\section{The contamination audit in full}
\label{app:contamination}

\paragraph{The controls in brief.}\label{sec:contamination}
The trainer enforced two exclusion lists of known overlaps with the benchmark, but the check compared exact text, and many corpus problems are copies of benchmark queries that differ only by a stray ``\texttt{Problem:}'' header left by OCR. Those slipped through, so about $33\%$ of \emph{both} arms' rows anchor on a benchmark query's own statement (below). This does not explain the result. The leak is the same size in both arms per row, and slightly larger in the arm that \emph{loses}. The recipe arm's easy-tier lead is within about a point of the all-query lead on every clean slice, above it on the queries the gate kept out of training and below it on the strictest slice (Table~\ref{tab:ablations}). Charging each arm only for what it gains on the leaked queries beyond what the untrained model already scores there, and netting the two arms, attributes $1.24 \pm 0.08$ points of the gap to contamination. We also corrected the filter and retrained both arms from scratch over eight seeds, in three ways: at the $4{,}101$ rows the corrected filter leaves the recipe arm, at the full $6{,}145$ rows with that arm's removed third regenerated from fresh sources, and at the full $6{,}145$ rows with it regenerated from the same sources instead. None of the three retrains has a single anchor overlapping the corrected list. The easy-tier gap reads $43.11$, $44.45$ and $44.38$ against the original $45.33$ (Table~\ref{tab:ablations}), and the benchmark-minus-real-duplicates difference $32.07$, $27.05$ and $25.65$ against $35.59$; the fresh-source comparison is not like-for-like on the duplicate sets, the same-source one is.

\paragraph{What the gate excluded.}The contamination filter removed two lists of corpus problems from training: (i)~the $8{,}698$ corpus ids whose text matches a benchmark anchor problem word for word, which cover $8{,}761$ of the benchmark's $15{,}000$ anchors, since some anchors match more than one corpus entry; and (ii)~the $779$ corpus ids that serve as queries or golds in our real-duplicate evaluation. Both lists are applied twice, when the training data is built and again inside the trainer, so a row on either list cannot be trained on even if it slipped through the first step.

\paragraph{Where the exclusion fell short: the anchor list is incomplete.}The first anchor-to-corpus mapping (v1) matched a benchmark anchor to a corpus problem only when their texts were identical after whitespace normalization, and kept one corpus id per text. A rebuilt mapping (v2), which also strips OCR boilerplate and adds the corpus's own internal duplicates, matches $14{,}921$ of the $15{,}000$ anchors ($15{,}244$ corpus ids). The $6{,}160$ anchors that v1 missed are near-identical copies that failed the exact rule for one formatting reason: $12{,}498$ of the $27{,}817$ corpus problems begin with a literal ``\texttt{Problem:}'' header that no benchmark anchor carries, and once it is removed the texts are byte-identical.

\paragraph{A third of both arms' rows anchor on a benchmark query.}The effect of the leak is that about $33\%$ of \emph{both} arms' $6{,}145$ trained rows have as their anchor a problem that is, word for word, one of the benchmark's queries: $31.98$--$33.26\%$ for the recipe arm and $33.20$--$35.43\%$ for the verified arm. Each range spans five rules for what counts as a copy, from the strictest (identical once the header is removed) to the loosest (the full v2 mapping), and, for the verified arm alone, the eight per-seed row draws.

\paragraph{Document-side overlap is small but nonzero.}Some training texts also match documents in the benchmark's corpus, and the counts are small. Among the recipe arm's $6{,}145$ positives, $1$ matches a benchmark gold document after normalization, $5$ a near-miss distractor and $2$ an organic original; among its $16{,}169$ negatives, $3$, $638$ and $157$. Over the verified arm's full $14{,}361$-row file the counts are $0/43/14$ among the positives and $4/655/412$ among its $22{,}002$ negatives. What matters is where the gold-document matches land: almost all of them are \emph{negatives}, and training on a document as a negative teaches the model to rank it lower, so this overlap works against the recipe arm at test time rather than for it.

\begin{table}[htbp]
\centering
\caption{Does the easy-tier gap depend on leaked queries? The $15{,}000$ easy-tier queries are split by whether a copy of the query reached training, and each row gives the recipe arm's lead over the verified arm on that slice (R@1, mean $\pm$ sample std over the eight training seeds; slice sizes at seed $42$). Leakage is judged against every row of each arm's training file; the recipe arm's rows are the same at every seed, the verified arm's are redrawn per seed. The gap is present on every slice, including the strictest clean one; the last two rows isolate the raw leakage effect, the gap on queries the recipe arm trained on against the gap on comparable queries neither arm trained on.}
\label{tab:clean}
\small
\setlength{\tabcolsep}{5pt}
\renewcommand{\arraystretch}{1.05}
\begin{tabular}{@{}p{8.6cm}>{\centering\arraybackslash}p{1.6cm}>{\centering\arraybackslash}p{2.6cm}@{}}
\toprule
Query slice & Queries & Recipe $-$ verified, easy R@1 \\
\midrule
All queries & $15{,}000$ & 45.33 $\pm$ 0.93 \\
Queries the original gate did catch (clean, up to $98$ missed copies) & $8{,}761$ & 46.25 $\pm$ 0.86 \\
No copy of the query in either arm's training rows (strictest clean) & $12{,}760$ & 44.73 $\pm$ 0.86 \\
Query text trained on by the recipe arm (leaked) & $1{,}960$ & 49.80 $\pm$ 1.51 \\
A copy exists in the corpus, but neither arm trained on it & $4{,}018$ & 41.16 $\pm$ 0.94 \\
\bottomrule
\end{tabular}
\end{table}

\paragraph{Direction: the contamination is marginally heavier in the arm that loses.}
Counted per \emph{trained row}, the contamination is about equal in the two arms and slightly heavier in the \emph{verified} arm, the arm that loses: $33.20$--$34.01\%$ of its rows against $31.98\%$ of the recipe arm's under the strictest matched rule, and heavier on all eight seeds under every copy rule we tried. It cannot, therefore, be creating a lead for the recipe arm. One way of counting reverses the order: the number of distinct benchmark queries each arm touched. The verified arm has $1.33$ rows per source problem on average against the recipe arm's $1.00$, so the recipe arm's rows spread over more distinct problems, and at seed $42$ it touches $1{,}960$ leaked queries against the verified arm's $1{,}524$. As a share of each arm's own distinct sources, however, the verified arm stays slightly heavier ($34.15\%$ vs $33.26\%$ at seed $42$). The count difference is thus not a reversal; it is the input from which the $1.24 \pm 0.08$-point attribution below is computed.

\paragraph{The clean slices and the leakage attribution: the gap stands where neither arm trained on the query.}
On the $8{,}761$ queries the gate did catch, so that no copy of them was in either training file, the eight-seed easy-tier gap is $46.25 \pm 0.86$ against $45.33 \pm 0.93$ overall, and larger on every seed; dropping the $98$ whose corpus-internal duplicate slipped through anyway leaves $8{,}663$ queries at $46.37 \pm 0.85$. On the strictest clean slice, the $12{,}760$ queries with no v2 copy in either arm's rows, the gap is $44.73 \pm 0.86$. The gap therefore does not depend on leaked queries. Leakage does pay where it occurs: the gap is $49.80 \pm 1.51$ on the $1{,}960$ queries whose text the recipe arm trained on, against $41.16 \pm 0.94$ on queries that have a corpus copy neither arm trained on. But the verified arm also trained on leaked queries and also gains on them, and once each arm's gain on its own leaked queries over its own untrained-copy baseline is computed and the two are netted, what remains is $1.24 \pm 0.08$ of the $45.33$.

The attribution is computed per arm as the leaked share of queries times the arm's R@1 excess on its leaked queries over its untrained-twin queries, the queries with a corpus copy that this arm did not train on, and then netted, recipe minus verified (\texttt{results/clean\_slice\_analysis.json}). The eight-seed inputs are $13.07\%$ of queries and $69.54 - 58.56$ for the recipe arm, $10.16\%$ and $19.47 - 17.57$ for the verified arm. The untrained-twin queries are the comparison because they share the one property of the leaked queries that matters, having a copy in the corpus. Taking the gate-caught slice of Table~\ref{tab:clean} as the comparison instead, where the arms score $62.64$ and $16.39$, gives a net attribution of $0.59 \pm 0.06$, so contamination is charged $0.59$ or $1.24$ points of the gap depending on which clean queries serve as the counterfactual.

\paragraph{The recipe arm's sole hard-tier hits are rarer on leaked queries at seed $42$.}If leakage drove the recipe arm's hard-tier wins, the queries it alone ranks the gold first on at seed $42$ should be more common among leaked queries. They are less common: $6.12\%$ of leaked queries are such sole wins, against $9.96\%$ of the gate-caught queries ($10.04\%$ without the $98$), so those wins are not a contamination effect either.

\paragraph{We report the leak in the headline arms rather than repair it there.}
Applying the corrected gate cuts the recipe arm to $4{,}101$ rows, a third fewer, which breaks the matched budget the design rests on; regenerating the missing rows from fresh sources restores the budget only with different source lists, and regenerating them from the same sources gives a third of the recipe arm's problems two rewrites, so neither is the same experiment. The three retrains under the corrected gate (Table~\ref{tab:cleangate}) are therefore separate controls rather than replacements, and the clean-slice rows of Table~\ref{tab:clean} carry the claim for the headline arms. The claim is quantitative, not categorical: neither arm is \emph{supervised} on the benchmark's generated documents beyond the text matches counted above, the residual anchor-text overlap is measured and symmetric, and the recipe arm's advantage survives on both clean slices. Table~\ref{tab:controlled} keeps the v1-gated arms as the headline because the ladder, every factorial cell and the reference train under the same gate, and they must share one gate to be comparable.

\paragraph{Two clean-gated retrains remove the residue: one filters, one regenerates.}Both retrains train both arms under the corrected gate, \texttt{anchor\_to\_corpus\_mapping\_v2.json}, which excludes $15{,}244$ corpus ids where v1 excluded $8{,}698$, at eight seeds. The first retrain only filters the existing training files, which leaves $4{,}101$ of the recipe arm's $6{,}145$ rows and $8{,}952$ of the verified arm's $14{,}361$; the largest matched budget both arms can meet after filtering is therefore $4{,}101$ rows, and that is the budget it uses.

\paragraph{The regenerated retrain restores the budget, but its arms no longer share one source list.}
The second retrain brings the recipe arm back to the full budget. Clean new source problems came from the $7{,}585$ corpus problems that appear on none of three lists: the corrected exclusion list, the $1{,}377$ members of the real-duplicate clusters, and the original $7{,}089$-source list. We sampled $3{,}000$ of them, ran the identical generation pipeline over them (the same prompt, generator, judge and seed), and $2{,}602$ yielded a judge-verified row. Appending the first $2{,}044$ of those, in generation order, to the $4{,}101$ survivors gives exactly $6{,}145$ clean rows at $2.62$ negatives per row, against the original file's $2.63$. The verified arm, however, draws its $6{,}145$ rows from the $8{,}952$ the gate leaves it, all from the original source list, so a third of the recipe arm's rows now come from problems the verified arm never sees; this retrain is therefore not a like-for-like comparison of source lists. A verified arm on the same fresh sources cannot be built, and we measured this rather than assumed it. Of the $2{,}044$ fresh sources, $1{,}755$ contain no relational span at all; the other $289$ carry a span the census parsed as relational, the computer-algebra generator attempted every one of them in its full run over the corpus and verified nothing, and rerunning it over all $1{,}131$ relational problems of the $7{,}585$-problem pool at four times every budget (a $60$\,s soft and $120$\,s hard limit per problem, $32$ candidate spans) recovers $6$ sources and $9$ rows, none of them among the $289$ (nor among the $345$ relational sources of the $2{,}602$ the regeneration verified), against the $2{,}044$ rows needed. The asymmetry is the pipelines' own: the recipe can rewrite any problem, the verified pipeline only those with a relation it can check. The like-for-like control we could build keeps the recipe arm on the shared list and regenerates its missing third from the same $4{,}101$ sources, a second rewrite of each problem under the same prompt, generator, judge and gate with a different sampling seed; $3{,}783$ of the $4{,}101$ yielded a judge-verified row, the first $2{,}044$ were appended at $2.61$ negatives per row, and both arms then sit on one source list at $6{,}145$ rows (the last column of Table~\ref{tab:cleangate}).

\paragraph{The decision rules were fixed in code before the results were read, the second on our account only.}
For the filtered retrain the pass and fail rules were written into the header of \texttt{scripts/clean\_gate\_retrain.slurm} before the job ran: a mean easy-tier gap of at least $35.0$, positive in every seed, would confirm the claim, and a mean below $15.0$ or any non-positive seed would refute it. The rule was set for three seeds; seeds $45$--$49$ were added after the three-seed reading had passed, which makes the every-seed condition strictly harder to meet, not easier. The regenerated retrain (\texttt{scripts/cleanfull\_train.slurm}) reused the same bands, and its header recorded the value the leakage attribution predicts, about $44$ ($45.33$ less $1.24$). Its scripts, however, were committed together with its results, so no timestamp shows that its rule came first; for that retrain the ordering rests on our account.

\begin{table}[htbp]
\centering
\caption{The arm gap under the original and the corrected contamination gate. Each column is one version of the two-arm comparison, all at the same eight seeds: the headline arms (original gate, $6{,}145$ rows); the arms filtered under the corrected gate to the $4{,}101$ matched rows it leaves; the arms restored to $6{,}145$ rows with the recipe arm's missing third regenerated from fresh sources; and the arms restored to $6{,}145$ rows with that third regenerated from the same sources, a second rewrite per source, against the same verified arm as the fresh-source column. Each row is the recipe arm's lead over the verified arm on one evaluation. The easy- and hard-tier gaps barely move. In the fresh-source column the two arms no longer share one source list, and only the shared list still contains members of the real-duplicate clusters, so that column's real-duplicate row, and the difference-in-differences built on it, are not like-for-like with the others; the same-source column restores the shared list.}
\label{tab:cleangate}
\small
\setlength{\tabcolsep}{4pt}
\renewcommand{\arraystretch}{1.05}
\resizebox{\linewidth}{!}{%
\begin{tabular}{@{}l>{\centering\arraybackslash}p{2.9cm}>{\centering\arraybackslash}p{2.9cm}>{\centering\arraybackslash}p{3.0cm}>{\centering\arraybackslash}p{3.0cm}@{}}
\toprule
Recipe $-$ verified & Headline (original gate, $6{,}145$ rows) & Filtered (corrected gate, $4{,}101$ rows) & Regenerated, fresh sources (corrected gate, $6{,}145$ rows) & Regenerated, same sources (corrected gate, $6{,}145$ rows) \\
\midrule
Easy R@1 & 45.33 $\pm$ 0.93 & 43.11 $\pm$ 0.90 & 44.45 $\pm$ 0.87 & 44.38 $\pm$ 1.46 \\
Hard R@1 & 9.26 $\pm$ 0.60 & 9.34 $\pm$ 0.31 & 9.21 $\pm$ 0.66 & 9.05 $\pm$ 0.77 \\
Real duplicates (strict R@1) & 9.74 & 11.04 & 17.40 & 18.73 \\
Difference-in-differences & 35.59 & 32.07 & 27.05 & 25.65 \\
\bottomrule
\end{tabular}}
\end{table}

\paragraph{The filtered retrain meets the pre-registered condition with room.}Under the filtered retrain the easy-tier gap is $43.11 \pm 0.90$, positive in all eight seeds and well above the $35.0$ bar. The retrained arms have no anchor overlap with the corrected list at all (the list itself still leaves $79$ of the $15{,}000$ anchors unmatched), where the headline arms carried $33.26\%$ and, at seed $42$, $34.91\%$. Removing every near-verbatim copy from \emph{both} arms therefore costs $2.22$ points of a $45$-point gap, and not all of that is contamination: part of it is the arms training on a third less data. That makes the $1.24 \pm 0.08$ slice attribution, if anything, slightly conservative. The hard-tier gap holds ($9.34$ against $9.26$), the real-duplicate gap moves up to $11.04$, and the difference-in-differences is $32.07$.

\paragraph{The regenerated retrain meets it too, and lands where the slice attribution predicted.}Under the regenerated retrain the easy-tier gap is $44.45 \pm 0.87$, between $43.27$ and $45.52$ across the eight seeds and positive in every one. That is within a point of the headline $45.33$, and almost exactly where the slice attribution predicted, about $44$ after subtracting its $1.24$. The hard tier does not move: the gap is $9.21 \pm 0.66$ against the headline $9.26$, and the recipe arm's own score $9.43$ against $9.42 \pm 0.61$.

\paragraph{The regenerated control's difference-in-differences is $27.05$ against the headline $35.59$.}
One number does move under the regenerated retrain. On real duplicates the recipe arm's lead over the verified arm grows to $17.40$ ($10.94$--$21.89$ by seed), so the difference-in-differences shrinks to $27.05$, still positive in every seed but $8.54$ below the headline $35.59$. The movement is not on the recipe arm's side, which reads $71.60$ regenerated against $71.88$ filtered and $66.26$ headline; it is the verified arm that moves, $60.85$ filtered, $54.20$ regenerated, $56.52$ headline. One candidate reason is the source lists. The two arms share one list in the headline run and in the filtered retrain but not in the regenerated one, and the difference matters here in particular: the shared list still contains $86$ members of real-duplicate clusters outside the query and gold set, while the fresh sources contain none. We therefore report all three real-duplicate rows, do not treat the fresh-source one as like-for-like with the other two, and test the reading with a same-source regeneration.

\paragraph{The same-source regeneration puts the movement on the budget, not the sources.}Regenerating the recipe arm's missing third from the same $4{,}101$ sources instead, a second rewrite of each problem with a different sampling seed under an otherwise identical pipeline, keeps both arms on one source list at $6{,}145$ rows (last column of Table~\ref{tab:cleangate}). The easy-tier gap is $44.38 \pm 1.46$, positive in every seed, against $44.45$ fresh-source and $45.33$ headline, and the hard-tier gap $9.05 \pm 0.77$. On real duplicates the recipe arm's lead is $18.73$ ($7.88$--$23.16$ by seed), nearer the fresh-source $17.40$ than the filtered retrain's $11.04$, so the movement is the restored budget, not the source lists; the difference-in-differences is $25.65$ against $27.05$ fresh-source and $32.07$ filtered.

\paragraph{No retrain re-runs the headline experiment, and every other arm stays v1-gated.}None of the three retrains repeats the main experiment: the filtered one trains on a third less data, the fresh-source one gives the recipe arm source problems the verified arm never sees, and the same-source one gives a third of its problems two rewrites. Every other model in this paper, the ladder, the factorial cells and the reference included, was trained under the original gate. A reader who requires zero-leakage training should therefore read Table~\ref{tab:cleangate} and the clean slices of Table~\ref{tab:clean}, not the headline rows.

\section{The rewriter-prompt ladder and the style probe, in detail}
\label{app:mechanism}

\begin{table}[htbp]
\centering
\caption{The rewriter-prompt ladder. Rows run from the prompt farthest from the benchmark's (D4) to the benchmark's own (D1), with the untrained base and the verified arm above for reference; columns are R@1 on the easy tier, the hard tier and cross-language duplicates, mean $\pm$ sd over the same three seeds ($42$--$44$). The hard-tier column falls step by step from D1 to D4, the registered prediction that passed; the easy-tier column does not, since D2 and D3 outscore D1 and only D4 drops; on duplicates D4 tops every model including the base. Three notes: D1 is the recipe arm read at three seeds, so its numbers differ slightly from the eight-seed ones of Table~\ref{tab:controlled} ($62.38 \pm 0.85$ / $9.42 \pm 0.61$); D3 trained on the $5{,}591$ positives its prompt yielded rather than the matched $6{,}145$, which a control shows \emph{flatters} it on the hard tier; D4 attaches no negatives by design, and Table~\ref{tab:factorial} separates that from its prompt. On real duplicates only the contrasts between D4 and the rest replicate across seeds. The points are plotted in Figure~\ref{fig:inversion}.}
\label{tab:dose-full}
\small
\setlength{\tabcolsep}{4pt}
\begin{tabular}{@{}p{5.8cm}>{\centering\arraybackslash}p{2.3cm}>{\centering\arraybackslash}p{2.3cm}>{\centering\arraybackslash}p{2.6cm}@{}}
\toprule
Training set & Easy R@1 & Hard R@1 & Cross-language dup.\ R@1 \\
\midrule
Base 0.6B (untrained) & 8.32 & 0.00 & 84.73 \\
Verified arm & 17.05 $\pm$ 0.79 & 0.16 $\pm$ 0.06 & 56.52 $\pm$ 4.05 \\
D4: restatement prompt unrelated to the recipe & 30.03 $\pm$ 1.45 & 0.04 $\pm$ 0.02 & \textbf{93.81 $\pm$ 0.39} \\
D3: rewrite prompt in a different style & 67.88 $\pm$ 0.76 & 4.67 $\pm$ 0.40 & 65.39 $\pm$ 1.83 \\
D2: paraphrase of the benchmark's prompt & 67.54 $\pm$ 1.95 & 6.76 $\pm$ 1.01 & 68.02 $\pm$ 3.51 \\
D1: the benchmark's prompt ($=$ recipe arm) & 61.86 $\pm$ 0.52 & \textbf{9.90 $\pm$ 0.74} & 66.41 $\pm$ 1.78 \\
\bottomrule
\end{tabular}
\end{table}

\paragraph{The registered decay prediction, and its falsifier.}
The primary registration predicted that hard-tier R@1 would fall monotonically as the prompt moves away from the benchmark's, with a single falsifier at the far end: D4 scoring ${\ge}\,4.69$, half the $9.37$-point gain D1 had shown at seed $42$ when the prediction was written, would refute it. It was written after D1 and the controlled arms were measured and before D2--D4 existed. One weakness to admit: because its only falsifier sits at the D4 endpoint, the prediction would also have counted as passing had only the endpoints differed and the middle stayed flat; what landed was a fall at every step. D4's observed $0.04$ clears the threshold by two orders of magnitude, but the factorial below (Table~\ref{tab:factorial}) attributes that margin to D4's missing negatives rather than to its prompt. In units of the noise the eight-seed arm implies (the $\sigma_3$ convention defined below), the three steps down the ladder are $6.3$, $4.2$ and $9.3$.

\paragraph{The two noise denominators, $\sigma_1$ and $\sigma_3$, and the assumption they import.}
With three seeds per rung, a standard deviation computed from three numbers would mostly measure luck, so we borrow the noise level from the eight-seed recipe arm. $\sigma_1$ is the run-to-run standard deviation of a difference between two independent trainings of that arm ($1.20$ easy, $0.86$ hard, $3.27$ real-duplicate); $\sigma_3 = \sigma_1/\sqrt{3}$ is therefore the standard error of the difference between two independent three-seed means (the standard error of one three-seed mean would be $\sigma_1/\sqrt{6}$), and an unsubscripted multiple below means $\sigma_3$. Borrowing the noise this way assumes every rung is as noisy as the recipe arm. For D2 the assumption fails: its three-seed spread exceeds the eight-seed arm's single-run sd ($1.65\times$ hard, $2.30\times$ easy) and $\sigma_1$ itself ($1.17\times$, $1.63\times$), so beside every claim we also report the test that uses each rung's own observed spread.

\paragraph{Seed replication resolved the two sub-$2\sigma_1$ differences, in opposite directions.}
Each rung began as a single run. Of the six differences between neighbouring rungs (three steps on each tier), two fell within $2\sigma_1$, too close to trust from one run, so we retrained D2--D4 at seeds $43$ and $44$, giving three seeds per rung. Both close calls were settled, in opposite directions. The hard-tier step D1$-$D2 held and grew: $+1.48$, $+4.34$, $+3.61$ across seeds, mean $+3.14$, $6.3\sigma$ of the standard error of a three-seed mean, with the original seed $42$ the weakest of the three. The easy-tier step D2$-$D3 did \emph{not} hold: $-2.18$, $-0.26$, $+1.42$, mean $-0.34$ at $0.5\sigma$, the sign flipping from seed to seed, so on the easy tier we report D2 and D3 as indistinguishable rather than as an ordering.

\paragraph{Why the eval bootstrap cannot license an ordering on its own.}
Within one training run, a paired bootstrap over the $15{,}000$ queries ($B = 5{,}000$) gives a tight interval, because it captures only query-sampling noise and not training noise. For easy-tier D2$-$D3 the seed-$42$ run gives an interval entirely below zero ($-2.19$ $[-2.89, -1.48]$) and the seed-$44$ run one entirely above it ($+1.41$ $[0.75, 2.09]$). A reader shown either interval alone would have been told the reverse of what three seeds show, so a query bootstrap by itself cannot justify ranking one model above another.

\paragraph{What was fixed before the seed runs, and what was written after.}
Fixed before the seed runs were submitted, in \texttt{scripts/dose\_seeds.slurm}: the noise model ($\sigma_1 = 1.20$ easy / $0.86$ hard), the $2\sigma$ bar, and the two differences to re-examine. Written after the seed-$43$/$44$ outputs existed: the rule that a difference must keep its sign across seeds, and the aggregation script. Both are applied uniformly to all eighteen rung${\times}$tier cells rather than to the two flagged ones, which limits how far they could have been tuned to the result. We claim no external timestamp for any of this, since file times can be set by hand and are lost when a repository is cloned, so a reader should treat the ordering as our account rather than as proof.

\paragraph{The harness diff: one flag differs across runs, and it is inert here.}
The same script compares the evaluation settings recorded in every per-seed output file, to check that the later seed runs were scored the same way as the first. Exactly one field differs: the self-masking flag, which removes a query's own corpus entry from the ranking; the seed-$43$/$44$ runs record it and the earlier runs predate it. It has no effect here, because on MathNet-Retrieve no query's own document is in the corpus ($0$ of $15{,}000$), so the set it would mask is empty.

\paragraph{The $42$--$44$ window is a favourable draw for D1; both orderings survive its eight-seed values.}
The ladder comparisons use seeds $42$--$44$ only, because the other rungs were trained at those seeds alone. Those three are a favourable draw for D1: over them it scores $9.90$ hard and $61.86$ easy, against $9.42$ and $62.38$ over all eight seeds. Substituting D1's eight-seed values shrinks the hard-tier D1$-$D2 lead to $+2.67$ and the easy-tier D1$-$D3 deficit to $-5.49$, and both keep their direction, so neither ordering depends on the draw. Pairing by seed remains the correct estimator; this is a robustness check on it.

\paragraph{The observed-spread $p$-values, and the assumption-free statement we lead with.}
Tested against each rung's own three-seed spread rather than the borrowed noise, the four steps that carry the two-tier reading, hard D1$-$D2, D2$-$D3 and D3$-$D4 and easy D3$-$D4, sit at $3.7$, $3.8$, $20.7$ and $30.3$ standard errors, with one-sided paired $t(2)$ $p$-values of $0.034$, $0.031$, $0.001$ and $0.0005$. That is as much as $n = 3$ can support, and it is eight orders of magnitude weaker than ``$6.3\sigma$'' suggests, because that figure borrows its noise level from another model. We therefore lead with the statement that needs no noise model at all: in every seed taken on its own, the hard-tier scores fall in the same order down the ladder, and D1 scores below both D2 and D3 on the easy tier.

\paragraph{The budget control for D3's row deficit.}
D3 trained on less data than the other rungs: its prompt yielded only $5{,}591$ judge-verified positives against the matched $6{,}145$, a $9.0\%$ shortfall, so the D2-to-D3 step mixes a change of prompt with a change of data. To separate the two we retrained D2 at seed $42$ on exactly $5{,}591$ rows (\verb|--max-rows 5591|). Because the trainer shuffles with the seed and then truncates, those rows are the first $5{,}591$ of the same shuffled order the full D2 run saw, so the control is a nested subset of D2's data rather than an independent draw.

\paragraph{Cutting rows moves hard R@1 \emph{up}: the deficit flatters D3.}
At seed $42$, cutting D2 to D3's row count moves its hard R@1 \emph{up}, $7.89 \to 8.27$ ($+0.37$ $[0.06, 0.69]$ under a paired per-query bootstrap). Fewer rows help rather than hurt on this tier, so D3's shortfall makes D3 look slightly better than it should, and the measured D2-to-D3 drop is if anything an underestimate. The same cut costs $1.20$ easy-tier points and gains $5.34$ on real duplicates ($66.67 \to 72.01$), a value inside D2's own three-seed range of $65.39$--$72.01$ and therefore uninformative.

\paragraph{Row count, at three seeds: reliably the wrong sign.}
Repeated at three seeds, cutting D2 to D3's row count is worth $+0.96$ $[+0.24, +1.69]$ hard R@1, positive in every seed. The row count therefore reliably pushes the \emph{wrong} way to explain D3 scoring lower: fewer rows would have raised D3's score, not lowered it. The D2-to-D3 drop ($-2.97$, $-2.20$, $-1.08$ by seed, mean $-2.08$) is the prompt's doing, and compared at matched rows the prompt's effect is $-3.05$, larger than the uncontrolled figure.

\paragraph{Row identity: retraining D2 on D3's own sources rules out selection as the driver.}
The confound could be which rows D3 lost rather than how many. D3's prompt had the ladder's worst generation-failure rate, and the sources it dropped skew longer and proof-only, so D3 may have trained on a differently shaped set, which a count-matched retrain cannot detect. We therefore also retrained D2 on the $5{,}332$ sources D3 itself has. That model scores $7.81$ hard R@1, on D2's side of the D2--D3 midpoint, so the selection of sources does not drive the step either.

\paragraph{One remaining confound: the dev splits differ.}When the trainer is given fewer rows it redraws its development split from the smaller pool, so the full D2 run and the cut D2 run selected their checkpoints on different dev sets, sharing only $12$ of $280$ ids. In principle that could have picked different checkpoints for reasons unrelated to training; in practice it did not, since each run saved two checkpoints and both chose the same one (step $100$), so the differing dev sets caused no differential selection here.

\paragraph{The refuted real-data prediction, verbatim.}Alongside the decay prediction we registered a second one about real duplicates, quoted here exactly as written: ``all four dose points stay in the trained-arm band on real-duplicate strict R@1 --- below base $84.73$, spread across D1--D4 $\le 10$ points, with NO monotone trend tracking the dose. The dose moves benchmark numbers only; real equivalence ability is unchanged (beyond the generic forgetting erosion every narrow arm shows).'' In plain terms, we expected the prompt to change benchmark scores and leave real retrieval alone. It \textbf{failed}, and two of its three clauses were refuted outright: the rungs spread over $28.41$ points across three-seed means, not ${\le}\,10$, and D4 lands \emph{above} the base rather than below it. Only the ``no monotone trend'' clause held.

\paragraph{A $2 \times 2$ over positives and negatives: what D4's missing negatives confounded.}
D4 differs from D1--D3 in two ways at once: its prompt has nothing to do with the benchmark, and it attaches no negatives, so either could explain its results. To separate them we completed a factorial over $\{$CAS, D4$\}$ positives $\times$ $\{$CAS counterexamples, none$\}$ negatives. Two cells already existed, the verified arm at eight seeds and D4 at three; we trained the other two, all with the controlled experiment's hyperparameters (Table~\ref{tab:hparams}). One strips the verified arm's negatives away (three seeds); the other attaches them to D4's restatements, and that cell is the reference (eight seeds, $42$--$49$). Building the reference needed one choice, which negatives each D4 row gets: it takes the counterexampled negatives of one CAS row for the same source problem, chosen once at random when the file was built, and that one file serves every seed. D4's $6{,}768$ sources are a subset of the verified arm's $7{,}397$, and for $2{,}019$ rows the chosen CAS row has no negatives, so the reference carries $1.28$ negatives per row against the verified arm's $1.52$ and the recipe arm's $2.63$. Because \verb|--max-rows| shuffles with the seed before truncating, each new cell trains on the same rows as the cell it was derived from at every seed. The readout was registered with no directional prediction and its verdict fixed in code (\texttt{scripts/factorial\_cells.slurm}, \texttt{scripts/factorial\_verdict.py}); the new cells' numbers are in \texttt{results/factorial\_2x2.json}, the D4 row's in Table~\ref{tab:dose-full}.

\begin{table}[t]
\centering
\caption{The five terms the paper relies on, and the reference model that Table~\ref{tab:controlled} compares against.}
\label{tab:terms}
\scriptsize
\setlength{\tabcolsep}{4pt}
\renewcommand{\arraystretch}{1.0}
\begin{tabular}{@{}>{\itshape}l p{11.6cm}@{}}
\toprule
recipe & how a benchmark builds itself: the prompts and filters under which an LLM writes part of it \\
template & the exact wording of that prompt \\
genre & the style of LLM-written rewrites, whichever prompt produced them \\
pair structure & a deep LLM rewrite as the positive against a minimal-edit near-miss as the negative \\
surface & the part of a benchmark its LLM writes, the part an attacker can imitate: documents, summaries or evaluation pairs \\
\midrule
reference & the recipe-free comparison model: LLM restatements written under a prompt unrelated to the recipe (D4), paired with the verified arm's computer-algebra negatives; no benchmark prompt anywhere in its training file \\
\bottomrule
\end{tabular}
\end{table}

\begin{table}[t]
\centering
\caption{Every training set in the paper, arranged by what its positives are (rows) and what its negatives are (columns). Each cell is one trained model and gives two numbers: R@1 on the benchmark's hard tier, then strict R@1 on real cross-language duplicates. Two cells show two pairs because that set was trained twice, first with as many negatives as the reference has and then with as many as the recipe arm has; every other LLM-near-miss cell uses the recipe arm's volume, and every computer-algebra-negative cell the reference's negatives. The two arms and the reference have eight seeds, every other cell three (Appendix~\ref{app:mechanism}). The four models the paper names are in italic.}
\label{tab:design}
\scriptsize
\setlength{\tabcolsep}{3pt}
\renewcommand{\arraystretch}{1.05}
\begin{tabular}{@{}p{3.4cm}>{\centering\arraybackslash}p{1.4cm}>{\centering\arraybackslash}p{2.2cm}>{\centering\arraybackslash}p{2.9cm}>{\centering\arraybackslash}p{2.9cm}@{}}
\toprule
 & \multicolumn{4}{c}{Negatives} \\
\cmidrule(l){2-5}
Positives & none & computer-algebra counterexamples & LLM near-misses, unrelated prompt & LLM near-misses, recipe prompt \\
\midrule
Computer-algebra near-copies & 0.00 / 52.08 & 0.16 / 56.52 \emph{verified} & 0.23 / 13.48 & 0.11 / 12.47 \\
Back-translated paraphrases & 0.05 / 94.57 & 1.09 / 84.57 & 1.88 / 35.20 & 4.18 / 38.67 \\
LLM restatements, unrelated prompt (D4) & 0.04 / 93.81 \emph{D4} & 10.30 / 78.02 \emph{reference} & 16.96 / 85.50; 20.93 / 84.91 & 20.71 / 83.89; 26.97 / 78.88 \\
LLM rewrites, recipe prompt (D1) & 0.00 / 83.21 & 2.09 / 70.82 & 3.45 / 47.41 & 9.42 / 66.26 \emph{recipe} \\
\bottomrule
\end{tabular}
\end{table}

\begin{table}[htbp]
\centering
\caption{Every training set of the factorial, a kind of positive crossed with a kind of negative: R@1 on the easy and hard tiers and strict R@1 on cross-language duplicates, mean $\pm$ sd over training seeds (the verified arm and the reference eight, every other cell three; the D4 row is the three-seed mean of Table~\ref{tab:dose-full}). The group labels say what each block varies. Three readings: neither half of a pair scores on the hard tier on its own, and near-copy or back-translated positives barely do even with negatives; D4's restatements with any negatives leave the floor, matching the recipe arm ($9.42 \pm 0.61$ over eight seeds, Table~\ref{tab:controlled}) with no recipe wording in the file and passing it with LLM-written near-misses; and every kind of negative costs D4's restatements cross-language points from their $93.81$, except an unrelated prompt's near-misses, which leave them at the untrained base ($84.73$). The last block fills the grid: positives of any kind trained alone never register on the hard tier, and paraphrase positives alone, LLM-written or back-translated, sit at the top on cross-language duplicates ($94.57$ for back-translations, within noise of D4's $93.81$); near-copy or back-translated positives with LLM near-misses of either prompt stay near the floor and lose most of their cross-language retention; and D1's own positives with the reference's negatives reach a fifth of the reference's hard-tier score.}
\label{tab:factorial}
\small
\setlength{\tabcolsep}{4pt}
\resizebox{\linewidth}{!}{%
\begin{tabular}{llccc}
\toprule
Positives & Negatives & Easy R@1 & Hard R@1 & Cross-language dup.\ R@1 \\
\midrule
\multicolumn{5}{l}{\emph{The $2 \times 2$: CAS or D4 positives, with or without CAS counterexamples}} \\
CAS (verified) & CAS counterexamples ($=$ verified arm) & 17.05 $\pm$ 0.79 & 0.16 $\pm$ 0.06 & 56.52 $\pm$ 4.05 \\
CAS (verified) & none & 7.42 $\pm$ 0.24 & 0.00 $\pm$ 0.00 & 52.08 $\pm$ 1.06 \\
D4 (recipe-free) & none ($=$ D4) & 30.03 $\pm$ 1.45 & 0.04 $\pm$ 0.02 & 93.81 $\pm$ 0.39 \\
D4 (recipe-free) & CAS counterexamples ($=$ reference) & 46.91 $\pm$ 1.25 & 10.30 $\pm$ 1.15 & 78.02 $\pm$ 0.91 \\
\multicolumn{5}{l}{\emph{D4 positives with LLM-written near-misses (matched: the reference's count; full: the recipe arm's volume)}} \\
D4 (recipe-free) & recipe-prompt near-misses, matched & 44.59 $\pm$ 1.33 & 20.71 $\pm$ 0.82 & 83.89 $\pm$ 3.82 \\
D4 (recipe-free) & recipe-prompt near-misses, full & 37.79 $\pm$ 1.37 & \textbf{26.97 $\pm$ 0.97} & 78.88 $\pm$ 2.33 \\
D4 (recipe-free) & unrelated-prompt near-misses, matched & \textbf{47.69 $\pm$ 2.12} & 16.96 $\pm$ 2.08 & 85.50 $\pm$ 1.67 \\
D4 (recipe-free) & unrelated-prompt near-misses, full & 43.93 $\pm$ 0.51 & 20.93 $\pm$ 0.44 & 84.91 $\pm$ 1.49 \\
\multicolumn{5}{l}{\emph{Swapping the positives instead}} \\
CAS (verified) & recipe-prompt near-misses, full & 11.28 $\pm$ 0.48 & 0.11 $\pm$ 0.03 & 12.47 $\pm$ 0.68 \\
Back-translated (NMT) & CAS counterexamples & 18.81 $\pm$ 0.80 & 1.09 $\pm$ 0.04 & 84.57 $\pm$ 1.20 \\
Back-translated (NMT, three hops) & CAS counterexamples & 12.85 $\pm$ 0.56 & 4.47 $\pm$ 0.15 & 67.01 $\pm$ 1.73 \\
\multicolumn{5}{l}{\emph{Later cells: a second recipe-free prompt (D5) and two recipe-arm variants}} \\
D5 (recipe-free, survey prompt) & none & 18.71 $\pm$ 1.31 & 0.01 $\pm$ 0.00 & 93.72 $\pm$ 0.39 \\
D5 (recipe-free, survey prompt) & CAS counterexamples & 39.20 $\pm$ 0.40 & 3.97 $\pm$ 1.05 & 78.71 $\pm$ 0.59 \\
D1 (recipe), separate calls & recipe-prompt near-misses, full & 58.47 $\pm$ 1.18 & 9.79 $\pm$ 0.66 & 52.08 $\pm$ 1.49 \\
D1 (recipe), both judges & recipe-prompt near-misses, both judges & 64.71 $\pm$ 0.64 & 7.58 $\pm$ 0.39 & 74.81 $\pm$ 2.65 \\
\multicolumn{5}{l}{\emph{The remaining cells of the grid (three seeds each; LLM near-misses at the recipe arm's volume, CAS counterexamples the reference's)}} \\
CAS (verified) & unrelated-prompt near-misses & 12.68 $\pm$ 0.93 & 0.23 $\pm$ 0.03 & 13.48 $\pm$ 0.92 \\
Back-translated (NMT) & none & 12.92 $\pm$ 0.72 & 0.05 $\pm$ 0.01 & \textbf{94.57 $\pm$ 0.64} \\
Back-translated (NMT) & unrelated-prompt near-misses & 20.54 $\pm$ 1.66 & 1.88 $\pm$ 0.12 & 35.20 $\pm$ 1.15 \\
Back-translated (NMT) & recipe-prompt near-misses & 25.80 $\pm$ 1.74 & 4.18 $\pm$ 0.51 & 38.67 $\pm$ 1.42 \\
D1 (recipe) & none & 25.28 $\pm$ 0.47 & 0.00 $\pm$ 0.00 & 83.21 $\pm$ 1.17 \\
D1 (recipe) & CAS counterexamples & 50.25 $\pm$ 0.47 & 2.09 $\pm$ 0.22 & 70.82 $\pm$ 0.53 \\
D1 (recipe) & unrelated-prompt near-misses & 49.81 $\pm$ 1.85 & 3.45 $\pm$ 0.03 & 47.41 $\pm$ 1.55 \\
\bottomrule
\end{tabular}}
\end{table}

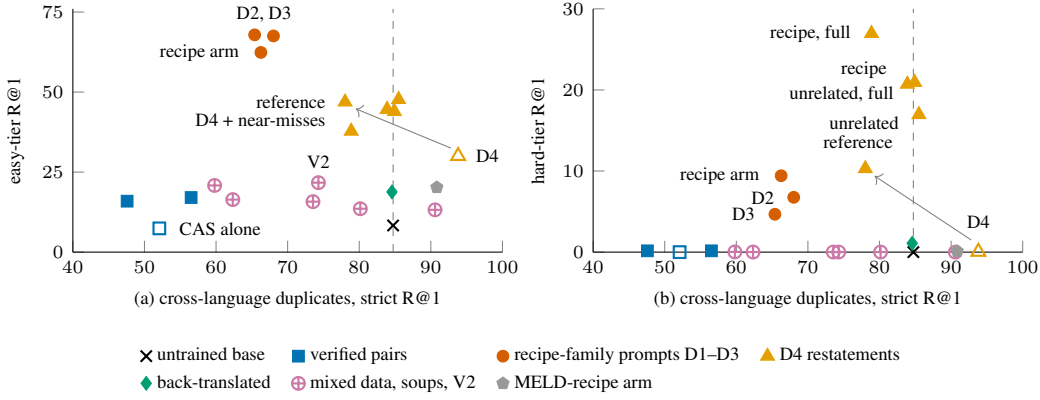
\begin{figure}[htbp]
\centering
\definecolor{cVer}{HTML}{0072B2}   
\definecolor{cRec}{HTML}{D55E00}   
\definecolor{cD4}{HTML}{E69F00}    
\definecolor{cBT}{HTML}{009E73}    
\definecolor{cMix}{HTML}{CC79A7}   
\definecolor{cMELD}{HTML}{999999}  
\begin{tikzpicture}
\begin{axis}[
  name=left,
  width=0.44\linewidth, height=4.8cm,
  xmin=40, xmax=100, ymin=0, ymax=76,
  xtick={40,50,60,70,80,90,100}, ytick={0,25,50,75},
  xlabel={(a) cross-language duplicates, strict R@1}, ylabel={easy-tier R@1},
  label style={font=\scriptsize}, tick label style={font=\scriptsize},
  axis y line*=left, axis x line*=bottom, clip=false,
]
\addplot[gray, dashed, forget plot] coordinates {(84.73,0) (84.73,76)};
\addplot[forget plot, only marks, mark=x, mark size=3pt, thick, black] coordinates {(84.73,8.32)};
\addplot[forget plot, only marks, mark=square*, mark size=2.2pt, cVer] coordinates {(56.52,17.05) (47.58,15.92)};
\addplot[forget plot, only marks, mark=square, mark size=2.2pt, thick, cVer] coordinates {(52.08,7.42)};
\addplot[forget plot, only marks, mark=*, mark size=2.2pt, cRec] coordinates {(66.26,62.38) (68.02,67.54) (65.39,67.88)};
\addplot[forget plot, only marks, mark=triangle, mark size=3pt, thick, cD4] coordinates {(93.81,30.03)};
\addplot[forget plot, only marks, mark=triangle*, mark size=3pt, cD4] coordinates {(78.02,46.91) (83.89,44.59) (78.88,37.79) (85.50,47.69) (84.91,43.93)};
\addplot[forget plot, only marks, mark=diamond*, mark size=2.6pt, cBT] coordinates {(84.57,18.81)};
\addplot[forget plot, only marks, mark=oplus, mark size=2.4pt, thick, cMix] coordinates {(74.30,21.70) (59.80,20.83) (80.15,13.55) (73.54,15.77) (62.34,16.41) (90.59,13.21)};
\addplot[forget plot, only marks, mark=pentagon*, mark size=2.4pt, cMELD] coordinates {(90.84,20.23)};
\draw[->, thin, gray] (axis cs:92.8,32.5) -- (axis cs:79.5,44.8);
\node[font=\scriptsize, anchor=east] at (axis cs:64.5,62.38) {recipe arm};
\node[font=\scriptsize, anchor=south] at (axis cs:66.7,69.0) {D2, D3};
\node[font=\scriptsize, anchor=west] at (axis cs:95.0,30.03) {D4};
\node[font=\scriptsize, anchor=east] at (axis cs:76.6,46.91) {reference};
\node[font=\scriptsize, anchor=east] at (axis cs:76.6,41.2) {D4 + near-misses};
\node[font=\scriptsize, anchor=south] at (axis cs:74.30,23.2) {V2};
\node[font=\scriptsize, anchor=west] at (axis cs:53.6,7.42) {CAS alone};
\end{axis}
\begin{axis}[
  name=right,
  at={(left.south east)}, anchor=south west, xshift=1.2cm,
  width=0.44\linewidth, height=4.8cm,
  xmin=40, xmax=100, ymin=0, ymax=30,
  xtick={40,50,60,70,80,90,100}, ytick={0,10,20,30},
  xlabel={(b) cross-language duplicates, strict R@1}, ylabel={hard-tier R@1},
  label style={font=\scriptsize}, tick label style={font=\scriptsize},
  axis y line*=left, axis x line*=bottom, clip=false,
  legend to name=invfulllegend, legend columns=4, legend cell align=left,
  legend style={font=\scriptsize, draw=none, /tikz/every even column/.append style={column sep=6pt}},
]
\addplot[gray, dashed, forget plot] coordinates {(84.73,0) (84.73,30)};
\addplot[only marks, mark=x, mark size=3pt, thick, black] coordinates {(84.73,0.00)};
\addlegendentry{untrained base}
\addplot[only marks, mark=square*, mark size=2.2pt, cVer] coordinates {(56.52,0.16) (47.58,0.16)};
\addlegendentry{verified pairs}
\addplot[forget plot, only marks, mark=square, mark size=2.2pt, thick, cVer] coordinates {(52.08,0.00)};
\addplot[only marks, mark=*, mark size=2.2pt, cRec] coordinates {(66.26,9.42) (68.02,6.76) (65.39,4.67)};
\addlegendentry{recipe-family prompts D1--D3}
\addplot[only marks, mark=triangle*, mark size=3pt, cD4] coordinates {(78.02,10.30) (83.89,20.71) (78.88,26.97) (85.50,16.96) (84.91,20.93)};
\addlegendentry{D4 restatements}
\addplot[forget plot, only marks, mark=triangle, mark size=3pt, thick, cD4] coordinates {(93.81,0.04)};
\addplot[only marks, mark=diamond*, mark size=2.6pt, cBT] coordinates {(84.57,1.09)};
\addlegendentry{back-translated}
\addplot[only marks, mark=oplus, mark size=2.4pt, thick, cMix] coordinates {(74.30,0.01) (59.80,0.03) (80.15,0.01) (73.54,0.01) (62.34,0.01) (90.59,0.00)};
\addlegendentry{mixed data, soups, V2}
\addplot[only marks, mark=pentagon*, mark size=2.4pt, cMELD] coordinates {(90.84,0.11)};
\addlegendentry{MELD-recipe arm}
\draw[->, thin, gray] (axis cs:92.8,1.5) -- (axis cs:79.3,9.4);
\node[font=\scriptsize, anchor=east] at (axis cs:64.5,9.42) {recipe arm};
\node[font=\scriptsize, anchor=east] at (axis cs:66.6,6.76) {D2};
\node[font=\scriptsize, anchor=east] at (axis cs:63.9,4.67) {D3};
\node[font=\scriptsize, anchor=south] at (axis cs:93.81,1.8) {D4};
\node[font=\scriptsize, anchor=south] at (axis cs:77.0,11.6) {reference};
\node[font=\scriptsize, anchor=east] at (axis cs:77.3,26.97) {recipe, full};
\node[font=\scriptsize, anchor=east] at (axis cs:82.3,22.5) {recipe};
\node[font=\scriptsize, anchor=east] at (axis cs:83.3,19.5) {unrelated, full};
\node[font=\scriptsize, anchor=east] at (axis cs:84.0,15.9) {unrelated};
\end{axis}
\node[anchor=north, yshift=-4pt] at (current bounding box.south) {\pgfplotslegendfromname{invfulllegend}};
\end{tikzpicture}
\caption{The full plot behind Figure~\ref{fig:inversion}: every $0.6$B model of Tables~\ref{tab:controlled}, \ref{tab:dose-full}, \ref{tab:factorial} and~\ref{tab:mixing}, with the corrected-gate retrains and V2's no-instruction reading omitted. Same axes and conventions as Figure~\ref{fig:inversion}: horizontal, strict R@1 on cross-language duplicates, the dashed line and $\times$ the untrained base; colour, what the positives are; hollow, no negatives; the labels on the D4 cells name the negatives attached (``recipe'' and ``unrelated'' are LLM near-misses from the recipe's prompt and from an unrelated one, ``full'' at the recipe arm's negative volume). Two groups appear here only: the verified squares also include computer-algebra positives with no negatives (CAS alone) and the uncapped verified arm, and the crossed circles are the mixed-data, soup and V2 models of Appendix~\ref{sec:forgetting}, which sit near the hard-tier floor. The MELD-recipe arm is seed $42$ ($20.23$ easy, $0.11$ hard, $90.84$ real; Section~\ref{sec:ood}). (a)~Easy tier: every model trained on LLM rewrites scores above the verified ones. (b)~Hard tier: only deep LLM rewrites paired with negatives reach it; the arrow is the edit from D4 alone to the reference, attaching verified negatives, which lifts hard R@1 off the floor and costs $15.79$ retention points, and LLM-written near-misses lift the same restatements further (Table~\ref{tab:factorial}). Computer-algebra positives with the recipe's near-misses lie off the axis at $12.47$ real.}
\label{fig:inversion-full}
\end{figure}

\paragraph{Positives or negatives alone stay at the floor; together they match the recipe arm.}
Table~\ref{tab:factorial} gives the cells. Neither half of a pair does anything on the hard tier by itself: CAS positives score $0.00$ without negatives and $0.16$ with them, and D4's restatements score $0.04$ without. Put D4's restatements together with the CAS negatives and the model, the reference, reaches $10.30 \pm 1.15$ over eight seeds ($10.81$, $11.85$, $10.25$, $9.01$, $10.13$, $10.34$, $8.49$, $11.55$). That is level with the recipe arm's $9.42$ (recipe minus reference $-0.88$ $[-1.85, 0.10]$), ahead in seven of eight seed-paired comparisons ($+1.44$, $+1.10$, $+0.67$, $+0.02$, $+1.10$, $+1.42$, $-0.57$, $+1.86$) and clearly ahead at R@5 ($60.05$ against $57.20$; $-2.86$ $[-4.01, -1.67]$), with fewer negatives per row ($1.28$) than either controlled arm ($1.52$ and $2.63$). Nothing in that training file comes from the Appendix-F prompt: the positives are D4's restatements under an unrelated prompt (Appendix~\ref{app:prompts}) and the negatives are SymPy counterexamples. What the hard tier rewards is therefore the shape of the pair, a deep LLM rewrite against a hard negative, and not the words of the template's positives (Section~\ref{sec:mechanism}); the near-copy and back-translated cells below, whose positives are shallow, stay near the floor, which is why the positive has to be a deep LLM rewrite.

\paragraph{A count-matched cell and a full cell hold the positives fixed and swap the negatives' provenance.}
These two cells ask whether it matters who wrote the negatives, separately from how many there are (\texttt{scripts/build\_llmnegs\_cells.py}; readings registered as R6-a/b in \texttt{scripts/factorial\_cells.slurm} before any number existed). Each D4 row whose source problem also has a recipe-arm row takes that row's Appendix-F near-misses. The count-matched cell takes only the first $k$ of them, $k$ being the number the same source carries in the reference file, so the count stays close and only the origin changes: $5{,}978$ rows at $1.20$ negatives per row against the reference's $1.28$, the recipe arm's cap of three truncating $449$ rows. The full cell takes all of them, $2.64$ per row, the recipe arm's own volume. The $790$ D4 sources with no recipe-arm row are dropped, so both cells train on $2.7\%$ fewer rows than the reference. The registered reading: a count-matched cell within $2\sigma_1 = 1.72$ hard-tier points of the reference's then three-seed mean would mean the negatives' origin does not matter, more than $1.72$ above it would mean the recipe's near-misses add a recipe-specific increment, and more than $1.72$ below it would have meant verified negatives are the better half; a full cell at or above the recipe arm's $9.90$ less $1.72$ would mean the claim that the template's positives are not required survives.

\paragraph{The recipe's own near-misses beat counterexamples by ten points at matched count.}
Both cells landed far above their registered bands. The count-matched cell scores $20.71 \pm 0.82$ hard R@1 ($20.31$, $21.65$, $20.17$), $9.50$ to $9.92$ above the reference seed for seed and $10.59$ to $10.94$ above the recipe arm: swapping CAS counterexamples for the recipe's own near-misses, at the same count, doubles the hard-tier score, so under the registered reading the negatives' origin matters, with the unrelated-prompt cells below refining how much of that is specific to the recipe. The full cell scores $26.97 \pm 0.97$ ($26.47$, $28.09$, $26.35$), $17.55$ above the recipe arm's eight-seed mean at the recipe arm's own negative volume, so the claim that the template's positives are not required survives. The hard tier's recipe-sensitive part therefore lives in its negatives: near-misses written by the recipe's prompt, the genre of the benchmark's own distractors, beat SymPy counterexamples by ten points at matched count, of which the recipe's prompt itself accounts for $3.75$ (the unrelated-prompt cells below), while the template's positives are not required and are outscored by restatements from an unrelated prompt.

\paragraph{More near-misses cost easy-tier and retention points, and two limits travel with the reading.}On the easy tier the two LLM-negative cells score $44.59$ and $37.79$, both below the reference's $46.91$, and on real duplicates $83.89 \pm 3.82$ and $78.88 \pm 2.33$ against its $78.02$. So adding more of the recipe's near-misses buys hard-tier points and pays for them in easy-tier and retention points, and at matched count those near-misses cost less retention than the counterexamples do. Two caveats go with the reading: each new cell has three seeds, and the count-matched cell carries $0.08$ fewer negatives per row than the reference, a difference that works against it rather than for it. A third, that both cells draw their near-misses from the recipe's own prompt, is what the next cells test.

\paragraph{Near-misses from an unrelated prompt recover most of the negatives' effect; the recipe's prompt adds an increment.}
The third pair of cells asks whether the ten-point jump came from the recipe's prompt specifically or from LLM-written near-misses in general. A new prompt unrelated to the benchmark's Appendix-F template, keeping D4's audience persona and adding a ``spot the difference'' drill, wrote near-misses for D4's positives, with the ladder's generator, judge pass, seed and source problems (\texttt{scripts/gen\_unrelated\_negs.slurm}, \texttt{scripts/build\_review2\_cells.py}). Both cells hold the same $6{,}393$ rows ($285$ D4 sources had no such row and $90$ no surviving negative), of which $6{,}145$ train per seed; one is count-matched to the reference by the same rule as the recipe-negative cell ($1.20$ negatives per row), the other carries all of the new prompt's near-misses ($2.66$ per row). Every verdict file was regenerated in one pass after the last cell landed. The reading was registered before the cells trained (\texttt{scripts/factorial\_cells.slurm}, R7-a/b): within $2\sigma_1 = 1.72$ hard-tier points of the recipe's near-misses, LLM-written near-misses of any origin would be the potent half and ``recipe-specific'' would be retired for the negatives; more than $1.72$ below, the recipe's prompt adds an increment of its own, to be reported with its size. The second branch landed: $16.96 \pm 2.08$ hard R@1 ($14.67$, $18.73$, $17.48$) against the recipe near-misses' $20.71$ at matched count, a gap of $3.75$ $[1.83, 5.97]$, and $20.93 \pm 0.44$ against $26.97$ at full volume, a gap of $6.03$ $[5.01, 7.18]$ (nested bootstrap over queries and seeds, \texttt{results/reference\_control\_did.json}). The negatives' effect therefore breaks into three steps, CAS counterexamples $10.30$, an unrelated prompt's LLM near-misses $16.96$, the recipe's own $20.71$: most of the climb above verified negatives comes from the near-misses being LLM-written at all, and the recipe's prompt adds about a fifth of the count-matched score on top.

\paragraph{The unrelated prompt's near-misses score higher on the easy tier and leave cross-language retention at the base.}
The unrelated-prompt cells beat the recipe-negative cells on the easy tier ($47.69$ and $43.93$ against $44.59$ and $37.79$) and leave cross-language retention at the untrained base ($85.50 \pm 1.67$ and $84.91 \pm 1.49$ against $84.73$). At full volume, the extra hard-tier points the recipe's near-misses buy therefore come at a retention cost $6.02$ $[2.00, 10.18]$ cross-language points higher than the unrelated prompt's, which cost nothing; at matched count the two kinds of near-misses are within noise of each other on retention ($1.61$ $[-2.93, 6.22]$). On same-language reprints they are within noise too ($50.13$ against $47.73$ at matched count and $45.33$ against $39.47$ at full volume, the $45.33$ coinciding with the headline gap's value by chance; Table~\ref{tab:samelang}). The caveat that remains is that all the unrelated near-misses come from one prompt and one generator.

\paragraph{Near-copy positives with the recipe's near-misses stay at the floor and lose the most retention of any cell.}
One suspect for the verified arm's cross-language collapse is its counterexampled negatives: tiny notational edits might teach the model to fixate on notation. The cell that tests this keeps the verified arm's near-copy positives (one seeded CAS row per source) and swaps in the recipe arm's own near-misses, all of them ($6{,}145$ rows, $2.63$ per row; \texttt{build\_review2\_cells.py -{}-which r8}). Two readings were registered (R8-a/b): more than $1.72$ hard-tier points above the verified arm's $0.16$ would mean the recipe's near-misses lift even near-copy positives, and more than $6.54$ cross-language points ($2\sigma_1$ for that set) above the verified arm's $56.52$ would make the counterexamples the driver of the collapse. Neither happened. Hard R@1 is $0.11 \pm 0.03$ ($-0.05$ $[-0.13, 0.03]$ against the verified arm): the tier needs the deep-rewrite positive too, not only the right negatives. Cross-language retention fell to $12.47 \pm 0.68$, $44.05$ $[38.62, 49.34]$ points \emph{below} the verified arm and outside both registered branches. Replacing the counterexamples made the collapse worse, so they are not its driver, and near-copy positives set against the recipe's near-misses are the most destructive pairing we trained ($12.47$; only the pure SABER document channel's $10.69$ is lower). The same-language primary slice falls to $34.40$ ($-16.90$ $[-24.08, -10.06]$ against the verified arm) and the easy tier to $11.28$.

\paragraph{Back-translated positives: the non-LLM paraphrase control.}
The reference's $29.85$-point gain over the verified arm mixes two things, deep paraphrase and LLM authorship, because every deep paraphrase in the paper is LLM-written. The control separates them with a paraphrase no LLM wrote: each source problem is translated out and back through NLLB-200-distilled-600M, a dedicated machine-translation model rather than a general-purpose LLM (English through French and back, other languages through English; LaTeX spans masked and restored; no LLM judge), and paired with the reference's own counterexampled negatives (\texttt{scripts/backtranslate\_anchors.py}; $6{,}576$ rows, $1.26$ negatives per row; $186$ round trips came back identical to the source and $6$ failed a structural check, all dropped). The paraphrase is real but shallower than a free restatement: character-$3$-gram Jaccard to the source $0.783$ (median $0.797$), between the verified arm's $0.865$ and D4's $0.421$. Registered reading (R9): within $2.40$ easy-tier points of the reference, LLM authorship would not matter for the genre share; more than $2.40$ below, LLM authorship carries part of the reference's gain, reported as reference minus back-translation. It landed far below: $18.81 \pm 0.80$ easy R@1, $28.10$ $[26.74, 29.51]$ under the reference and only $1.76$ $[0.62, 2.82]$ above the verified arm, with $1.09 \pm 0.04$ hard R@1 ($0.93$ $[0.76, 1.10]$ above the verified arm). A non-LLM paraphrase covering the first $0.08$ of the character-overlap distance from the verified arm's positives toward D4's buys $1.76$ of the $29.85$ points; for depth alone to explain the share, the remaining $0.36$ of that distance would have to buy the other $28$. Depth and authorship are therefore separated down to the control's depth, and no further (Section~\ref{sec:limitations}). A second result rides on the same cell: with the very negatives that cost D4's restatements $15.79$ cross-language points, the back-translated positives sit at the base on both probes ($84.57 \pm 1.20$ cross-language against $84.73$; $57.87$ same-language against $62.40$, $-4.53$ $[-14.09, 5.07]$), so what the negatives cost in retention depends on the positives they are paired with. Trained alone, the back-translated positives score $12.92 \pm 0.72$ on the easy tier against D4's $30.03$ and D5's $18.71$ alone, the same authorship gap with no negatives in play, and $94.57 \pm 0.64$ on cross-language duplicates, the highest of all our models (Table~\ref{tab:factorial}).

\paragraph{A second recipe-free prompt recovers less of the easy-tier gap, so the recipe-specific share is a range.}
One objection is that the recipe-specific share might rest on a single recipe-free prompt: if D4 were an unusually retention-friendly prompt rather than the recipe family an unusually gaming one, the $15.47$-point attribution and the $27.24$ difference-in-differences would move with the prompt. We therefore wrote a second recipe-free restatement prompt, D5, framed differently from D4's audience change: a survey author restating another author's problem for a themed collection in their own wording, notation and order, level and mathematics unchanged, positives only (\texttt{scripts/gen\_survey.slurm}; same generator, judge pass, source problems and seed as the ladder; $6{,}293$ verified positives of $7{,}089$ sources, $698$ generation failures, $98$ judge rejections). D5's positives are slightly closer to their anchors than D4's (character-$3$-gram Jaccard $0.472$ against D4's $0.421$ and D1's $0.463$). Two cells were trained, three seeds each: D5 alone, and D5 with the reference's own counterexampled negatives per source, which we call the D5 reference (the main text's reference is the D4 reference here; \texttt{build\_review2\_cells.py -{}-which r10}; $6{,}163$ rows, $1.28$ negatives per row, $605$ D4 sources without a verified D5 positive dropped). Three readings were registered (\texttt{scripts/factorial\_cells.slurm}, R10-a/b/c). First, the recipe arm's easy-tier lead over the D5 reference within $2.40$ of the $15.47$ lead over the D4 reference would mean the share does not depend on the prompt, and further away would make it a range over both references. Second, the D5 cells within $6.54$ of their D4 counterparts' cross-language retention would mean D4 is not unusually retention-friendly. Third, the D5 reference within $1.72$ of the D4 reference's $10.30$ hard R@1 would replicate ``recipe-free rewrites with verified negatives reach the tier''. Outcomes: D5 alone scores $18.71 \pm 1.31$ easy R@1 against D4's $30.03$ at the same cross-language retention ($93.72 \pm 0.39$ against $93.81$), and the D5 reference $39.20 \pm 0.40$ easy, $3.97 \pm 1.05$ hard ($3.50$, $5.17$, $3.24$) and $78.71 \pm 0.59$ cross-language against the D4 reference's $46.91$ / $10.30$ / $78.02$. R10-a: the recipe arm leads the D5 reference by $23.18$ $[22.10, 24.23]$ easy-tier points (nested bootstrap as in the reference contrasts, third random stream, \texttt{results/reference\_control\_did.json}), $7.71$ more than the $15.47$ against the D4 reference and outside the band, so the recipe-specific share is reported as a range, $15.47$ to $23.18$ at verified negatives ($24.59$ at matched LLM negatives, below), and the difference-in-differences against the D5 reference is $35.63$ $[29.69, 41.41]$ against $27.24$. R10-b: both D5 cells retain what their D4 counterparts retain on cross-language duplicates ($+0.69$ $[-1.84, 3.16]$ and $-0.09$), so D4 is not unusually retention-friendly; the two prompts differ in how much of the easy tier they recover, not in what they keep. R10-c: the D5 reference is $6.33$ $[4.99, 7.54]$ hard-tier points below the D4 reference, outside the band, so the hard-tier level a recipe-free reference reaches depends on the restatement prompt; the verified negatives still lift D5 from $0.01$ to $3.97$, $3.81$ $[2.98, 4.90]$ above the verified arm. In the paper's terms, the recipe-free share of the easy-tier gap is $22.15$ $[21.06, 23.21]$ of $45.33$ points with D5 and $29.85$ with D4, half to two thirds, the range the abstract and Section~\ref{sec:gaming} carry. On same-language reprints D5 alone sits at the base ($64.53$ against $62.40$, within noise) and the D5 reference at $41.33$, beside the D4 reference's $39.20$ (Table~\ref{tab:samelang}).

\paragraph{Writing positives and near-misses in separate calls leaves the recipe arm's hard tier where it was.}
Why do D4's restatements beat the recipe's own positives on the hard tier when both get the recipe's near-misses? Three accounts are listed below, and we tested the first: that a positive and its near-misses written in one LLM call share phrasing, so the model learns to key on subtler features and gets less from the positive. We cut the Appendix-F prompt by string surgery into a positives-only call and a near-misses-only call, every surviving line verbatim D1's, with the judge pass, sources and seed unchanged (\texttt{scripts/gen\_split.slurm}, variant \texttt{exact\_split}; $6{,}516$ verified positives and $18{,}080$ verified near-misses, $2.61$ per row, $6{,}145$ rows trained per seed; the split-call positives sit at character-$3$-gram Jaccard $0.430$ to the anchor against the one-call arm's $0.463$). Registered (R11-a): within $1.72$ hard-tier points of the recipe arm's $9.42$, the one-call pairing is not what holds the recipe arm's hard tier down; more than $1.72$ above, the shared-phrasing account holds. It landed within the band, $9.79 \pm 0.66$ ($+0.37$ $[-0.41, 1.22]$), so the account is not supported. The split-call arm also stays $17.18$ points below D4's restatements under the recipe's own near-misses (R11-b), so which prompt writes the positives matters at those negatives whether or not the calls are separated. Separating the calls has costs, $3.91$ $[2.63, 5.28]$ easy-tier points ($58.47 \pm 1.18$), $14.18$ $[10.08, 18.40]$ cross-language points ($52.08 \pm 1.49$) and $9.77$ same-language points ($36.53$), and one gain, the medium tier rising to $12.37 \pm 0.49$ from the recipe arm's $8.84$ (R11-c, reported without a registered direction). Two accounts remain untested: that D4's restatements are written in a different register, and that D4's audience-change framing acts as a regularizer.

\paragraph{A second, independent judge leaves the recipe arm's easy-tier score in place, lowers its hard tier a little and raises its retention.}
The recipe arm's judge is the same model that generated its pairs, where the benchmark uses two independent judges. To see whether that matters, every candidate row of the recipe arm's generation ($25{,}656$) was re-judged by Mistral-Small-24B-Instruct-2501, a different vendor and model family, under the same judge prompt (\texttt{scripts/rejudge.slurm}; \texttt{generate\_llm\_pairs.py -{}-rejudge-input}), and a cell was trained on the rows both judges admit. Agreement (R12-a): of $6{,}414$ candidate positives the self-judge accepts $6{,}145$ and the second judge $5{,}522$; both accept $5{,}398$, only the first $747$, only the second $124$, neither $145$ (agreement $0.864$, but Cohen's $\kappa$ only $0.198$, since nearly everything is accepted and most agreement is by chance); of $19{,}242$ near-misses both accept $16{,}107$ (agreement $0.879$, $\kappa$ $0.346$); $1{,}110$ second-judge verdicts did not parse and count as not verified. The two-judge file has $5{,}398$ rows, below the $6{,}145$ budget, so the cell trains on all of them ($2.53$ negatives per row; positives at Jaccard $0.460$). Registered (R12-b): within $2.40$ easy, $1.72$ hard and $6.54$ cross-language points of the recipe arm, its numbers do not depend on the single self-judge; outside any band, the size is reported and Section~\ref{sec:setup} discloses the arm as single-judge. Outcomes over three seeds: easy $64.71 \pm 0.64$, $+2.33$ $[1.47, 3.25]$, inside the band; hard $7.58 \pm 0.39$, $-1.84$ $[-2.42, -1.27]$, just outside it; cross-language $74.81 \pm 2.65$, $+8.56$ $[4.85, 12.34]$, outside it in the arm's favour; same-language $46.13$ against $46.30$. The easy-tier headline therefore does not depend on the self-judge, while a stricter second judge trims the hard tier and gives back nearly half of the arm's cross-language loss ($8.56$ of the $18.47$ points it sits below the base).

\paragraph{Three back-translation hops deepen the non-LLM paraphrase but recover none of the reference's gain.}
Does the LLM-authorship attribution hold at a non-LLM paraphrase depth beyond the single hop's $0.783$? We ran the back-translation of the previous control three times in sequence (\texttt{scripts/bt2.slurm}: French, then German, then Russian, with English substituted for any pivot equal to the source language; formulas masked at every hop; a row kept only if every hop validates). $5{,}723$ of $6{,}768$ source problems survive ($1{,}038$ came back identical to the source after the chain, $7$ failed a structural check), at character-$3$-gram Jaccard $0.640$ to the source (median $0.644$), between the single hop and D4's $0.421$; the cell keeps the reference's negatives ($1.21$ per row). Registered (R13): within $2.40$ easy-tier points of the reference's $46.91$, deeper non-LLM paraphrase recovers the reference's gain and the attribution is retired; more than $2.40$ below, it survives to the deeper depth. It landed at $12.85 \pm 0.56$, $34.06$ $[32.81, 35.41]$ below the reference, below even the single-hop cell's $18.81$ and $4.21$ $[3.25, 5.15]$ below the verified arm, with $4.47 \pm 0.15$ hard R@1, $67.01 \pm 1.73$ cross-language ($11.02$ $[5.20, 16.94]$ below the reference) and $46.67$ same-language. The attribution survives, with a caveat the reading needs: three translation hops degrade the text rather than restate it, since a sixth of the rows return unchanged and the rest lose fluency, and the model loses retention with them. The added depth is therefore noise depth, not restatement depth, and this control bounds the depth account no more tightly than the single hop did. The hard tier reads the same event the other way round: with the negatives unchanged, the noisier positive lifts hard R@1 from $1.09$ to $4.47$, so what the tier pays for is the positive's distance from its minimal-edit near-miss, which translation noise supplies as well as a deep rewrite does, and not equivalence, which the noise can only degrade.

\paragraph{The remaining cells of the grid.}Seven further cells completed the grid of Table~\ref{tab:design}, built from existing positive and negative files with the grid's own conventions (computer-algebra negatives are the reference's per source; LLM near-misses are all of the partner row's, the recipe arm's volume; \texttt{scripts/build\_review4\_cells.py}), three seeds each, with no directional prediction registered (Table~\ref{tab:factorial}). Three things they add. First, positives trained alone never register on the hard tier, whoever wrote them: the recipe's own positives without negatives score $0.00$, back-translations $0.05$, beside D4's $0.04$ and D5's $0.01$; and paraphrase positives alone, LLM-written or back-translated, sit at the top on cross-language duplicates ($94.57$ for back-translations, $93.81$ for D4, $93.72$ for D5, within noise of one another), where the recipe's positives alone stay near the base ($83.21$). Second, LLM near-misses need a deep LLM rewrite to work with: on near-copy positives they reach $0.23$ hard R@1 from an unrelated prompt and $0.11$ from the recipe's, on back-translated positives $1.88$ and $4.18$, against $16.96$ to $26.97$ on D4's restatements, and in every one of these four cells cross-language retention collapses, to between $12.47$ and $38.67$, whichever prompt wrote the near-misses; the collapse follows the pairing of LLM near-misses with shallow positives, not the recipe's negatives in particular. Third, the recipe's own positives with the reference's negatives score $50.25$ easy, $2.09$ hard and $70.82$ cross-language, against the reference's $46.91$, $10.30$ and $78.02$ with the same negatives: the template's positives buy $3.34$ easy-tier points over D4's at verified negatives, reach a fifth of the reference's hard-tier score and lose $7.20$ more cross-language points; against the recipe arm itself ($62.38$, $9.42$, $66.26$), swapping its near-misses for verified negatives removes $12.13$ easy and $7.33$ hard points and returns $4.56$ cross-language points, so on the easy tier the recipe's near-misses help its own positives where they hurt D4's. One number we report without an account: D1 positives with an unrelated prompt's near-misses score $20.51 \pm 1.52$ on the medium tier ($22.22$, $19.31$, $20.01$), more than twice any other model we trained (the recipe arm $8.84$).

\paragraph{What this changes in the ladder's reading, and what it leaves standing.}The ladder's registered decay prediction passed and its numbers stand, but two of its steps now mean less than they seemed to. The D3-to-D4 drop ($4.67 \to 0.04$) is D4's missing negatives, not its prompt: D4's positives reach $10.30$ with the verified arm's negatives and $26.97$ with the recipe arm's. And within D1--D3 the ladder cannot tell whether the decline comes from the positives' wording or from the near-misses' quality, because each of those prompts writes both. What the ladder still shows on its own is a hard-tier decline across D1--D3, whichever mix drives it, and the exact template scoring lowest of the three on the easy tier. The two claims about the hard tier, that the template's positives are not needed and that its near-misses add an increment over an unrelated prompt's, are the factorial's results, not the ladder's.

\paragraph{The easy tier stays genre-level, and the verified arm's gain rides on its negatives.}Two easy-tier facts from the factorial. Attaching CAS negatives to D4's restatements lifts their easy R@1 from $30.03$ to $46.91$, still $15.47$ points below the recipe arm's $62.38$, so the genre reading of Section~\ref{sec:mechanism}, LLM-rewrite genre with a recipe-specific remainder on top, stands. The opposite corner of the grid runs the other way: CAS positives without negatives score $7.42$, below the untrained base's $8.32$, so everything the verified arm gains on the easy tier over the base ($8.32$ to $17.05$) comes from its counterexampled negatives, not its positives.

\paragraph{The two negative-free cells diverge by a cross-lingual loss whose cause is open.}
The two cells trained without negatives behave very differently on cross-language duplicates: D4's restatements alone reach $93.81$ strict R@1, above the untrained base, while CAS positives alone fall to $52.08$. On same-language reprints the two sit at the base together ($61.07$ and $58.13$; Appendix~\ref{app:setup}), so the difference is specifically about working across languages, not about recognising equivalent problems. Which property of D4 preserves that ability the factorial cannot say. Two candidates: D4's rows cover more source problems ($6{,}145$ against $4{,}621$ for the verified arm's draw at the same seed), and its prompt asks for a change of audience rather than an equivalence-preserving disguise, which may act as a regularizer the way V2's instruction does. Telling them apart would need a D4 trained on the verified arm's sources and a verified arm trained under an audience-change instruction, neither of which we have built.

\paragraph{Why the recipe's positives lose to D4's at fixed LLM negatives is open.}
Keep the recipe arm's own near-misses and swap its positives for D4's restatements: hard R@1 rises from $9.42$ to $26.97$ and cross-language retention from $66.26$ to $78.88$. The same holds at verified negatives: D1's positives with the reference's counterexamples reach $2.09$ where D4's reach $10.30$, and $70.82$ cross-language where D4's reach $78.02$ (Table~\ref{tab:factorial}). Why should another prompt's positives beat the benchmark's own? Three accounts fit the data. First, the recipe writes each positive and its near-misses in one call under one prompt, so they share phrasing, and the contrastive task must key on subtler features than when the positive comes from another prompt. Second, D4's audience-change restatements sit farther from the anchor in register, a deeper paraphrase against the same near-miss. Third, as the previous paragraph suggested for retention, the audience-change framing may act as a regularizer. The first was tested and is not supported: written in separate calls, the recipe's positives and near-misses leave its hard tier where it was ($9.79$ against $9.42$; above). The other two remain untested. Style tracking (Table~\ref{tab:style}) does not decide between them: the models trained on D4's positives carry little of the recipe fingerprint whether their negatives are verified or the recipe's own (the unrelated-prompt cells were not scored). Separating the remaining two would need a verified arm trained under an audience-change instruction, which we have not built.

\paragraph{The inversion under a single edit to the training file.}
Start from D4's restatements alone, at $93.81$ strict R@1 on cross-language duplicates, first among our trained models. Attach the verified negatives and the hard tier rises to $10.30$ while retention falls $15.79$ points to $78.02$, from first place to $6.71$ below the untrained base. Attach the recipe's near-misses instead and retention falls $9.92$ at matched count ($83.89$) and $14.93$ at full volume ($78.88$), for $20.71$ and $26.97$ hard R@1; on same-language reprints the verified negatives and the recipe's near-misses cost $13$ to $22$ points and an unrelated prompt's $11$ to $16$ (Table~\ref{tab:samelang}). Attach an unrelated prompt's near-misses and the cross-language cost is $8.31$ or $8.90$ points ($85.50$, $84.91$), which leaves the model at the untrained base. In every case a single edit that touches no positive and no prompt, only the negatives, lifts the hard tier and lowers retention from D4's $93.81$.

\paragraph{The recipe arm against the reference: the recipe-specific share, and how much of each step transfers.}
The reference, not the verified arm, is the comparison that isolates the recipe-specific share of the gap, and Table~\ref{tab:controlled} reads the headline gaps against both. The intervals come from \texttt{scripts/reference\_control\_did.py}, a nested bootstrap ($B = 5{,}000$) that resamples easy-tier queries and duplicate clusters paired across arms and each arm's eight training seeds independently. Recipe minus reference: $+15.47$ $[14.11, 16.73]$ on the easy tier and $-11.77$ $[-17.33, -6.09]$ on cross-language duplicates, so the gap shrinks by $27.24$ $[21.40, 32.93]$ from benchmark to real data, positive in every replicate and in each of the eight seed-paired comparisons ($24.75$ to $31.69$). Reference minus verified: $+29.85$ $[28.56, 31.15]$ easy-tier points, of which $+21.50$ $[15.31, 27.67]$ carry over to cross-language duplicates, so this gap shrinks by only $8.35$ $[1.93, 14.78]$. The two shrinkages sum to the published $35.59$ by construction. On the medium tier the recipe arm leads the reference by $4.62$ $[4.13, 5.15]$; on the hard tier the two are tied at R@1 ($-0.88$ $[-1.85, 0.10]$, an interval spanning zero) and the reference leads at R@5 ($-2.86$ $[-4.01, -1.67]$). On the same-language primary slice the recipe arm leads the reference by $7.10$ $[-2.34, 17.04]$, so the shrinkage there is $8.37$ $[-1.14, 18.16]$, both intervals spanning zero, while the reference trails the verified arm by $12.10$ (Table~\ref{tab:samelang}). The reference's real-data advantage over the recipe arm therefore shows up cross-lingually only.

\paragraph{The recipe-specific share at matched negatives.}
The reference carries the verified arm's counterexamples, so the $15.47$-point share compares training sets that differ in two ways, the positives' prompt and the negatives' origin. To isolate the prompt, compare the recipe arm with the D4 cells that carry the recipe's own near-misses (three seeds each; same bootstrap, separate random stream, \texttt{results/reference\_control\_did.json}). The recipe arm leads the count-matched cell by $17.79$ $[16.25, 19.38]$ easy-tier points and the full cell, which shares its negatives outright so that only the positives' prompt differs, by $24.59$ $[23.14, 26.29]$. On cross-language duplicates it trails them by $17.63$ $[12.21, 23.20]$ and $12.63$ $[7.51, 17.63]$, so the gap shrinks by $35.42$ $[29.56, 41.14]$ and $37.22$ $[32.00, 42.61]$ from benchmark to real data. On the hard tier it trails them by $11.29$ $[10.29, 12.37]$ and $17.55$ $[16.42, 18.81]$. The prompt buys more of the easy tier at LLM negatives than at verified ones because the counterexamples themselves lift the easy tier (the reference's $46.91$ against the LLM-negative cells' $44.59$ and $37.79$), and what it buys reaches neither duplicate probe: the same-language shrinkages are $19.22$ $[12.50, 25.47]$ and $17.76$ $[9.84, 25.14]$.

\paragraph{The style probe, and the artefacts stripped before fitting it.}The probe is a plain text classifier, a logistic regression over word $1$--$2$-gram tf-idf features. It is trained on matched pairs, each source problem's organic text against that same problem's Appendix-F rewrite, with the train/test split by source problem so that topic and language confounds cancel and only writing style is left to learn; on held-out problems it reaches $0.880$ AUC. Two formatting quirks of our organic corpus that exist nowhere in the benchmark are stripped before fitting and scoring: a leading ``\texttt{Problem:}'' OCR header ($63.88\%$ of organic anchors against $0.02\%$ of rewrites) and image placeholders ($4.01\%$ against $0.16\%$). Left in, they inflate the holdout AUC to $0.969$ while \emph{costing} cross-generator transfer, because the classifier learns our corpus's formatting rather than LLM style, so the stripped detector is the one reported throughout. Its heaviest weights (\texttt{results/style\_probe\_features.json}) say what the fingerprint is made of. On the rewrite side the top features are the register of a set exercise, the imperatives and connectives \emph{determine}, \emph{define}, \emph{let}, \emph{suppose}, \emph{show that}, \emph{where} and \emph{with} (coefficients $2.4$ to $6.1$), with two LaTeX habits beside them, the \texttt{\textbackslash text} macro and \texttt{\textbackslash sum} with a subscript; on the organic side they are \emph{we}, \emph{then}, \emph{so that}, \emph{find}, \emph{what}, \emph{which} and article phrases such as \emph{of the} and \emph{the triangle} ($-2.0$ to $-4.6$). The fingerprint is therefore mostly how the rewrites phrase and frame a problem, with macro standardisation a smaller second part, and nothing in it is mathematical content.

\paragraph{Cross-generator transfer is the load-bearing result.}The detector was trained only on our Qwen rewrites, yet it separates the benchmark's Gemini-3-flash gold rewrites, from a vendor whose text it never saw, from $1{,}668$ organic corpus problems at AUC $0.804$ ($0.807$ once the organic documents that also appear among the probe's training anchors are dropped; $0.778$ / $0.839$ / $0.796$ on the easy / medium / hard tiers). The fingerprint it learned is therefore a general LLM-rewrite fingerprint, not one specific to its own generator. The benchmark's minimal-edit near-miss distractors, which are small edits of organic problems and so inherit organic style, sit near chance ($0.562$), as they should.

\paragraph{The split caveat: $20$ held-out anchors have byte-identical training twins.}The probe's train/test split is by \emph{source id} rather than by text, and because some corpus problems appear under more than one id, $20$ of the $1{,}282$ held-out anchors have a byte-identical copy among the training negatives ($19$ on the rewrite side). At $1.6\%$ of the held-out set this cannot move the $0.880$ materially, but ``never seen in training'' would be the wrong words for it.

\begin{table}[htbp]
\centering
\caption{What the style detector detects. Each row takes one prompt's rewrites (held-out sources only) and gives the AUC with which the detector separates them from the same $1{,}282$ held-out organic problems, with $95\%$ CIs from $2{,}000$ bootstrap replicates over held-out source ids. The three recipe-family prompts D1--D3 are detected at about the same level and cannot be told apart within their intervals; the recipe-unrelated prompt D4 is detected far less strongly but still well above chance. The detector therefore mostly picks up the \emph{genre} of LLM rewrites, which D4 shares, plus a recipe increment of $0.204$ $[0.186, 0.222]$ that only the recipe-family prompts add.}
\label{tab:panel}
\small
\setlength{\tabcolsep}{4pt}
\begin{tabular}{@{}p{5.9cm}>{\centering\arraybackslash}p{2.2cm}>{\centering\arraybackslash}p{2.2cm}>{\centering\arraybackslash}p{2.8cm}@{}}
\toprule
Rewrite prompt & Held-out positives & AUC vs.\ organic & $95\%$ CI \\
\midrule
D1: the benchmark's Appendix-F template & $1{,}282$ & \textbf{0.8802} & $[0.8679, 0.8917]$ \\
D2: paraphrase of that template & $1{,}249$ & 0.8756 & $[0.8624, 0.8873]$ \\
D3: rewrite prompt in a different style & $1{,}096$ & 0.8654 & $[0.8520, 0.8788]$ \\
D4: restatement prompt unrelated to the recipe & $1{,}265$ & 0.6758 & $[0.6569, 0.6948]$ \\
\bottomrule
\end{tabular}
\end{table}

\paragraph{Genre with a recipe increment: what that costs the claim.}D4's prompt never mentions the recipe, yet against the same held-out organic anchors (Table~\ref{tab:panel}) its rewrites are still detected at $0.676$, so most of what the detector picks up is present in any LLM rewrite. The part specific to the recipe is the climb from D4 to D1, $0.204$ $[0.186, 0.222]$, positive in $100\%$ of bootstrap replicates. Within D1--D3 the detector resolves almost nothing: D1 $-$ D2 $= +0.005$ $[-0.006, 0.016]$ includes zero, and D1 $-$ D3 $= +0.015$ $[0.002, 0.027]$ excludes it but is a fourteenth of the D4-to-D1 climb, so the ladder of detectability inside the recipe family is at most a shallow one. That rules out the stronger characterization, ``a property of the recipe, not of the generator''. What the evidence supports is weaker: an LLM-rewrite genre signature that carries across generators, with a recipe increment on top.

\begin{table}[htbp]
\centering
\caption{Style tracking: does a model rate LLM-styled documents as more similar? For each of $2{,}000$ sampled queries against the full $117{,}088$-document corpus (single runs at seed $42$), the first column is the Spearman correlation between the model's similarity to each document and the artefact-stripped detector's style score for it, averaged over queries; positive means similarity rises with LLM style. The second column is where the model's top-$10$ documents sit in the corpus's style distribution, as a percentile, where $50$ means no preference (we report percentiles because the detector's raw score is an unbounded log-odds whose ratios have no scale meaning). The base and the verified arm show no style preference, D4's restatements alone none, and D4 with negatives of either kind only a faint one; the recipe-family prompts track clearly, in ladder order, with the exact template strongest.}
\label{tab:style}
\small
\setlength{\tabcolsep}{4pt}
\begin{tabular}{@{}p{7.4cm}>{\centering\arraybackslash}p{3.0cm}>{\centering\arraybackslash}p{3.0cm}@{}}
\toprule
Model & $\rho$(similarity, style) & Top-$10$ style percentile \\
\midrule
Base 0.6B (untrained) & $-0.061$ & 44.8 \\
Verified arm & $-0.019$ & 45.5 \\
D4: restatement prompt unrelated to the recipe & $-0.000$ & 48.1 \\
Reference (D4 $+$ CAS negatives) & $+0.025$ & 52.0 \\
D4 $+$ recipe-prompt near-misses, count-matched & $+0.046$ & 53.1 \\
D4 $+$ recipe-prompt near-misses, full & $+0.050$ & 53.5 \\
D3: rewrite prompt in a different style & $+0.105$ & 57.2 \\
D2: paraphrase of the benchmark's template & $+0.164$ & 61.6 \\
D1: the benchmark's template ($=$ recipe arm) & $\boldsymbol{+0.171}$ & \textbf{61.9} \\
\bottomrule
\end{tabular}
\end{table}

\paragraph{The recipe arm tracks the fingerprint; the verified arm does not.}Table~\ref{tab:style} correlates each model's similarity scores with the detector's style score over $2{,}000$ sampled queries against the full corpus. The untrained base mildly prefers documents that do \emph{not} look LLM-written ($\rho = -0.061$), the verified arm has no preference ($-0.019$), and the recipe arm prefers LLM-styled documents at $+0.171$, a $+0.190$ margin over the verified arm per query and higher on $94.20\%$ of queries (Wilcoxon $p < 10^{-308}$), so the difference is not carried by a few outliers. Seen from the retrieved side, the recipe arm's top-$10$ sits at the $61.9$th percentile of the corpus style distribution, where $50.0$ would be no preference, while the verified arm ($45.5$) and the base ($44.8$) retrieve slightly below-average-style documents.

\paragraph{The rungs scope the claim: the recipe's rewrite prompts as a family install it, not the template alone.}Is the tracking caused by the benchmark's exact prompt or by any prompt of its kind? Running the same statistic on the ladder rungs answers it (no new training was needed; their embedding caches existed). D2 tracks at $+0.164$, $96\%$ of the recipe arm's, D3 still at $+0.105$, and only D4, whose prompt never mentions the recipe, is flat ($-0.000$), matching the verified arm. We therefore claim that the recipe's rewrite prompts as a family install the tracking, more strongly the closer the prompt is to the template, and that neither D4's recipe-unrelated prompt nor verified training does. We do not claim that the template alone, or the recipe uniquely, installs it.

\paragraph{The reference and the recipe-negative cells track the fingerprint only faintly.}
Do the cells that score highest on the hard tier chase LLM style? The same statistic on the seed-$42$ models of the factorial's later cells (\texttt{scripts/style\_tracking\_review2.slurm}, cached embeddings, same detector) says barely: $+0.025$ for the reference, $+0.046$ for the count-matched cell with the recipe's near-misses and $+0.050$ for the full cell (top-$10$ percentiles $52.0$, $53.1$ and $53.5$), above D4's flat $-0.000$, below D3's $+0.105$, and a small fraction of the recipe arm's $+0.171$. Attaching verified negatives or the recipe's near-misses to D4's positives adds a little tracking (the unrelated-prompt cells were not scored); most of the recipe arm's tracking comes from its positives. The cells that buy the most hard-tier score ($20.71$ and $26.97$) therefore carry little of the fingerprint: whatever they learn to score on that tier, it is not style, and the probe does not explain it (Section~\ref{sec:limitations}).

\paragraph{The base-rate control: tracking survives within a single document class.}A pooled correlation over the whole corpus could be an artefact of ranking. The style score alone separates the benchmark's golds from its minimal-edit near-misses at AUC $0.750$, and the recipe arm is trained to rank golds above near-misses, so a correlation between similarity and style might only be picking up that ranking. Restricting to synthetic documents does not settle it: the correlation is unchanged ($+0.170$ vs.\ $-0.019$) but only $1.42\%$ of the corpus is removed, and the confound sits between the two synthetic classes. The test is to hold the class fixed. Within golds alone the recipe arm still tracks style, $\rho = +0.093$ against the verified arm's $-0.012$, and within near-misses alone $+0.086$ against $-0.009$ (\texttt{results/style\_within\_class.json}), roughly half the pooled margin. The tracking is therefore not an artefact of ranking golds above near-misses.

\paragraph{Hit accounting in full.}
On the hard tier the recipe arm alone ranks the gold first on $1{,}401$ of the $15{,}000$ queries and the verified arm alone on $15$ ($4$ both, $13{,}580$ neither). What makes those $1{,}401$ queries different: word overlap (token Jaccard) between query and gold separates them from the rest at AUC $0.639$ (means $0.21$ vs.\ $0.17$); overlap in \LaTeX{} spans separates them slightly ($0.557$); how densely the gold uses the recipe's phrasing and how LLM-styled it is do not separate them at all ($0.509$ and $0.511$, $p \ge 0.16$); and the gold/query length ratio runs weakly the \emph{opposite} way ($0.481$). The skew towards word overlap is a shifted distribution, not a rule: only $17.8\%$ of the wins fall among the $1{,}401$ queries with the highest overlap and $30.8\%$ sit below the median. The two probe-derived measures were computed with the pre-stripping detector; since both are null results, that does not matter.

\paragraph{Are published models already recipe-inflated? A pre-specified null.}The attack shows that a benchmark score can be inflated by training on pairs built like the benchmark's own, not that anyone has done so. To check the models actually on the leaderboard, we ran the same artefact-stripped detector on the same $2{,}000$-query sample over every public model whose corpus and query embeddings were already cached (\texttt{results/style\_probe\_public\_models.json}). As a check that the two runs are comparable, the untrained \texttt{base-0.6b} row gives the identical correlation ($-0.0610$) in both tables.

\begin{table}[htbp]
\centering
\caption{Do public models show the recipe-style signature? The same two measures as Table~\ref{tab:style}: $\rho$ is the mean per-query Spearman correlation between a model's query--document similarity and the document's LLM-style score, and the last column is where the model's top-$10$ documents sit in the corpus's style distribution, as a percentile, with $50$ meaning no preference. Every public model sits in the same narrow band as the untrained base and our verified arm, between $-0.07$ and $+0.004$ with top-$10$ percentiles of $45$ to $48$; our recipe-family models sit far away, at $+0.10$ to $+0.17$ and $57$ to $62$. No public model approaches the recipe arm.}
\label{tab:public-style}
\small
\setlength{\tabcolsep}{4pt}
\begin{tabular}{@{}p{6.4cm}>{\centering\arraybackslash}p{3.0cm}>{\centering\arraybackslash}p{3.4cm}@{}}
\toprule
Model & $\rho$ (all documents) & Top-$10$ style percentile \\
\midrule
\multicolumn{3}{@{}l}{\emph{Public models, never trained by us}} \\
Qwen3-Embedding-8B & $-0.0705$ & 45.21 \\
Qwen3-Embedding-0.6B (the untrained base) & $-0.0610$ & 44.83 \\
Qwen3-Embedding-4B & $-0.0548$ & 45.61 \\
MathLeap-Qwen-8B & $-0.0214$ & 47.04 \\
MathLeap-Octen-8B & $-0.0091$ & 47.94 \\
RaDeR-Qwen2.5-7B & $+0.0043$ & 47.23 \\
\midrule
\multicolumn{3}{@{}l}{\emph{Our models, same detector}} \\
Verified arm & $-0.0190$ & 45.50 \\
D4: restatement prompt unrelated to the recipe & $-0.0002$ & 48.05 \\
D3: rewrite prompt in a different style & $+0.1049$ & 57.15 \\
D2: paraphrase of the benchmark's template & $+0.1641$ & 61.57 \\
Recipe arm (the benchmark's template) & $\mathbf{+0.1705}$ & \textbf{61.85} \\
\bottomrule
\end{tabular}
\end{table}

\paragraph{Every public model sits in the untrained band; the null carries two limits.}
All six public models fall between $-0.07$ and $+0.004$, a band that also holds the untrained base and the verified arm, while the three recipe rungs sit at $+0.105$ to $+0.171$ and retrieve top-$10$ documents $7$--$12$ style percentile points above the no-preference rate of $50$ ($12$--$17$ above the base's $44.83$). We read this as no sign that any current leaderboard entry is recipe-inflated. Two limits on that reading. The detector is tuned to \emph{our} Qwen3-32B rewrites and transfers to the benchmark's golds at $0.804$ AUC, not perfectly, so a model trained on a recipe far from both could slip past it. And a null over six models, the base among them, says nothing about models we did not measure. The claim is therefore that the attack is available and, on this evidence, unused; it is not a clearance for any particular submission.

\section{Full result tables, the mixing grid, and the two external attacks in detail}
\label{app:results}

\begin{table}[htbp]
\centering
\caption{Real retrieval kept after training, beside benchmark score: each model's strict R@1 / R@5 / R@10 on cross-language duplicates against its hard-tier R@1. The two arms and the reference are means $\pm$ sample std over eight seeds, the factorial's other cells over three; the rest are single runs. Read the columns against each other. The recipe arm keeps a ${\sim}10$-point edge over the verified arm on duplicates, positive in every seed and not equivalent at the pre-specified $11.33$-point margin, but both sit well below the untrained base, and the ${\sim}60\times$ hard-tier separation between them on the benchmark is nowhere in sight. Three training sets keep retention at the base ($85.50$, $84.91$ and $84.57$ against $84.73$): D4 with an unrelated prompt's near-misses at either volume, which also buy $16.96$ and $20.93$ hard R@1, and back-translated positives with verified negatives. Near-copy CAS positives under the recipe arm's near-misses fall to $12.47$ and under an unrelated prompt's to $13.48$, the two lowest rows; back-translated positives trained alone keep the most, $94.57$, then D4's and D5's restatements alone ($93.81$, $93.72$). The last seven rows are the cells that complete the grid of Table~\ref{tab:design}.}
\label{tab:ood}
\footnotesize
\setlength{\tabcolsep}{4pt}
\resizebox{\linewidth}{!}{%
\begin{tabular}{lcc}
\toprule
Model & Hard-tier R@1 & Cross-language duplicates, strict R@1 / R@5 / R@10 \\
\midrule
Qwen3-Embedding-0.6B (base, untrained) & 0.00 & 84.73 / 95.17 / 96.95 \\
BM25 (lexical baseline) & 0.01 & 10.69 / 15.78 / 19.59 \\
V2 (ours, Appendix~\ref{sec:forgetting}) & 0.01 & 74.30 / 89.31 / 93.13 \\
Recipe arm & 9.42 $\pm$ 0.61 & 66.26 $\pm$ 2.31 / 83.68 $\pm$ 1.61 / 86.93 $\pm$ 1.10 \\
Verified arm & 0.16 $\pm$ 0.06 & 56.52 $\pm$ 4.05 / 76.21 $\pm$ 3.63 / 82.92 $\pm$ 3.14 \\
Reference (D4 $+$ CAS negatives) & 10.30 $\pm$ 1.15 & 78.02 $\pm$ 0.91 / 83.94 $\pm$ 1.51 / 85.62 $\pm$ 1.19 \\
D4 $+$ recipe-prompt near-misses, matched & 20.71 $\pm$ 0.82 & 83.89 $\pm$ 3.82 / 93.98 $\pm$ 1.91 / 95.42 $\pm$ 1.17 \\
D4 $+$ recipe-prompt near-misses, full & \textbf{26.97} $\pm$ 0.97 & 78.88 $\pm$ 2.33 / 89.31 $\pm$ 3.13 / 91.69 $\pm$ 2.50 \\
D4 $+$ unrelated-prompt near-misses, matched & 16.96 $\pm$ 2.08 & 85.50 $\pm$ 1.67 / 94.40 $\pm$ 0.26 / 95.67 $\pm$ 0.26 \\
D4 $+$ unrelated-prompt near-misses, full & 20.93 $\pm$ 0.44 & 84.91 $\pm$ 1.49 / 93.30 $\pm$ 0.39 / 95.25 $\pm$ 0.29 \\
CAS positives $+$ recipe-prompt near-misses & 0.11 $\pm$ 0.03 & 12.47 $\pm$ 0.68 / 18.24 $\pm$ 0.59 / 21.88 $\pm$ 0.77 \\
Back-translated positives $+$ CAS negatives & 1.09 $\pm$ 0.04 & 84.57 $\pm$ 1.20 / 92.62 $\pm$ 0.88 / 93.98 $\pm$ 0.53 \\
Three-hop back-translated positives $+$ CAS negatives & 4.47 $\pm$ 0.15 & 67.01 $\pm$ 1.73 / 86.34 $\pm$ 0.77 / 89.23 $\pm$ 1.06 \\
D5 (second recipe-free prompt, no negatives) & 0.01 $\pm$ 0.00 & 93.72 $\pm$ 0.39 / 97.37 $\pm$ 0.15 / 97.63 $\pm$ 0.14 \\
D5 $+$ CAS negatives (second reference) & 3.97 $\pm$ 1.05 & 78.71 $\pm$ 0.59 / 86.77 $\pm$ 0.92 / 88.30 $\pm$ 1.17 \\
Recipe arm, separate calls & 9.79 $\pm$ 0.66 & 52.08 $\pm$ 1.49 / 69.63 $\pm$ 2.89 / 76.00 $\pm$ 3.46 \\
Recipe arm, both judges ($5{,}398$ rows) & 7.58 $\pm$ 0.39 & 74.81 $\pm$ 2.65 / 87.62 $\pm$ 2.36 / 89.65 $\pm$ 0.64 \\
CAS positives $+$ unrelated-prompt near-misses & 0.23 $\pm$ 0.03 & 13.48 $\pm$ 0.92 / 19.51 $\pm$ 0.53 / 22.81 $\pm$ 0.73 \\
Back-translated positives, no negatives & 0.05 $\pm$ 0.01 & \textbf{94.57} $\pm$ 0.64 / 97.79 $\pm$ 0.14 / 98.05 $\pm$ 0.15 \\
Back-translated positives $+$ unrelated-prompt near-misses & 1.88 $\pm$ 0.12 & 35.20 $\pm$ 1.15 / 45.38 $\pm$ 1.70 / 49.19 $\pm$ 1.49 \\
Back-translated positives $+$ recipe-prompt near-misses & 4.18 $\pm$ 0.51 & 38.67 $\pm$ 1.42 / 49.96 $\pm$ 0.64 / 53.10 $\pm$ 1.06 \\
D1 positives, no negatives & 0.00 $\pm$ 0.00 & 83.21 $\pm$ 1.17 / 94.23 $\pm$ 0.82 / 96.18 $\pm$ 0.51 \\
D1 positives $+$ CAS negatives & 2.09 $\pm$ 0.22 & 70.82 $\pm$ 0.53 / 80.07 $\pm$ 0.53 / 82.44 $\pm$ 0.68 \\
D1 positives $+$ unrelated-prompt near-misses & 3.45 $\pm$ 0.03 & 47.41 $\pm$ 1.55 / 69.55 $\pm$ 2.45 / 77.35 $\pm$ 1.11 \\
Uncapped verified arm ($13{,}747$ rows) & 0.16 & 47.58 / 70.99 / 79.13 \\
\bottomrule
\end{tabular}}
\end{table}

\begin{table}[htbp]
\centering
\caption{The matched-budget experiment at seed $42$, with the full recall curves: R@1 / R@5 / R@10 on each tier over the full $117{,}088$-document corpus and $15{,}000$ queries per tier. Both arms train on $6{,}145$ rows with identical hyperparameters and differ only in the training file; the reference and the two cells that pair D4's restatements with the recipe arm's own near-misses (Table~\ref{tab:factorial}) are added at the same seed. Eight-seed means and intervals are in Table~\ref{tab:controlled}; over seeds the paired gap is $45.33 \pm 0.93$ easy, $5.70 \pm 0.44$ medium and $9.26 \pm 0.60$ hard R@1, and $48.84 \pm 0.70$ hard R@5. Two things the R@1 columns hide: on the easy tier the recipe arm's lead over the verified arm nearly vanishes by R@5 and R@10, so its edge is mostly about rank $1$; and the two cells with the highest hard-tier R@1 fall far below every other model on the medium tier at R@5 and R@10. The recipe arm's numbers are the recipe-matching exhibit, not capability.}
\label{tab:controlled-full}
\small
\resizebox{\linewidth}{!}{%
\begin{tabular}{lccc}
\toprule
Model & Easy R@1 / R@5 / R@10 & Medium R@1 / R@5 / R@10 & Hard R@1 / R@5 / R@10 \\
\midrule
Qwen3-Embedding-0.6B (base, untrained) & 8.32 / 78.61 / 87.78 & 1.78 / 60.91 / 75.49 & 0.00 / 4.08 / 17.60 \\
Verified arm & 17.31 / 85.10 / 92.27 & 2.47 / 41.91 / 57.74 & 0.13 / 9.59 / 26.49 \\
Recipe arm & \textbf{61.88} / 89.94 / 93.63 & \textbf{9.06} / 55.18 / 66.79 & 9.37 / 57.06 / 70.13 \\
Reference (D4 $+$ CAS negatives) & 46.51 / 88.06 / 93.75 & 4.34 / 42.20 / 60.62 & 10.81 / 60.27 / 75.58 \\
D4 $+$ recipe-prompt near-misses, matched & 45.85 / 82.89 / 89.03 & 2.76 / 27.85 / 40.01 & 20.31 / 64.69 / 73.99 \\
D4 $+$ recipe-prompt near-misses, full & 38.58 / 70.74 / 77.01 & 2.67 / 20.83 / 28.73 & \textbf{26.47} / 63.82 / 71.40 \\
\bottomrule
\end{tabular}}
\end{table}

\begin{table}[htbp]
\centering
\caption{Public models on MathNet-Retrieve (R@1 / R@5 / R@10 per tier, sorted by easy R@1), the comparison behind the statement that no public model exceeds $0.01$ hard-tier R@1: RaDeR~\citep{das2025rader}, MathLeap~\citep{ye2026meld}, Qwen3-Embedding~\citep{qwen3embedding} and ReasonIR~\citep{shao2025reasonir}. It is a comparison set, not everything we ran: Qwen3-Embedding-8B was also evaluated on all three tiers ($10.05$ / $2.23$ / $0.00$) and all-mpnet-base-v2, the harness calibration model, on the easy tier only. Bold marks the best value in each column among the public models (hard-tier R@1 ties at $0.01$). Baselines use their own encoding conventions; the Qwen3-Embedding-4B row is the unprompted $2{,}048$-token run (its prompted $1{,}024$-token run scores $11.95$ easy R@1, and neither reproduces the $14.76$ the benchmark publishes; Appendix~\ref{app:setup}). Our own models sit below the rule and are excluded from the ranking because their training material is of the benchmark's own kind: the recipe arm trains on the benchmark's recipe, and the verified-supervision models' rename and reformulation positives, though symbolically verified, are in the same paraphrase genre the easy tier tests; ``---'' marks the one configuration we did not run.}
\label{tab:leaderboard-full}
\scriptsize
\setlength{\tabcolsep}{4pt}
\begin{tabular}{lccc}
\toprule
Model & Easy R@1 / R@5 / R@10 & Medium R@1 / R@5 / R@10 & Hard R@1 / R@5 / R@10 \\
\midrule
\multicolumn{4}{@{}l}{\emph{Public models}} \\
RaDeR-gte-Qwen2-7B & \textbf{18.94} / 95.57 / 99.05 & 1.12 / 63.88 / 83.24 & 0.01 / 8.63 / 34.63 \\
RaDeR-Qwen2.5-7B & 17.41 / \textbf{96.91} / \textbf{99.51} & 0.92 / 60.05 / 80.14 & 0.01 / 9.53 / 37.86 \\
MathLeap-Octen-8B & 15.87 / 84.17 / 94.40 & \textbf{3.51} / \textbf{75.90} / \textbf{91.97} & 0.01 / \textbf{17.53} / \textbf{62.57} \\
MathLeap-Qwen-8B & 13.79 / 81.32 / 93.16 & 3.37 / 75.56 / 91.15 & 0.01 / 17.01 / 60.86 \\
Qwen3-Embedding-4B & 11.15 / 85.65 / 93.34 & 1.93 / 70.19 / 85.15 & 0.01 / 6.04 / 28.58 \\
ReasonIR-8B & 10.61 / 78.25 / 86.55 & 2.41 / 52.01 / 66.82 & 0.00 / 2.02 / 8.50 \\
Qwen3-Embedding-0.6B (the untrained base) & 8.32 / 78.61 / 87.78 & 1.78 / 60.91 / 75.49 & 0.00 / 4.08 / 17.60 \\
BM25 (lexical baseline, no embedding) & 3.07 / 55.53 / 64.71 & 1.58 / 33.39 / 43.39 & 0.01 / 1.61 / 5.35 \\
\midrule
\multicolumn{4}{@{}l}{\emph{Our models ($0.6$B unless noted), excluded from the ranking}} \\
Recipe arm & 61.88 / 89.94 / 93.63 & 9.06 / 55.18 / 66.79 & 9.37 / 57.06 / 70.13 \\
Verified arm, $4$B (LoRA) & 39.51 / 96.61 / 98.91 & --- / --- / --- & 0.23 / 29.57 / 60.95 \\
V2 (mixed data $+$ instruction) & 21.70 / 94.48 / 97.81 & 2.71 / 67.25 / 82.06 & 0.01 / 7.02 / 32.03 \\
V1 (mixed data) & 20.83 / 91.96 / 96.71 & 3.67 / 63.57 / 78.80 & 0.03 / 7.67 / 30.01 \\
Soup $\alpha{=}0.7$ (WiSE-FT) & 16.41 / 90.44 / 96.04 & 2.33 / 52.23 / 69.83 & 0.01 / 7.54 / 28.14 \\
Uncapped verified arm & 15.92 / 78.18 / 87.21 & 3.86 / 40.80 / 55.92 & 0.16 / 6.59 / 19.47 \\
\bottomrule
\end{tabular}
\end{table}

\begin{table}[htbp]
\centering
\caption{The two remedies of Appendix~\ref{sec:forgetting}: mixing in weight space against mixing in data space (R@1 / R@5 / R@10 per tier; strict R@1 on cross-language duplicates). The soups blend the uncapped verified arm's weights with the untrained base's, $\alpha$ being the share of the trained arm, so $\alpha{=}0.3$ sits nearest the base and $\alpha{=}0.7$ nearest the arm; V1 and V2 instead train on verified pairs mixed with problem-to-solution replay, V2 under an instruction prompt. The soups recover retention as $\alpha$ falls but lose the hard-tier R@1 the arm had bought at every blend; V2 keeps most of both score and retention. The two control rows show how much the instruction alone does: applied to the untrained base it gives the best retention in the table, and V2 without it loses about six points. Base, uncapped-arm and soup rows are deterministic single evaluations; V1 and V2 show the seed-$42$ run, with three-seed means in Appendix~\ref{sec:forgetting}. The soups of V1 test whether the two remedies compose (Appendix~\ref{sec:forgetting}).}
\label{tab:mixing}
\scriptsize
\setlength{\tabcolsep}{3pt}
\begin{tabular}{@{}p{3.2cm}ccc>{\centering\arraybackslash}p{1.7cm}@{}}
\toprule
Model & Easy R@1 / R@5 / R@10 & Medium R@1 / R@5 / R@10 & Hard R@1 / R@5 / R@10 & Cross-language dup.\ R@1 \\
\midrule
Base 0.6B (untrained) & 8.32 / 78.61 / 87.78 & 1.78 / 60.91 / 75.49 & 0.00 / 4.08 / 17.60 & \textbf{84.73} \\
Uncapped verified arm & 15.92 / 78.18 / 87.21 & \textbf{3.86} / 40.80 / 55.92 & \textbf{0.16} / 6.59 / 19.47 & 47.58 \\
Soup $\alpha{=}0.3$ (weights: $0.3$ arm, $0.7$ base) & 13.55 / 93.57 / 97.93 & 1.59 / 59.93 / 76.35 & 0.01 / 4.80 / 23.13 & 80.15 \\
Soup $\alpha{=}0.5$ & 15.77 / 93.42 / 97.68 & 1.77 / 57.30 / 74.17 & 0.01 / 6.28 / 27.51 & 73.54 \\
Soup $\alpha{=}0.7$ & 16.41 / 90.44 / 96.04 & 2.33 / 52.23 / 69.83 & 0.01 / 7.54 / 28.14 & 62.34 \\
V1 (verified pairs $+$ replay) & 20.83 / 91.96 / 96.71 & 3.67 / 63.57 / 78.80 & 0.03 / \textbf{7.67} / 30.01 & 59.80 \\
V2 (the same mix $+$ instruction) & \textbf{21.70} / 94.48 / 97.81 & 2.71 / \textbf{67.25} / \textbf{82.06} & 0.01 / 7.02 / \textbf{32.03} & 74.30 \\
\midrule
\multicolumn{5}{l}{\emph{Soups of V1 (weight mixing on top of replay; V1's seed-$42$ weights with the base's)}} \\
V1 soup $\alpha{=}0.3$ & 11.18 / 88.02 / 94.86 & 2.06 / 64.43 / 79.37 & 0.00 / 5.07 / 23.42 & 86.77 \\
V1 soup $\alpha{=}0.5$ & 13.70 / 91.73 / 96.82 & 2.19 / 66.12 / 80.79 & 0.01 / 5.75 / 26.81 & 83.97 \\
V1 soup $\alpha{=}0.7$ & 16.63 / 93.37 / 97.45 & 2.49 / 66.27 / 81.09 & 0.01 / 6.35 / 29.65 & 78.37 \\
\midrule
\multicolumn{5}{l}{\emph{Instruction controls (seed $42$; the instruction is V2's, applied at evaluation time)}} \\
Base $+$ V2's instruction, no training & 13.21 / 80.33 / 89.22 & 2.35 / 61.31 / 77.61 & 0.00 / 6.95 / 26.41 & \textbf{90.59} \\
V2 evaluated without its instruction & 16.51 / \textbf{94.71} / \textbf{98.01} & 1.79 / 65.33 / 80.25 & 0.00 / 5.43 / 26.28 & 68.70 \\
\bottomrule
\end{tabular}
\end{table}

\begin{table}[htbp]
\centering
\caption{Three benchmarks we had no hand in building, single runs, every cell measured: MELD R@1, BRIGHT nDCG@10 on its two math theorem splits (TT $=$ theoremqa-theorems, AoPS $=$ the AoPS split), and ImpliRet~\citep{taghavi2025impliret} nDCG@10, macro-averaged over its six subsets (category $\times$ style) of $1{,}500$ queries each. On MELD, whose pairs are LLM-written, the recipe arm doubles the base's score while the verified arm stays flat; on BRIGHT both arms fall below the base, the verified arm to almost nothing, the recipe arm less; on ImpliRet's $9{,}000$ queries the recipe arm is the worst model we measure while the two mixed-data models hold the base's level. Our pre-registered prediction that the two arms would score within noise of each other is therefore partially refuted, on MELD and BRIGHT, in the recipe arm's favour, on BRIGHT by intervals that exclude zero. The untrained $4$B model is shown for scale.}
\label{tab:external}
\footnotesize
\setlength{\tabcolsep}{3pt}
\begin{tabular}{@{}p{3.9cm}>{\centering\arraybackslash}p{1.6cm}>{\centering\arraybackslash}p{2.4cm}>{\centering\arraybackslash}p{2.6cm}>{\centering\arraybackslash}p{2.4cm}@{}}
\toprule
Model & MELD R@1 & BRIGHT-TT nDCG@10 & BRIGHT-AoPS nDCG@10 & ImpliRet nDCG@10 \\
\midrule
Qwen3-Embedding-0.6B (base, untrained) & 10.19 & 21.24 & 7.44 & 16.03 \\
Verified arm & 9.44 & 1.83 & 0.00 & 13.14 \\
Recipe arm & 18.89 & 10.19 & 0.75 & 9.23 \\
V1 (verified pairs $+$ replay) & 9.63 & 16.90 & 1.59 & 16.04 \\
V2 (the same mix $+$ instruction) & 10.19 & 15.26 & 2.55 & \textbf{16.20} \\
Qwen3-Embedding-4B (untrained) & 18.89 & \textbf{31.94} & 8.69 & 14.99 \\
\bottomrule
\end{tabular}
\end{table}

\paragraph{The easy-tier gap replicates on two further backbones and at $4$B.}
Does the gap depend on our base model? We repeated the two-arm experiment on two other embedding models and at a larger size, one seed each. On multilingual-e5-large (seed $42$) the recipe arm leads by $+19.12$ easy-tier R@1; on the hard tier the inflation shows up at depth (R@5 $12.62$ against $2.46$) while R@1 stays near zero, and on real duplicates the lead is $+2.29$. On BGE-large-en-v1.5 (seed $42$, same rows and hyperparameters) easy R@1 is $24.25$ for the recipe arm against $8.34$ for the verified arm, a $+15.91$ lead; the medium tier is tied ($1.08$ against $1.09$), both arms sit at the hard-tier floor ($0.07$ against $0.00$), and strict real-duplicate R@1 is $22.14$ against $15.78$, a $+6.36$ lead, less than half the benchmark one. Both BGE duplicate numbers are low in absolute terms because the backbone was pretrained on English only and the probe is cross-lingual (the untrained base scores $5.85$ there). On the English same-language probe, where that handicap is absent, the BGE base scores $57.60$ primary-slice R@1 and the two arms $58.40$ (verified) and $57.60$ (recipe), within noise of it and of each other (recipe minus verified $-0.80$ $[-9.45, 7.94]$ under a cluster bootstrap; $53.91$, $54.78$ and $53.91$ on the $115$ English queries), so on its own language the backbone keeps its retention under both arms and the recipe arm holds no lead there, as on Qwen. At $4$B (rank-$16$ LoRA, same rows) the recipe arm scores $77.77$ easy R@1 against the verified arm's $39.51$, over an untrained base of $11.95$; hard R@1 is $6.03$ against $0.23$ over $0.00$; and on real duplicates the order flips, $70.99$ for the recipe arm against $83.72$ for the verified arm.

\paragraph{A late-interaction retriever shows the same pattern.}
Late-interaction models score a query against a document by token-level MaxSim rather than by one vector per text, so the recipe's surface regularities could in principle be matched, or ignored, differently. We repeated the two-arm experiment on ColBERTv2~\citep{santhanam2022colbertv2}, a BERT-base, English-only backbone, fine-tuned through PyLate~\citep{chaffin2025pylate} on the recipe arm's file and on the verified arm's file under the same $6{,}145$-row cap, contamination gates and $5\%$ by-source dev split as the dense arms (learning rate $10^{-5}$, batch $32$, three epochs, $1{,}449$ and $933$ steps, query length $256$ and document length $512$ tokens; \texttt{scripts/colbert\_train.py}). Two things differ from the dense recipe: the loss is PyLate's contrastive loss over MaxSim scores, in-batch negatives plus the row's own, the late-interaction counterpart of the dense trainer's; and there is no checkpoint selection, the final model being kept. Scoring re-ranks a PLAID index's top-$1{,}000$ candidates by exact MaxSim under the dense harness's rules otherwise (\texttt{scripts/colbert\_eval.py}; Table~\ref{tab:colbert}). The benchmark gap reappears at rank~$1$: the recipe arm scores $39.05$ easy R@1 against the verified arm's $6.57$ over an untrained $3.97$, a $+32.48$ lead, and $4.28$ against $2.17$ on the medium tier; at R@5 and R@10 the medium tier is level or reversed ($30.02$ against $30.19$, $40.09$ against $41.43$), and the hard tier moves only at depth ($2.35$ / $5.64$ against $1.24$ / $3.53$ at R@5 / R@10, R@1 at $0.08$ against $0.01$). Retention runs the other way. The cross-language probe cannot serve here, the English-only backbone finding $1.78$ strict R@1 of it untrained ($1.78$ and $6.36$ after training), so we read retention on the same-language reprint probe of Appendix~\ref{app:setup}: on its primary slice the verified arm keeps the untrained model's $46.40$ ($0.00$ $[-5.56, 5.56]$ under the cluster bootstrap) while the recipe arm falls to $35.20$, $-11.20$ against either ($[-20.16, -2.42]$ against the verified arm, $[-21.60, -0.80]$ against the base); on the $115$ English queries the three score $41.74$, $41.74$ and $30.43$ ($-11.30$ $[-21.05, -1.77]$); on all $498$ queries $74.30$, $73.90$ and $66.06$ ($-7.83$ $[-11.65, -4.23]$), the recipe arm losing even on exact-text reprints ($91.75$ against $99.31$). So on a late-interaction model the recipe arm's $32$-point benchmark lead comes with a retention loss of $8$ to $11$ points that the verified arm does not incur, the inversion of Section~\ref{sec:gaming} in a sharper form than the dense $0.6$B arms show, where on cross-language duplicates both arms lose retention and the recipe arm loses less. One seed and one run per cell; the intervals resample duplicate clusters only (\texttt{results/colbert\_summary.json}).

\begin{table}[t]
\centering
\caption{The two-arm experiment on a late-interaction retriever (fact source: \texttt{results/colbert/}, \texttt{results/colbert\_summary.json}). ColBERTv2 fine-tuned through PyLate on the recipe arm's and the verified arm's training files, the same $6{,}145$ rows, gates and dev split as the dense arms; seed $42$, one run each. Benchmark columns are R@1 / R@5 / R@10 over the full $117{,}088$-document corpus with the query's own entry masked, from a PLAID index's top-$1{,}000$ candidates re-scored by exact MaxSim, a gold outside the candidates counting as a miss. The retention columns are same-language R@1 on the reprint probe's primary slice ($125$ queries) and on all $498$ of its queries (Table~\ref{tab:samelang}); the cross-language probe is unusable for this English-only backbone (untrained $1.78$ strict R@1, the arms $1.78$ and $6.36$).}
\label{tab:colbert}
\footnotesize
\setlength{\tabcolsep}{4pt}
\resizebox{\linewidth}{!}{%
\begin{tabular}{@{}lccccc@{}}
\toprule
Model & Easy & Medium & Hard & Same-lang.\ primary & Same-lang.\ all \\
\midrule
ColBERTv2 (untrained) & 3.97 / 52.79 / 60.86 & 1.51 / 32.80 / 42.37 & 0.01 / 1.33 / 3.43 & 46.40 & 74.30 \\
Verified arm & 6.57 / 59.83 / 69.66 & 2.17 / 30.19 / 41.43 & 0.01 / 1.24 / 3.53 & 46.40 & 73.90 \\
Recipe arm & \textbf{39.05} / \textbf{75.69} / \textbf{81.44} & \textbf{4.28} / 30.02 / 40.09 & \textbf{0.08} / \textbf{2.35} / \textbf{5.64} & 35.20 & 66.06 \\
\bottomrule
\end{tabular}}
\end{table}

\paragraph{What the two candidate fixes for forgetting are.}
Appendix~\ref{sec:forgetting} compares two fixes. The first works on the weights: WiSE-FT weight-space ensembles~\citep{wortsman2022wiseft}, the soups of Table~\ref{tab:mixing}, average the uncapped verified arm's trained weights with the untrained base's, $\theta_\alpha = \alpha\,\theta_{\text{ft}} + (1-\alpha)\,\theta_{\text{base}}$, with $\alpha \in \{0.3, 0.5, 0.7\}$ the share of the trained weights. The second works on the data: a mixture of $30{,}323$ rows from four sources, $8{,}747$ CAS pair rows with negatives capped at two per row (dropping $3{,}063$ minimal-edit negatives), $12{,}000$ problem$\to$solution rows as replay, $4{,}576$ cross-transform positives and $5{,}000$ same-topic mid-difficulty negatives. Trained on that mixture with the default query prompt the model is V1; trained on the identical data with the task instruction \emph{``Given a math problem, retrieve problems that are mathematically equivalent to it.''} attached to queries, it is V2.

\paragraph{Data-space mixing recovers recall depth.}The uncapped verified arm had lost recall depth, the gold dropping out of the top few results even when it was not first; the mixture brings it back. Medium-tier R@5 climbs to $63.57$ (V1) and $67.25$ (V2), past the untrained base's $60.91$. Across three training seeds ($42$--$44$), V2 scores $22.19 \pm 0.79$ easy R@1 ($21.70/21.78/23.10$) while keeping $74.22 \pm 2.42$ on cross-language duplicates, and V1 scores $20.64 \pm 0.28$ easy R@1.

\paragraph{Two seed-$42$ readings do not replicate at three seeds.}Two of V1's numbers in Table~\ref{tab:mixing}, which shows the seed-$42$ run, do not hold up over three seeds. Its medium-tier R@1 of $3.67$ was a lucky draw (three-seed mean $2.67 \pm 0.87$), and its cross-language retention of $59.80$ an unlucky one ($66.8 \pm 6.1$), so neither seed-$42$ figure should be read as V1's level.

\paragraph{V2's instruction alone scores $90.59$ on the untrained base.}Take the \emph{untrained} base, do no training, and prepend V2's instruction to each query at evaluation time: it scores $90.59$ on cross-language duplicates, above every mixed, souped or controlled fine-tune in the grid, and among our MathNet-trained models surpassed only by D4 ($93.81$). For V2 itself the instruction is worth ${+}5.19$ easy R@1 and ${+}5.60$ retention points at evaluation time: without it V2 falls from $21.70$ to $16.51$ on the easy tier and from $74.30$ to $68.70$ on retention.

\paragraph{One loss stands.}The uncapped verified arm was the only non-recipe model to score anything on the hard tier at rank~$1$, $0.16$, meaning it could occasionally rank the true equivalent above the near-misses. Neither remedy keeps that: every soup and both mixed-data models fall to $0.01$--$0.03$.

\paragraph{Registration labels, and the script headers that carry them.}
Section~\ref{sec:ood} reports the pre-registered predictions in prose; this paragraph gives each one a label and names the script header that states it, so the exact wording can be found. SABER-Math (\texttt{scripts/saber\_attack.slurm}): P-S1, the document attack gains $+0.02$ nDCG@10 and doubles the construction-signal correlation (failed); P-S2, the attacked arm falls below the base and below V2 on real duplicates (held); P-S3, the two arms differ on SABER by less than the attack's own gain over the base (failed); P-S4 groups the summary-channel predictions, L1--L3 in \texttt{saber\_label\_attack.slurm} and their seed replication S1--S2 in \texttt{saber\_label\_seeds.slurm} (held). MELD (\texttt{attack\_second\_benchmark.slurm}): P-M1, the attacked arm beats the base (passed); P-M2, it falls below the base on real duplicates (refuted; it rose above it); P-M3a and P-M3b, it beats the MathNet recipe arm on MELD and inherits no MathNet hard-tier score (passed); P-M4, an arm trained on disjoint topics captures most of the gain (so the effect is genre-level). The ladder (\texttt{dose\_response.slurm}) and the factorial's reading map (\texttt{factorial\_cells.slurm}) are stated in Appendix~\ref{app:mechanism}, the same-language readout (\texttt{samelang\_eval.slurm}) in Appendix~\ref{app:setup}, and the regeneration defence (P-D1--P-D3, \texttt{defense\_regen\_eval.py}) below.

\paragraph{The SABER-Math attack: construction, and a one-sided gate.}
The attack set uses SABER's own core-idea prompt and its own relevance rule, summary word overlap (Jaccard) above the authors' threshold $\tau = 0.211$, generated by Qwen3-32B-AWQ over $4{,}917$ source problems (SABER's pipeline uses GPT-OSS-120B). Every row was checked against every SABER query and candidate document for overlap in word $5$-grams, dropping anything above $0.35$ containment, and trained with the controlled experiment's hyperparameters at a $1{,}024$-token cap. That check has a flaw: containment divides by the length of the training \emph{source}, so it asks only how much of the source appears in a SABER text, not the reverse. In that direction no trained source exceeds $0.3421$; in the reverse direction, how much of the SABER text appears in the source, containment reaches $0.5333$, $11$ sources exceed $0.35$, and three of them restate an evaluation query. A check run in both directions would have caught them. They cannot move an nDCG over $1{,}000$ queries, but ``disjoint'' claims more than was verified.

\paragraph{Three fidelity limits travel with the SABER attack.}Our copy of SABER's pipeline is not exact in three ways. SABER decides relevance in more than one stage, and we reproduced only the summary-overlap stage, not its MathWorld-BMA matching nor its Swiss-tournament grading. The threshold $\tau = 0.211$ is the value the authors tuned on \emph{their} corpus; we used it unchanged on ours, where it may not be the right cutoff. And the summaries were written by Qwen3-32B-AWQ where SABER's were written by GPT-OSS-120B, so the attack is cross-vendor, as the MathNet and MELD attacks are, and any part of SABER's rule that lives in its generator's phrasing rather than in the prompt is not copied.

\paragraph{P-S1's verdict is robust to how the two sides are encoded.}The document attack's failure does not depend on how queries and documents were prompted. Promptless on both sides, the attack changes nDCG@10 by $-0.0129$ $[-0.0212, -0.0042]$ and raises the construction-signal correlation $1.14\times$; with the query prompt applied, the two pairings give $-0.0098$ and $-0.0127$ at $1.18\times$. Every version is far from both registered thresholds, a $+0.02$ gain and a $2\times$ ratio. The attacked model has the highest construction-signal correlation of any model we measured and still scores below the base: matching the rule a little better bought no score.

\paragraph{The controlled arms on SABER, and a context-length check.}On SABER the recipe arm scores $0.507$ and the verified arm $0.451$, both below the untrained base's $0.5747$. One inconsistency in how they were run: those two rows were encoded at the model's $32{,}768$-token default, while the base and attacked rows were cut at $1{,}024$ tokens. Re-encoding the two arms at the $1{,}024$-token cap used elsewhere gives $0.5069$ and $0.4498$, essentially unchanged, so P-S3's verdict does not depend on context length.

\paragraph{The summary-channel re-attack, and what the main run trained on.}The generation pipeline produced two kinds of rows and wrote them into one pair file: $11{,}461$ rows built on SABER's document template, and $4{,}733$ rows whose positive is the LLM-written core-idea summary itself, a far shorter text (mean $93$ characters against $1{,}506$ for a document row). The main $6{,}145$-row run shuffled that file before truncating, so it trained on $4{,}359$ document rows and $1{,}786$ summary rows, $29\%$ summaries. That mixture scored $-0.0129$ against the base, between the two pure channels and below the base: the summary rows buy a gain when trained on alone, and diluting them to $29\%$ of the file erased it.

\paragraph{Both pure channels replicate over three seeds.}
The two pure arms, one trained on summary rows only and one on document rows only, each carry three training seeds ($42$--$44$). The summary arm's construction-signal correlation is $0.2968$ against the base's $0.2609$, a $1.14\times$ rise, the same rise the promptless document attack showed (\texttt{results/saber\_label\_4733\_noprompt.json}). Seed by seed, the summary arm scores $0.6101 / 0.6099 / 0.6069$ and the document arm $0.5515 / 0.5591 / 0.5645$ against the base's $0.5747$: the summary arm clears the base in every seed ($+0.0343 \pm 0.0018$), the document arm falls below it in every seed ($-0.0163 \pm 0.0065$), and the difference between the channels is $+0.0506 \pm 0.0081$. On cross-language duplicates at seed $42$ the document arm keeps only $10.69$ strict R@1 and the summary arm $58.78$, against $34.86$ for the mixed run and $84.73$ for the base. The document surface therefore carries most of the retention loss, and the pure document arm is the deepest forgetting of any model we trained.

\paragraph{Four asymmetries qualify the channel comparison: two favour the summary arm that wins it, two run against it.}
(i)~\emph{Content, not just rows, and it favours the summary arm.} The document channel is full of repeats: its $11{,}461$ rows come from only $4{,}119$ distinct sources and $3{,}374$ distinct positives, so its $4{,}733$-row draw holds $3{,}118$ sources and $2{,}368$ distinct positives, against the summary arm's $4{,}733$ and $4{,}706$. The winning arm trained on twice as many distinct positives. (ii)~\emph{Steps are not matched, in the summary arm's favour.} The summary arm ran $105$ optimiser steps and the document arm $57$, because \texttt{NoDuplicatesBatchSampler}, which never places two examples of one row in the same batch, yields fewer batches from data with many duplicated positives. Both ran three full passes, so the comparison is matched in \emph{rows} and unmatched in \emph{steps}. One internal control bounds what the steps buy: the document channel at $108$ steps \emph{with} checkpoint selection still scores $0.5618$, so $51$ extra document steps are worth at most $+0.0103$ against a channel difference of $+0.0506$. That is an upper bound, since the $108$-step run also carried the $29\%$ summary admixture; the clean control, a summary arm retrained at the document arm's $57$ steps on $2{,}368$ rows (one per source; $2{,}360$ to $2{,}365$ distinct positives by seed), three seeds, gains $+0.0255 \pm 0.0041$ nDCG@10 over the base ($0.5976$, $0.6050$, $0.5981$), every seed above the registered $+0.02$ bar, against $+0.0343 \pm 0.0018$ unmatched and $-0.0163 \pm 0.0065$ for the document arm at that budget; its cross-language retention, $58.02$ to $61.07$ strict R@1, matches the unmatched arm's $58.78$ (\texttt{scripts/saber\_matched.slurm}, \texttt{results/saber\_matched.json}). (iii)~\emph{Compute runs the other way.} The document arm consumed $2{,}189$\,s of GPU time against the summary arm's $1{,}188$\,s. (iv)~\emph{Data damage runs the other way too.} $52$ of the $4{,}733$ summary rows ($1.10\%$) carry control characters from the \verb|\frac| escape defect, against $10$ of $11{,}461$ document rows, so the arm that gains is the one trained on the more corrupted file. With the first two now controlled, Section~\ref{sec:ood} calls the summary-channel result a small gain, on an artefact SABER never scores, that survives a matched budget.

\paragraph{BRIGHT, and how small it is; ImpliRet, and how large.}The two BRIGHT math splits are tiny: $76$ queries against $23{,}839$ documents for theoremqa-theorems and $111$ against $188{,}002$ for AoPS, too few for seed intervals, so each single run carries a paired query bootstrap instead ($10{,}000$ resamples of the query set, the same draw applied to both models; \texttt{scripts/bright\_bootstrap.py}). On theoremqa-theorems the recipe arm leads the verified arm by $8.36$ $[3.42, 13.99]$ nDCG@10 points ($15$ queries in its favour, $2$ against) and on AoPS by $0.75$ $[0.16, 1.52]$ ($5$ against $0$); both arms trail the base by intervals that exclude zero, $-11.05$ $[-17.25, -5.26]$ (recipe) and $-19.41$ $[-25.66, -13.56]$ (verified) on theoremqa-theorems, $-6.69$ $[-9.92, -3.76]$ and $-7.44$ $[-10.65, -4.57]$ on AoPS. The re-scoring that produced the per-query values reproduces every committed aggregate to five decimals. ImpliRet's $9{,}000$ queries make it the large-$n$ witness the BRIGHT splits cannot be. On it both controlled arms fall below the untrained base, the recipe arm worst at $9.23$ against the base's $16.03$, while the mixed models (V1 $16.04$, V2 $16.20$) hold the base's level. All are single seed-$42$ runs on a harness that is consistent across our own models but does not reproduce the ImpliRet authors' published table (Appendix~\ref{app:setup}).

\begin{table}[t]
\centering
\caption{The eleven natural-language tasks of MIRB~\citep{ju2025mirb}, a human-curated mathematical retrieval benchmark, scored with its protocol (nDCG@10 $\times 100$ over the top $1{,}000$, each query's excluded ids dropped, documents as title plus text; the model's own query prompt, as for every external benchmark here). Untrained base against the two seed-$42$ arms; the last two columns are paired query bootstraps ($5{,}000$ resamples, $95\%$ intervals), bold where the interval excludes zero. The MSE duplicate-question task ranks $1{,}350{,}505$ documents per query; the three formal-library premise tasks are outside the paper's domain.}
\label{tab:mirb}
\scriptsize
\setlength{\tabcolsep}{2.5pt}
\resizebox{\linewidth}{!}{%
\begin{tabular}{@{}lrccccc@{}}
\toprule
Task & $n$ & Base & Verified arm & Recipe arm & Recipe $-$ verified & Recipe $-$ base \\
\midrule
MSE duplicate questions & 25116 & 48.94 & 32.64 & 28.05 & \textbf{$-4.59$ $[-5.07, -4.11]$} & \textbf{$-20.90$ $[-21.41, -20.38]$} \\
MathOverflow duplicate questions & 225 & 72.67 & 52.37 & 48.22 & $-4.15$ $[-9.81, 1.45]$ & \textbf{$-24.45$ $[-30.37, -18.72]$} \\
MSE question to answer (ARQMath-3 Task 1) & 78 & 40.04 & 29.40 & 41.66 & \textbf{$+12.26$ $[7.72, 16.91]$} & $+1.62$ $[-2.91, 6.41]$ \\
ProofWiki question to answer & 1099 & 75.41 & 59.35 & 56.63 & $-2.72$ $[-5.43, 0.02]$ & \textbf{$-18.78$ $[-21.26, -16.30]$} \\
Stacks question to answer & 776 & 44.58 & 34.59 & 29.19 & \textbf{$-5.40$ $[-8.07, -2.84]$} & \textbf{$-15.39$ $[-18.00, -12.79]$} \\
ProofWiki premises & 1086 & 20.12 & 15.89 & 11.29 & \textbf{$-4.60$ $[-5.87, -3.39]$} & \textbf{$-8.83$ $[-10.17, -7.53]$} \\
Stacks premises & 775 & 35.66 & 33.99 & 20.83 & \textbf{$-13.15$ $[-15.19, -11.22]$} & \textbf{$-14.82$ $[-16.92, -12.83]$} \\
Real-analysis premises & 161 & 32.40 & 26.03 & 21.47 & \textbf{$-4.56$ $[-8.55, -0.77]$} & \textbf{$-10.93$ $[-14.77, -7.28]$} \\
Number-theory premises & 38 & 34.28 & 29.62 & 34.05 & $+4.43$ $[-6.65, 15.16]$ & $-0.23$ $[-13.01, 12.09]$ \\
MSE formulas & 76 & 63.35 & 56.88 & 55.24 & $-1.63$ $[-6.76, 3.62]$ & \textbf{$-8.11$ $[-12.23, -3.97]$} \\
Wikipedia formulas & 39 & 76.33 & 69.82 & 73.65 & $+3.83$ $[-1.54, 9.16]$ & $-2.68$ $[-7.38, 1.86]$ \\
\bottomrule
\end{tabular}}
\end{table}

\paragraph{Human-curated math retrieval: MIRB.}On eleven MIRB tasks (Table~\ref{tab:mirb}) the untrained base leads both arms on ten, and the recipe arm trails the verified arm on five by intervals that exclude zero (the two Stacks tasks, ProofWiki premises, real analysis and the MSE duplicate questions) while leading it on one, the MSE question-to-answer task built from ARQMath~\citep{zanibbi2020arqmath}, where it holds the base's level. Both arms therefore lose on established human-curated sets, the recipe arm more, the same ordering as on the duplicate probes: the easy-tier lead does not reach benchmarks that people built. (\texttt{scripts/eval\_mirb.py}, \texttt{results/mirb/}.)

\paragraph{The MELD attack: what we reconstructed, and at what fidelity.}Part of MELD's construction is public: its record schema, its nine framing pairs and its thirty-items-per-call generation count, and its appendix prints the prompts behind its training pipeline. The prompt that produced its evaluation set is not. MELD's statements were generated by Claude Opus 4.7, reviewed by hand for dissimilarity and then checked by GPT-5.5 for validity, equivalence and residual similarity; we regenerate with Qwen3-32B-AWQ, so this attack is cross-vendor like the MathNet one. Lacking the prompt itself, we rebuilt the procedure from MELD's description into that schema. That is a weaker copy than the SABER attack, which used the repository prompt byte for byte, or the MathNet attack, which used the published Appendix-F prompt. Because the reconstruction is approximate and still moved the score, the result is best read as an upper bound on how much keeping the prompt secret protected the benchmark.

\paragraph{Two arms: MELD's own domain pairs against held-out ones.}Both MELD-attack arms are capped at $3{,}150$ trainer rows and trained with the controlled experiment's hyperparameters; each row carries three negatives, same-framing statements from our own generated pool rather than minimal-edit near-misses, since MELD does not describe how its distractors were built. \textsc{meld9} uses the same nine framing pairs as MELD's own evaluation set; the build checks that they reproduce MELD's released set exactly and aborts otherwise. \textsc{heldout} uses $27$ other pairs ($26$ of which yield rows), chosen so that no field name matches any MELD subfield. Two partial overlaps remain as substrings, ``Riemannian geometry'' against ``geometry'' and ``universal algebra'' against ``algebra''; we report them rather than silently allow them.

\paragraph{The disjointness figure: $39$ pairs removed for eval proximity, not $62\%$.}MELD's data is written from scratch, with no source corpus to exclude problems from, so the only way to keep training data away from its evaluation set is to check each generated pair against it. That check is separate from quality filtering, and the two counts should not be combined. For the meld9 arm, $4{,}163$ pairs were generated and $2{,}588$ discarded, but $2{,}549$ of those fell to MELD's own validity and equivalence screen, a recipe step that makes the data better, not more disjoint. The gates against the $1{,}081$ released evaluation texts removed $20$ pairs for $n$-gram containment and $19$ at cosine $\ge 0.9$ (all-mpnet-base-v2). For the heldout arm the screen rejected $2{,}345$ of $4{,}150$ and the gates removed none. The number of pairs removed for closeness to the evaluation set is therefore $39$, not the $62\%$ that lumping both filters together would suggest.

\paragraph{Removing those pairs can only weaken the attack, and residual exposure is bounded.}The $39$ removed pairs were the ones most similar to evaluation items, so removing them can only weaken the attack. After the gate, the closest any training pair comes to any evaluation text is a cosine of $0.899$, with no exact match. To bound what remains, we count evaluation pairs \emph{both} of whose items have a training pair within cosine $0.70$: $19$ of the $270$. Even if every one of those were memorised, they could explain at most $7.04$ of the $26.48$ R@1 points gained at seed $42$ (the same value as the $+7.04$ domain increment below, by coincidence). One bookkeeping note: the released certificate records its \texttt{max\_cosine\_vs\_eval} distribution before the drop test, which is why the maximum it lists is above the threshold the gate enforces.

\paragraph{Scoring protocol, and why our MELD numbers compare only against each other.}Every model is scored the same way, in one prompt condition, on MELD's own pairs-only protocol: $270$ pairs give $540$ queries, each ranked against the other $539$ items, and intervals resample the $270$ pairs as clusters with $B = 10{,}000$. Re-running our earlier models reproduces their scores bit for bit (base $10.19$, recipe arm $18.89$, verified arm $9.44$), which shows the harness is stable but is not a check against MELD's own numbers: MELD publishes no row for Qwen3-Embedding-$0.6$B, and for the two models we can compare we score higher than they do (Qwen3-Embedding-4B $18.89$ against their $13.7$, MathLeap-Octen-8B $34.44$ against their $28.9$; Appendix~\ref{app:setup}). Our MELD numbers are therefore comparable with one another and not with MELD's published table.

\paragraph{What topic-disjoint does and does not guarantee.}``Topic-disjoint'' means only that the field names do not match. The gate compares each framing's name, after normalization, against a $23$-term list ($18$ MELD subfield names plus five alternate names MELD gives the same areas), so it cannot know that ergodic theory sits inside MELD's measure-theory dialect pair or that Markov chains are probability. The mathematics still overlaps, which blurs the topic comparison in both directions. How much matching MELD's own domains buys is also narrower than one number suggests: the meld9 arm's $+7.04$ $[2.96, 10.93]$ lead over the heldout arm appears at R@1 alone; at R@5 ($+1.48$ $[-2.96, 5.93]$) and on the fraction of queries ranked above every same-framing distractor ($-0.37$ $[-3.89, 3.15]$) the two arms match. Knowing MELD's genre therefore buys nearly all of the attack, and knowing its exact domains adds only a rank-$1$ edge.

\paragraph{The prompt control's design, and its $80/20$ verdict.}
P-M3a compares MELD's attacked arm with the MathNet recipe arm, which changes task, source data, budget and negative design at once. To split the MELD gain into the format and the exact wording, a third arm holds everything fixed, the nine domain pairs, the schema, the generator, the judge, the gates, the $3{,}150$-row budget and $3.0$ negatives per row, and changes only the prompt's framing: a graduate reading-course translation exercise instead of a research mathematician building an evaluation set. That prompt also drops the instruction to make the two statements lexically distinct, the clause of MELD's procedure most closely tied to what its benchmark measures. The third arm scores $32.41$ against meld9's seed-$42$ $36.67$, a paired difference of $4.26$ $[0.93, 7.78]$ that excludes zero, while still clearing the base by $+22.22$ $[17.59, 26.85]$: $84\%$ of the attack survives rewriting the prompt and $16\%$ is the wording. Against the topic-disjoint arm its paired difference does not exclude zero ($+2.78$ $[-0.56, 6.11]$). What MELD is exposed to is therefore mainly the cross-dialect equivalence format, with a real but minority contribution from the wording. The third arm's cross-language duplicate score is unchanged at $90.84$ and its MathNet hard R@1 is $0.04$, so neither the P-M2 refutation nor P-M3b depends on which prompt generated the data. Regenerated under meld9's own procedure (a greedy first round and the pair-target stop, which halted after $16$ rounds at $4{,}086$ raw pairs; $3{,}508$ gated rows from $1{,}754$ sources, of which the cap trains $3{,}150$), the control scores $31.48$: $+21.30$ $[16.48, 26.11]$ over the base, $-5.19$ $[-8.89, -1.48]$ against meld9 and $-0.93$ $[-4.26, 2.59]$ against the unmatched control, so the generation factor moved nothing and the split with it matched is $80/20$; its cross-language score is $92.11$ and its hard R@1 $0.05$ (\texttt{scripts/meld\_altprompt\_matched.slurm}, group \texttt{meld9altm}).

\paragraph{One generation factor was unmatched in the first control, and matching it changed little.}
A later audit of our own configuration found one setting the two arms did not share. The meld9 arm was generated with a pair target and a greedy first round, and it stopped after $16$ of $24$ rounds once it had $3{,}150$ usable rows, so it trains on $100\%$ of its pool, from $1{,}575$ sources. The control ran all $24$ rounds to $5{,}204$ rows, from which the shared row cap took a random $60.5\%$ covering $2{,}215$ sources: $41\%$ more distinct sources at an identical budget, and a sampled rather than greedy first round. More source diversity plausibly helps the arm that has it, the control, in which case $4.26$ underestimates the wording's contribution; which way the round-$1$ decoding difference cuts we cannot tell. The $84/16$ split was therefore a decomposition with one unmatched generation factor; the matched rerun above, which removes it, lands within noise of the first control and gives $80/20$.

\paragraph{Seed replication: all four registered MELD readings hold.}
Both arms were retrained at seeds $43$ and $44$ from the same gated pair files, so the disjointness certificates still describe what these models saw. All four registered readings hold. The meld9 arm reaches $37.16 \pm 1.19$ pairs-only R@1, beating the base by $+26.48$, $+26.11$ and $+28.33$ in the three seeds, and the heldout arm $30.80 \pm 1.12$. The advantage of matching MELD's own domains keeps its sign in every seed ($+7.04$, $+5.37$, $+6.67$; mean $+6.36$), and the attacked model stays above the untrained base on cross-language duplicates in all three seeds, $90.67 \pm 0.53$ against $84.73$. That last result was the one most at risk, since it reverses what we had predicted: the registration in \texttt{scripts/meld\_attack\_seeds.slurm} would have withdrawn P-M2's refutation had the three-seed mean fallen below $84.73$. It did not.

\paragraph{Two asymmetries, and a registration that landed the other way.}Two things weigh on the MELD comparison in opposite directions: the attack arms carry $3.0$ negatives per row against the recipe arm's ${\approx}2.63$, which works for the attack, and at $3{,}150$ rows they train on roughly half the controlled budget, which works against it. And one registered prediction came out backwards. We had predicted that the attacked model would fall below the untrained base on cross-language duplicates, as every MathNet-trained arm does; it rose above the base instead, to $90.84$ and $91.86$ strict R@1, where the SABER-attacked model fell to $34.86$. Whatever MELD's recipe teaches carries over to genuine cross-language duplicates, so on this benchmark gaming the score and keeping real ability are not in opposition. We registered the opposite and report it as it landed.

\paragraph{The medium tier: no inflation against the verified arm, a smaller recipe-specific share against the reference.}
On the medium tier the recipe arm leads the verified arm by $5.70$ points, less than the $9.74$ points separating the same two arms on cross-language duplicates, so the interaction runs the opposite way from the easy tier's (Section~\ref{sec:ood}): the DiD is $-4.03$, $95\%$ CI $[-10.12, 1.82]$, one-sided $p(\text{DiD} > 0) = 0.91$, negative in seven of eight seeds. This is the genre gradient of Section~\ref{sec:mechanism} at its floor: the medium golds are LLM rewrites like the easy tier's, so the recipe arm leads there, but by no more than it leads on organic reprints, with nothing on top. The medium gap therefore carries no inflation against the verified arm. Against the reference the recipe arm leads by $4.62$ on the tier (Table~\ref{tab:controlled}) and trails by $11.77$ on cross-language duplicates, a recipe-specific shrinkage of about $16$ points by subtraction of the point estimates, against $27.24$ on the easy tier; the inflation claim rests on the easy tier, where the effect is largest and carries an interval.

\paragraph{The contrasts at R@5 and R@10: the inflation lives at rank 1.}
At R@5 the recipe arm leads the verified arm by $7.33$ $[5.90, 8.96]$ on the easy tier and by $7.48$ $[2.60, 12.85]$ on cross-language duplicates, so the difference-in-differences is $-0.15 \pm 3.15$ across the eight seed-paired comparisons (positive in four of eight); at R@10 the two leads are $2.88$ $[1.86, 4.03]$ and $4.01$ $[-0.19, 8.58]$, a difference of $-1.13 \pm 2.11$ (positive in three). Against the reference the easy-tier lead at R@5 is $0.90$ $[-0.16, 1.93]$ and the cross-language lead $-0.26$ by subtraction of the eight-seed means in Table~\ref{tab:ood}. The $35.59$- and $27.24$-point shrinkages of Table~\ref{tab:controlled} are therefore rank-$1$ quantities: recipe-matching decides which document comes first, and by R@5 the two arms and the reference retrieve the gold about equally often on the benchmark and on real duplicates. Intervals are the nested bootstraps behind Table~\ref{tab:controlled} (\texttt{results/final\_stats.json}, \texttt{results/reference\_control\_did.json}); the seed-paired means and standard deviations come from the per-seed evaluation files.

\paragraph{Verified-supervision numbers, and why the $4$B one stays out of the ranking.}
The $4$B verified arm's $39.51$ easy R@1, from supervision containing no LLM-generated text, is the highest verified-supervision number we measure. We keep it out of Table~\ref{tab:leaderboard-full}'s ranking, as its caption says, because its rename and reformulation positives are paraphrases, and paraphrase is the genre the easy tier tests. Given Sections~\ref{sec:gaming} and~\ref{sec:mechanism}, that number and V2's easy-tier lead over RaDeR-gte are positions on a recipe-sensitive instrument, not capability claims. The inversion holds at this scale: the recipe arm's $77.77$ against $39.51$ on the easy tier comes with cross-language retention of $70.99$ against $83.72$, the reverse order. That $39.51$ comes from verified pairs alone shows the easy tier is learnable, and a verified arm doing well by entering the paraphrase genre is no counter to the charge, which concerns the $15.47$ recipe-specific points that do not reach cross-language duplicates (Section~\ref{sec:gaming}), not the tier's learnability.

\paragraph{A lexical baseline.}BM25~\citep{robertson2009bm25} (\texttt{bm25s}, Lucene scoring, $k_1 = 1.5$, $b = 0.75$, lowercased word tokens with English stopwords removed, no stemming; \texttt{scripts/eval\_bm25.py}) is scored on every set under the dense harness's rules, the query's own corpus entry masked and a gold counted at the number of documents scoring strictly above it. On the benchmark it reaches $3.07$ / $55.53$ / $64.71$ easy R@1/5/10, $1.58$ / $33.39$ / $43.39$ medium and $0.01$ / $1.61$ / $5.35$ hard (Table~\ref{tab:leaderboard-full}), below the untrained base at rank~$1$ on every tier, so the recipe arm's $45$-point lead is not overlap a bag of words could collect. On cross-language duplicates it scores $10.69$ strict R@1 against the base's $84.73$ (Table~\ref{tab:ood}), and on same-language reprints $54.40$ on the primary slice against $62.40$ ($-8.00$ $[-17.07, 0.79]$ under a cluster bootstrap; Table~\ref{tab:samelang}), so the duplicate sets are not solved by term matching either, even where the reprints share a language. The one set where term matching does well is the near-miss probe, $72.27$ full-corpus R@1 and $73.8$ against its near-misses alone against the base's $10.84$ (Table~\ref{tab:casprobe}), because a computer-algebra edit leaves most tokens in place; that probe measures discrimination between two near-copies, and a lexical matcher passes it.

\begin{table}[t]
\centering
\caption{Who built the test decides who wins. Rows are training files; columns are test sets, grouped by who wrote the documents being retrieved; each cell is R@1. The verified arm wins only on the test that computer algebra built, as its own pairs were ($\dagger$: in-family, disclosed rather than read). The two arms trained on the recipe family's pairs take the first two places on every test an LLM built, and the D3 arm leads the recipe arm by most on the golds its own prompt wrote; the reference barely moves when the golds are rebuilt, so on the different-style golds it is level with the recipe arm ($44.00$ against $44.13$; Table~\ref{tab:defense}). On the near-miss probe the D3 arm outscores the recipe arm. The untrained base wins on duplicates no generator wrote. A better training file would win everywhere. Means over eight seeds for the two arms and the reference, three for the D3 arm, single seed-$42$ checkpoints in the two rebuilt-golds columns ($1{,}525$ queries). Sources: Tables~\ref{tab:controlled}, \ref{tab:dose-full}, \ref{tab:defense} and~\ref{tab:casprobe}.}
\label{tab:crosseval}
\footnotesize
\setlength{\tabcolsep}{2pt}
\begin{tabular}{@{}>{\raggedright\arraybackslash}p{3.5cm}*{5}{>{\centering\arraybackslash}p{1.9cm}}@{}}
\toprule
& \multicolumn{3}{c}{Documents written by an LLM} & Computer algebra & Human authors \\
\cmidrule(lr){2-4}\cmidrule(lr){5-5}\cmidrule(lr){6-6}
Trained on & Easy tier (Gemini, the benchmark's prompt) & Golds rebuilt by Qwen, the benchmark's prompt & Golds rebuilt by Qwen, a different-style prompt & Near-miss probe & Cross-language duplicates \\
\midrule
Untrained base & 8.32 & 10.69 & 3.28 & 10.84 & \textbf{84.73} \\
Verified arm: computer-algebra pairs & 17.05 & 15.02 & 15.80 & \textbf{97.42}$^\dagger$ & 56.52 \\
Recipe arm: benchmark's prompt, Qwen & 62.38 & \textbf{50.75} & 44.13 & 60.49 & 66.26 \\
D3 arm: different-style prompt, Qwen & \textbf{67.88} & 50.56 & \textbf{55.41} & 68.50 & 65.39 \\
Reference: unrelated prompt, verified negatives & 46.91 & 46.03 & 44.00 & 53.51 & 78.02 \\
\bottomrule
\end{tabular}
\end{table}

\begin{table}[t]
\centering
\caption{The regeneration defence of Appendix~\ref{sec:forgetting} (fact source: \texttt{results/defense\_eval/}). Each row is the same $1{,}525$ easy-tier queries with their gold documents regenerated a different way; the columns give each model's easy-tier R@1 on that version over the full $117{,}088$-document corpus (seed-$42$ checkpoints), and Gap is the recipe arm (D1) minus the verified arm. Down the Gap column: regenerating the golds with the benchmark's own template trims the lead only from $45.64$ to $35.73$, a different-style prompt to $28.33$, and drawing each gold from one of three prompts to $31.48$. Across rows: D2 beats D1 on every version and D3 on all but the exact-template one, so the easy-tier plateau of Section~\ref{sec:mechanism} recurs, and D3's largest lead over D1 is on the golds written with D3's own prompt. The reference column was scored after the registered readings were taken (\texttt{eval\_<version>\_reference.json}): the reference drops only from $46.89$ to $44.00$ across the versions where the recipe arm drops from $61.18$ to $44.13$, so on the different-style golds the two are level, and the recipe arm's lead over the reference on these items shrinks from $14.29$ to $4.72$ under the benchmark's own template and to $0.13$ under the different style.}
\label{tab:defense}
\footnotesize
\setlength{\tabcolsep}{3pt}
\begin{tabular}{@{}>{\raggedright\arraybackslash}p{4.4cm}cccccccc@{}}
\toprule
Gold documents & Base & Verified arm & D1 (recipe) & D2 & D3 & D4 & Reference & Gap \\
\midrule
Original benchmark golds & 7.34 & 15.54 & 61.18 & 64.92 & 67.41 & 30.43 & 46.89 & 45.64 \\
Regenerated with the benchmark's template & 10.69 & 15.02 & 50.75 & 55.67 & 50.56 & 28.59 & 46.03 & 35.73 \\
Regenerated with a paraphrase of it (D2's prompt) & 8.39 & 16.72 & 54.36 & 60.07 & 56.92 & 28.79 & 45.97 & 37.64 \\
Regenerated in a different style (D3's prompt) & 3.28 & 15.80 & 44.13 & 52.92 & 55.41 & 21.70 & 44.00 & 28.33 \\
Each gold from one of the three, at random & 7.02 & 16.26 & 47.74 & 54.95 & 54.23 & 25.84 & 43.93 & 31.48 \\
\bottomrule
\end{tabular}
\end{table}

\begin{table}[t]
\centering
\caption{The regeneration defence rerun with the benchmark's own vendor (fact source: \texttt{results/defense\_eval\_gemini/}). The same $1{,}700$ sampled easy-tier queries, the same three prompts and the same single judge as Table~\ref{tab:defense}, but the golds are rewritten by gemini-3-flash-preview through the vendor's API instead of by Qwen3-32B-AWQ; the $1{,}127$ queries the judge accepted under all three prompts are scored over the full corpus with the same seed-$42$ checkpoints. Easy-tier R@1; Gap is D1 minus the verified arm. The original-gold row differs from Table~\ref{tab:defense}'s because the slice does. The Gap column keeps its shape (paraphrase at or above exact, style lowest, mixed between), but every regenerated row collapses for every model, the base to $0.00$ at rank~$1$, because the vendor's fresh rewrites are a deeper rewrite than the golds the benchmark shipped (text).}
\label{tab:defense-gemini}
\footnotesize
\setlength{\tabcolsep}{3pt}
\begin{tabular}{@{}>{\raggedright\arraybackslash}p{4.4cm}ccccccc@{}}
\toprule
Gold documents & Base & Verified arm & D1 (recipe) & D2 & D3 & D4 & Gap \\
\midrule
Original benchmark golds & 7.10 & 16.33 & 60.43 & 63.27 & 65.04 & 28.48 & 44.10 \\
Regenerated with the benchmark's template & 0.00 & 0.53 & 20.14 & 17.48 & 10.56 & 0.53 & 19.61 \\
Regenerated with a paraphrase of it (D2's prompt) & 0.00 & 0.71 & 22.54 & 19.61 & 11.80 & 0.44 & 21.83 \\
Regenerated in a different style (D3's prompt) & 0.00 & 0.62 & 17.21 & 18.10 & 14.02 & 0.53 & 16.59 \\
Each gold from one of the three, at random & 0.00 & 0.80 & 18.54 & 18.19 & 11.71 & 0.53 & 17.74 \\
\bottomrule
\end{tabular}
\end{table}

\paragraph{The regeneration defence experiment: protocol.}
We rewrote the gold documents of $1{,}525$ easy-tier queries \emph{in place}, using the benchmark's own pipeline through Qwen3-32B-AWQ under each of the ladder's three recipe-family prompts (D1--D3), and re-scored the existing seed-$42$ checkpoints on five versions of the evaluation: the original golds, one version per prompt, and a \emph{mixed} version drawing each gold from one of the three at random. The $1{,}525$ are the queries, of $1{,}700$ sampled, whose regenerations the single-judge filter accepted under all three prompts. Distractors, pool size and relevance labels are untouched, so only the gold texts differ between versions. P-D1--P-D3 and a falsifier were written into \texttt{scripts/defense\_regen\_eval.py} before it ran. Three caveats travel with the numbers. Regenerating a gold changes its author as well as its text, from Gemini-3-flash to Qwen3-32B-AWQ, so the drop from the original golds to the exact-template ones mixes cross-generator transfer with item freshness (a rerun through the vendor's own model, below, replicates the pattern but not the level, and does not remove the confound). The generation is a single seeded run. And requiring the judge to accept all three regenerations selects toward items that are easy to judge acceptable.

\paragraph{The regeneration defence experiment: registered outcomes.}
\textbf{P-D1 passes}: rewriting the golds with the benchmark's own recipe through our generator leaves the recipe arm $35.73$ points ahead of the verified arm, of its original $45.64$, so same-recipe regeneration is no defence. \textbf{P-D2 fails}: we predicted the lead would shrink in steps from the exact template to its paraphrase to a different style, but the paraphrase version leaves it unchanged ($37.64$ against $35.73$, the point estimate moving the wrong way on a single run) and only the different-style prompt cuts it ($28.33$), which is what genre-level gaming of the easy tier predicts. \textbf{P-D3 passes}: the mixed version lands at $31.48$, strictly between the endpoints, so drawing eval golds from several prompts is a real but partial mitigation. The falsifier, the different-style lead at or above the exact-template lead, did not trigger. The sharpest cell was not registered: on the different-style version D3, the model trained under that same prompt, overtakes the recipe arm by $11.28$ points ($55.41$ against $44.13$), so which model tops the benchmark is decided by which prompt built the evaluation.

\paragraph{The regeneration defence rerun with the benchmark's own vendor.}
The first caveat above can be tested directly, by regenerating with the vendor itself. We sent the same $1{,}700$ queries under the same three prompts to gemini-3-flash-preview, the vendor's API model of that name at the time of writing (temperature $0.7$, thinking level \texttt{low}, one sample per query and prompt), and passed the candidates through the same Qwen3-32B-AWQ judge; it accepted $1{,}406$, $1{,}451$ and $1{,}422$ of about $1{,}698$ candidates under the exact, paraphrase and style prompts (against $1{,}614$, $1{,}652$ and $1{,}627$ of about $1{,}690$ for the Qwen rewrites), and $1{,}127$ queries passed under all three, $1{,}032$ of them also in the Qwen slice. Table~\ref{tab:defense-gemini} scores the same six checkpoints on that slice. The shape of the Gap column replicates: paraphrase at or above exact ($21.83$ against $19.61$, so P-D2 fails again), the different style lowest ($16.59$), the mixed version between the endpoints ($17.74$, as P-D3 predicts), and no version taking the lead to zero (P-D1). The level does not: every regenerated row collapses for every model. The untrained base scores $0.00$ at rank~$1$ on all four regenerated versions (against $10.69$ on the Qwen exact-template version) while still finding $22.80$ / $38.95$ of the exact-template golds at R@5 / R@10, and the verified arm and the recipe-free D4 sit between $0.44$ and $0.80$. The texts explain why. On the $1{,}032$ queries the two slices share, a gold from the vendor rerun shares $0.21$ of its word tokens with its query (Jaccard over lowercased letter and digit runs), against $0.44$ for the golds the benchmark shipped and $0.49$ for the Qwen rewrites (\texttt{results/defense\_gemini\_summary.json}). The shipped gold for a query asking for the number of distinct values of $\lfloor x\rfloor + \lfloor 2x\rfloor + \lfloor 5x/3\rfloor + \lfloor 3x\rfloor + \lfloor 4x\rfloor$ on $[0, 100]$ is that sentence with the function renamed and the wording changed; the vendor rerun's is ``Let $S = \{1, 2, \frac{5}{3}, 3, 4\}$. Determine the cardinality of the set $\{\sum_{c \in S} \lfloor cx \rfloor : x \in [0, 100]\}$.'' The same template, sent to the vendor's current model, returns a deeper rewrite than the benchmark's golds are, and whether the benchmark used another snapshot, no thinking or other sampling settings we cannot know, so this run does not hold the gold genre fixed and cannot isolate the vendor fingerprint as intended. What it does show is that the recipe arm's lead is not lexical overlap: on golds sharing a fifth of their tokens with the query, where the base, the verified arm and D4 find almost none at rank~$1$, the three recipe-family arms find between $10.56$ and $22.54$ of a hundred, and the arm trained with the exact template finds the most on the exact-template, paraphrase and mixed versions ($20.14$ against D2's $17.48$ and D3's $10.56$ on the exact one), an ordering by prompt distance that the Qwen run, where D2 led on every version, did not show; on the different-style version D2 edges D1 ($18.10$ against $17.21$) and D3 closes to $14.02$. These are single runs without intervals, like Table~\ref{tab:defense}.

\section{Forgetting, the mixing frontier, and the regeneration defence}
\label{sec:forgetting}
\label{sec:discussion}

\paragraph{Verified pairs and LLM near-misses erode real retrieval; a few training files escape.}
Every model we fine-tune on verified pairs, or on LLM near-misses, finds cross-language duplicates \emph{worse} than the untrained base, except D4's restatements with near-misses at the reference's count, which stay within noise of it with the recipe's near-misses and at it with an unrelated prompt's; and the more such rows a model sees, the more it loses. Two files escape without dropping their negatives, D4's restatements with an unrelated prompt's near-misses and back-translated positives with verified negatives, both at the base; the same unrelated-prompt near-misses paired with near-copy or back-translated positives collapse it ($13.48$ and $35.20$; Appendix~\ref{app:mechanism}). The base scores $84.73$ strict R@1. Both matched-budget arms fall below it; the uncapped verified arm, trained on more verified rows (Table~\ref{tab:ood}), falls further; SABER's document-heavy arms furthest, $34.86$ for the mostly-document $6{,}145$-row run and $10.69$ for the pure-document run (Appendix~\ref{app:results}). A few training files escape, and they score \emph{above} the base on cross-language duplicates and at it on same-language ones: D4's restatements with no negatives, back-translated positives with no negatives ($94.57$) and the MELD-attacked arms (Sections~\ref{sec:mechanism} and~\ref{sec:ood}); none of these uses minimal-edit negatives or verified pairs, the MELD arms' negatives being same-framing statements rather than near-misses (Appendix~\ref{app:results}). On same-language reprints no trained model beats the base beyond noise (Appendix~\ref{app:setup}).

\paragraph{Two remedies trade rather than dominate.}We tried two ways of keeping benchmark score without losing real retrieval. Weight averaging with the untrained base (WiSE-FT weight-space ensembles~\citep{wortsman2022wiseft}, the soups) buys back retention and breadth at the cost of benchmark score (Table~\ref{tab:mixing}). Mixing verified pairs with problem-to-solution replay under a task instruction (V2, the second verified variant) keeps most of both, $22.19$ easy R@1 at $74.22$ cross-language retention. Data-space mixing here \emph{is} the addition of $12{,}000$ replay rows, so its retention gain is by construction what those rows contribute: the uncapped verified arm, trained on verified pairs alone, keeps $47.58$ cross-language and $47.20$ same-language R@1, and V1, the same pairs plus replay, $59.80$ and $60.80$. The two remedies therefore differ in content as well as mechanism, and a control that composes them says which does what: soups of V1 with the base at the same $\alpha$ keep more retention than the verified arm's soups at a similar easy-tier score ($86.77$, $83.97$ and $78.37$ cross-language at $\alpha = 0.3$, $0.5$, $0.7$ against $80.15$, $73.54$ and $62.34$, with easy R@1 of $11.18$, $13.70$ and $16.63$ against $13.55$, $15.77$ and $16.41$), and hold same-language retention at the base ($63.20$ at every $\alpha$ against $62.40$). Weight mixing adds retention on top of replay rather than standing in for it. Neither remedy is free. The one skill the uncapped verified arm paid for, ranking a true equivalent above a near-miss at rank~$1$, survives neither remedy; in the soups it is present only at $\alpha=1$, the untouched fully trained model. And across the soups, the blends that keep the most retention are nearly the ones that score lowest on the benchmark.

\paragraph{Regenerating the golds is a measured, partial defence.}Beyond the four practices of Section~\ref{sec:conclusion}, a fifth lever we measured is regenerating the golds themselves: we rewrote $1{,}525$ easy-tier golds (Table~\ref{tab:defense}). Rewriting them with the benchmark's own recipe is no defence: the recipe arm keeps $35.73$ of its $45.64$-point lead on those items. Drawing each gold from one of three prompts shrinks the lead to $31.48$, and a single prompt in a different style to $28.33$; we had registered a step-by-step fall from template to paraphrase, and it failed, since the drops are small. One result shows how fragile the ranking is: on the golds regenerated in the different style, D3, the model trained under that same style, overtakes the recipe arm by $11.28$: the top model is decided by the prompt that built the test set. We therefore accept \emph{genre} as the easy tier's operative term. A builder whose golds must be LLM-written is left with a recipe-sensitivity number (practice~iv) and with drawing golds from several prompts, which removes under two fifths of the gap.

\paragraph{The hard tier is buyable, not yet shown to be measured, and no public model has bought it yet.}
When a benchmark's test documents come out of a procedure a model can train on, a leaderboard score mixes two things, how good the model is and how well it matches the procedure; on MathNet-Retrieve it does. The easy tier is not worthless: V2 and the $4$B verified arm lift the base with no LLM-written pairs, so the tier can be learned honestly. The charge concerns only the recipe-specific share of the score, and that share vanishes on cross-language duplicates, where the reference leads. The hard tier's incentives point \emph{backwards}: D4, with no benchmark wording and no negatives, sits at the top on cross-language duplicates and at the base on same-language ones, yet scores zero on that tier. Negatives are what make the tier register, and every kind costs D4 retention on both probes, though not by forgetting alone: the count-matched recipe-negative cell reaches $20.71$ hard R@1 at cross-language retention within noise of the base ($83.89$ against $84.73$), while its same-language retention falls ($47.73$ against $62.40$). We call the tier \emph{buyable} and cannot yet call it measured: recipe-free rewrites with verified negatives reach it; on a near-miss test that no LLM wrote, recipe-matching does buy some real discrimination, but the training sets that score highest on the tier do worse on that test than the reference, and no human has read the recipe arm's winning queries (Section~\ref{sec:limitations}).

No public model yet shows the signature: the style probe puts five public models in the same narrow band as the untrained base, far from the models trained on recipe prompts (Table~\ref{tab:public-style}), so nobody on the current leaderboard appears to have done this. Until leaderboards score models on material no generator wrote, or publish how much a model can gain by matching the recipe, a score close to that gain warrants scrutiny.

\section{The prompts, verbatim}
\label{app:prompts}

This appendix prints every prompt behind the paper's attacks word for word, so that a reader can judge how far apart they are rather than take our labels for it. First the four rungs of the ladder of Section~\ref{sec:mechanism}, which change one thing between rungs, the system prompt given to Qwen3-32B-AWQ, while the judge, the source problems and the training settings stay the same; each rung's user message carries only the source problem under a one-line header, and is printed beneath its system prompt. D1--D3 share one output format (D4 asks for restatements only and writes no negatives by design). D1 is MathNet-Retrieve's own published Appendix-F rewriter prompt, changed only so that its output matches our JSON format. Then the two further controls of Appendix~\ref{app:mechanism}, the survey-author prompt D5 and the unrelated near-misses prompt, the judge that filters every LLM-written pair, and the prompts of the MELD and SABER-Math attacks of Section~\ref{sec:ood}. Braces mark the slots the scripts fill.

\paragraph{Word overlap as a distance, and why we do not use it.}
One could measure the distance between prompts by how many words they share. The Jaccard similarity of word sets to D1 is D2 $= 0.411$, D3 $= 0.192$, D4 $= 0.259$, which does not follow the ladder's order: D4 shares more words with D1 than D3 does, yet D4 is the rung whose hard-tier score collapses, because it asks for a different task (restating a problem for another audience) and attaches no negatives. The factorial of Table~\ref{tab:factorial} shows the collapse comes from the missing negatives ($10.30$ hard R@1 over eight seeds once the verified arm's counterexamples are attached, $26.97$ with the recipe arm's own near-misses). Word overlap is therefore the wrong measure of prompt distance, and the hard-tier ordering of Table~\ref{tab:dose-full} does not follow it.

\paragraph{D1: the benchmark's own Appendix-F template (the recipe arm).}
\begin{Verbatim}[breaklines=true,breakanywhere=true,fontsize=\small]
You are a rigorous mathematical editor and rewriter.

## Step 1: Produce {n_pos} Equivalent Variant(s)
- Each is mathematically identical to the original and solvable by the same method.
- Make them look substantially different in structure/phrasing.
- Use at least 3 distinct transformation types across the whole set (may vary by item).
- Keep statements concise.
- For EACH variant, provide:
  > "problem": the LATEX statement,
  > "justification": a 1-sentence explanation of why it's equivalent,
  > "tags": 1-4 short labels naming the transformation(s) used (e.g., "Variable rename", "Modular rewrite").

// Some Transformation Ideas (use multiple or your own):
1. Variable/Parameter Changes (rename, reorder, replace constants with symbols)
2. Algebraic/Arithmetic Rewrites (exponent swap, divisibility <-> congruence, sum <-> product)
3. Language Restatements (synonyms, contrapositive, reorder assumptions)
4. Geometric Equivalences (relabel points, coordinates <-> vectors, area forms)
5. Combinatorial Rephrasings (choose <-> arrange <-> distribute, complement counting)
6. Number Theory Tricks (gcd, modular swaps, prime restatements)
7. Structural Shifts (lemma/theorem framing, reverse flow, regrouping)
8. Sequences <-> Functions (recurrence <-> functional form)
9. Ratios/Normalizations (normalized terms, integer ratios)
10. Duality/Symmetry (polynomial reciprocal, geometry duality, inequality flips)
11. Alternate Representations (closed form <-> recurrence, coordinates <-> trig)
12. Constraint Shifts (additive <-> multiplicative, quantifier swaps)
13. Meta-Level Changes (prove vs. counterexample, classification form)

## Step 2: Produce {n_neg} Near-Miss Variant(s)
- Should look deceptively similar but require a meaningfully different solving method.
- Make small, crucial mathematical changes (e.g., sign flip, add a square, exponent shift, modulus tweak, relation type change, altered bounds).
- Ensure each truly alters the strategy (not a trivial edit).
- For EACH near-miss, provide:
  > "problem": the LATEX statement,
  > "justification": a 1-sentence explanation of why it's a near miss,
  > "tags": 1-4 short labels naming the modification(s) (e.g., "Sign flip", "Modulus change", "Inequality direction").

## Step 3: Return Strict JSON
Return only a single valid JSON object using this exact schema:
{{
  "original_problem": "LaTeX string of the cleaned original problem",
  "equivalent_variants": [
    {{"problem": "...", "justification": "...", "tags": ["...", "..."]}}
  ],
  "near_miss_variants": [
    {{"problem": "...", "justification": "...", "tags": ["..."]}}
  ]
}}
// Rules:
- All math must be in LATEX.
- Exactly {n_pos} equivalent and {n_neg} near-miss variants.
- Each variant MUST include non-empty "tags".
- Output JSON only; no code fences, no extra text.
- If your draft is not valid JSON, fix it until it is valid and return only the JSON.
\end{Verbatim}

User message of D1:
\begin{Verbatim}[breaklines=true,breakanywhere=true,fontsize=\small]
Original problem:

{problem}
\end{Verbatim}

\paragraph{D2: a close paraphrase of that template.}
\begin{Verbatim}[breaklines=true,breakanywhere=true,fontsize=\small]
You act as a careful, exacting editor of mathematics problems.

Your assignment has two parts.

### Part A -- equivalent restatement(s)
Write {n_pos} variant(s) of the problem you are given that carry exactly the same mathematical content as the original: identical in substance and solvable by the very same method, yet visibly different in wording and surface structure. Across the whole set of variants, draw on at least 3 different kinds of transformation; options include renaming variables, reordering them, or swapping constants for symbols; rewriting the algebra or arithmetic (exchanging exponents, trading divisibility statements for congruences, turning sums into products); restating the language with synonyms, a contrapositive, or reordered assumptions; relabeling geometric points or moving between coordinates, vectors, and area formulas; recasting counting arguments (choosing vs. arranging vs. distributing, counting the complement); number-theoretic rewrites via gcd, moduli, or primes; reframing as a lemma or reversing the logical flow; trading recurrences for functional forms; normalizing ratios; exploiting duality or symmetry; switching to an alternate representation; shifting constraints between additive and multiplicative form or swapping quantifiers; or changing the meta-level (prove vs. find a counterexample, classification form). You may also invent transformations of your own. Keep every statement concise. Record for each variant:
  > "problem": the statement in LATEX,
  > "justification": a single sentence explaining why nothing mathematical changed,
  > "tags": 1-4 brief labels naming the transformation(s) applied (e.g., "Variable rename", "Modular rewrite").

### Part B -- near-miss lookalike(s)
Write {n_neg} problem(s) that closely resemble the original at a glance yet genuinely demand a different solving method. Achieve this with small but decisive mathematical alterations -- a flipped sign, an added square, a shifted exponent, an adjusted modulus, a changed relation type, or moved bounds -- and make certain each alteration truly forces a new strategy rather than being a cosmetic edit. Record for each near-miss:
  > "problem": the statement in LATEX,
  > "justification": a single sentence explaining why it merely looks similar,
  > "tags": 1-4 brief labels naming the modification(s) made (e.g., "Sign flip", "Modulus change", "Inequality direction").

### Part C -- answer in strict JSON
Return only a single valid JSON object using this exact schema:
{{
  "original_problem": "LaTeX string of the cleaned original problem",
  "equivalent_variants": [
    {{"problem": "...", "justification": "...", "tags": ["...", "..."]}}
  ],
  "near_miss_variants": [
    {{"problem": "...", "justification": "...", "tags": ["..."]}}
  ]
}}
Constraints: express all mathematics in LATEX; produce exactly {n_pos} equivalent and {n_neg} near-miss variant(s); never leave "tags" empty; emit nothing besides the JSON object -- no code fences, no commentary; and if your draft fails to parse as JSON, repair it until it is valid and return only the JSON.
\end{Verbatim}

User message of D2:
\begin{Verbatim}[breaklines=true,breakanywhere=true,fontsize=\small]
The problem to work from:

{problem}
\end{Verbatim}

\paragraph{D3: a rewriter prompt in a deliberately different style.}
\begin{Verbatim}[breaklines=true,breakanywhere=true,fontsize=\small]
Hi! I coach a competition-math training group, and I'm assembling this week's worksheet. I need your help with one exercise at a time.

Here's the game. My students have already seen the problem I'm about to show you. To test whether they recognize a problem by its mathematics rather than by its looks, the worksheet mixes disguises of it in among traps:

1. {n_pos} disguise(s): the same problem dressed up to look new. A disguise must be the original in every mathematical respect -- same content, same answer, same solution path -- but a returning student shouldn't recognize it at first sight. Dress it up however you like: new letters, a different setting or story, reshuffled givens, reworded conditions. Keep it lean; competition problems don't ramble.

2. {n_neg} trap(s): problems that LOOK like today's problem but are NOT it. A good trap differs by one quiet mathematical detail -- perhaps a sign, a power, a modulus, a bound, or the type of relation -- chosen so that the trap sends a solver down a genuinely different path. A trap that ends up solved the same way as the original is a failed trap; don't hand me those.

For every disguise and every trap, jot down:
  > "problem": the full statement (all mathematics in LATEX),
  > "justification": one line -- for a disguise, why it's still the same problem; for a trap, why it only looks like it,
  > "tags": 1-4 short notes on what you changed.

So my worksheet tool can ingest your work, reply with strict JSON only:
{{
  "original_problem": "LaTeX string of the cleaned original problem",
  "equivalent_variants": [
    {{"problem": "...", "justification": "...", "tags": ["...", "..."]}}
  ],
  "near_miss_variants": [
    {{"problem": "...", "justification": "...", "tags": ["..."]}}
  ]
}}
House rules: disguises go under "equivalent_variants" and traps under "near_miss_variants", exactly {n_pos} and {n_neg} of them; every "tags" list non-empty; all math in LATEX; nothing outside the JSON object -- no code fences, no chat; and if what you wrote isn't valid JSON, fix it before you answer. Thanks!
\end{Verbatim}

User message of D3:
\begin{Verbatim}[breaklines=true,breakanywhere=true,fontsize=\small]
Today's problem for the worksheet:

{problem}
\end{Verbatim}

\paragraph{D4: a restatement prompt unrelated to the recipe.}
\begin{Verbatim}[breaklines=true,breakanywhere=true,fontsize=\small]
You are an editor adapting mathematics problems for textbooks aimed at different audiences.

Rewrite the given problem for a student at a DIFFERENT level of mathematical maturity than its original audience. Pick whichever direction suits the problem: a younger student meeting the topic for the first time (friendlier vocabulary, gentler sentence rhythm, concrete framing), or an advanced undergraduate who prefers terse, formal statements. The restated problem must remain mathematically identical to the original -- the same question, the same given conditions, the same answer, solvable by the same reasoning. You may rename characters or objects in story contexts and adapt notation conventions to the audience, but you must not add, remove, weaken, or strengthen any mathematical condition.

Produce {n_pos} restatement(s). For each, record:
  > "problem": the restated problem (all mathematics in LATEX),
  > "audience": who you wrote it for (e.g., "middle-school student", "advanced undergraduate"),
  > "justification": one sentence confirming the mathematics is unchanged.

Return only a single valid JSON object using this exact schema:
{{
  "original_problem": "LaTeX string of the cleaned original problem",
  "restatements": [
    {{"problem": "...", "audience": "...", "justification": "..."}}
  ]
}}
// Rules:
- All math must be in LATEX.
- Exactly {n_pos} restatement(s).
- Output JSON only; no code fences, no extra text.
- If your draft is not valid JSON, fix it until it is valid and return only the JSON.
\end{Verbatim}

User message of D4:
\begin{Verbatim}[breaklines=true,breakanywhere=true,fontsize=\small]
Problem to adapt:

{problem}
\end{Verbatim}

\paragraph{D5: the survey-author prompt, the second recipe-free reference.}
A restatement prompt whose task framing is not an audience change: the model writes the problem as a survey author would present another author's problem. Like D4 it asks for restatements only.
\begin{Verbatim}[breaklines=true,breakanywhere=true,fontsize=\small]
You are a mathematician writing the problems section of a survey article that collects competition problems on one theme.

Restate the given problem in your own words as a self-contained statement, the way you would present another author's problem inside your survey: your own sentence structure, your own notation and your own order of presentation, with the mathematics unchanged. Do not simplify or generalise the problem, do not add hints, and do not change its level.

Produce {n_pos} restatement(s). For each, record:
  > "problem": the restated problem (all mathematics in LATEX),
  > "context": one short phrase naming the theme under which your survey files this problem (e.g., "bounds on sums of reciprocals"),
  > "justification": one sentence confirming the mathematics is unchanged.

Return only a single valid JSON object using this exact schema:
{{
  "original_problem": "LaTeX string of the cleaned original problem",
  "restatements": [
    {{"problem": "...", "context": "...", "justification": "..."}}
  ]
}}
// Rules:
- All math must be in LATEX.
- Exactly {n_pos} restatement(s).
- Output JSON only; no code fences, no extra text.
- If your draft is not valid JSON, fix it until it is valid and return only the JSON.
\end{Verbatim}

User message of D5:
\begin{Verbatim}[breaklines=true,breakanywhere=true,fontsize=\small]
Problem to restate for the survey:

{problem}
\end{Verbatim}

\paragraph{The unrelated near-misses prompt.}
D4's restatement task with a second step that writes minimal-edit companions, so that near-misses can be produced by a prompt that shares nothing with the recipe.
\begin{Verbatim}[breaklines=true,breakanywhere=true,fontsize=\small]
You are an editor adapting mathematics problems for textbooks aimed at different audiences.

First, rewrite the given problem for a student at a DIFFERENT level of mathematical maturity than its original audience. Pick whichever direction suits the problem: a younger student meeting the topic for the first time (friendlier vocabulary, gentler sentence rhythm, concrete framing), or an advanced undergraduate who prefers terse, formal statements. The restated problem must remain mathematically identical to the original -- the same question, the same given conditions, the same answer, solvable by the same reasoning. You may rename characters or objects in story contexts and adapt notation conventions to the audience, but you must not add, remove, weaken, or strengthen any mathematical condition.

Second, this textbook pairs every exercise with a short "spot the difference" drill. Write {n_neg} companion exercise(s) for that drill. Each companion should read like the original problem at a glance, in the same register and about the same length, yet ask something mathematically different, so that a student who solves it by copying the original's reasoning arrives at a wrong answer. Change one thing that matters (a given condition, a quantity, a relation between the objects, or what is asked for) and leave everything else as it was.

Produce {n_pos} restatement(s) and {n_neg} companion(s). For each restatement, record:
  > "problem": the restated problem (all mathematics in LATEX),
  > "audience": who you wrote it for (e.g., "middle-school student", "advanced undergraduate"),
  > "justification": one sentence confirming the mathematics is unchanged.
For each companion, record:
  > "problem": the companion exercise (all mathematics in LATEX),
  > "what_changed": one sentence naming the change and why the answer differs.

Return only a single valid JSON object using this exact schema:
{{
  "original_problem": "LaTeX string of the cleaned original problem",
  "restatements": [
    {{"problem": "...", "audience": "...", "justification": "..."}}
  ],
  "companions": [
    {{"problem": "...", "what_changed": "..."}}
  ]
}}
// Rules:
- All math must be in LATEX.
- Exactly {n_pos} restatement(s) and {n_neg} companion(s).
- Output JSON only; no code fences, no extra text.
- If your draft is not valid JSON, fix it until it is valid and return only the JSON.
\end{Verbatim}

User message:
\begin{Verbatim}[breaklines=true,breakanywhere=true,fontsize=\small]
Problem to adapt:

{problem}
\end{Verbatim}

\paragraph{The judge that keeps or drops every LLM-written pair.}
Ours, since the benchmark does not print its judge prompt; run through Qwen3-32B-AWQ, and for the two-judge control of Appendix~\ref{app:mechanism} through Mistral-Small-24B-Instruct-2501 as well.
\begin{Verbatim}[breaklines=true,breakanywhere=true,fontsize=\small]
You are an expert mathematical olympiad editor acting as a strict judge of mathematical equivalence.

You will be given Problem A and Problem B. Decide whether they are MATHEMATICALLY EQUIVALENT in the strict sense (Invariance): B is a reformulation of A such that the two problems have identical mathematical content — every complete solution of one converts to a complete solution of the other by mechanical translation (renaming variables, rewriting notation, restating language), and both are solved by the same method. Superficial similarity is NOT equivalence: a sign flip, changed exponent, altered modulus, different bound, or added/removed constraint that changes the answer or the solving strategy makes them NOT equivalent.

Think through the mathematics carefully, then return ONLY a single valid JSON object:
{"verdict": "equivalent" | "not_equivalent", "confidence": 0.0-1.0, "reason": "one-sentence justification"}
// Rules:
- Output JSON only; no code fences, no extra text.
- "confidence" is your subjective probability that your verdict is correct.
\end{Verbatim}

User message of the judge:
\begin{Verbatim}[breaklines=true,breakanywhere=true,fontsize=\small]
Problem A:

{anchor}

Problem B:

{candidate}
\end{Verbatim}

\paragraph{The MELD attack: the reconstructed generation prompt.}
MELD publishes the prompts of its training pipeline but not the one behind its evaluation set; this is our reconstruction of the procedure its paper describes, into the schema of its released records. The connection descriptions per subfield pair are ours.
\begin{Verbatim}[breaklines=true,breakanywhere=true,fontsize=\small]
You are a research mathematician building an evaluation set of mathematically equivalent but lexically different statement pairs.

You are given two complementary mathematical subfields and a description of the connection between them. Produce exactly {n_pairs} statement pairs. In every pair:
- entry_1 is written entirely in the language of {field_1};
- entry_2 is written entirely in the language of {field_2};
- the two statements are mathematically equivalent: they assert the same thing about the same underlying mathematics, and a reader who knows both dialects would call them the same statement;
- they cannot be matched by surface-level lexical similarity: share as few words, symbols and notational conventions as possible while staying faithful;
- the content belongs to a basic graduate curriculum in mathematics.

Each pair carries a short "topic" label naming the shared idea. All {n_pairs} topics must be distinct from each other.

Write each statement in natural language and LaTeX, in the voice of a textbook definition or theorem statement, one or two sentences long.

Output ONLY a single valid JSON object, no code fences and no commentary:
{{"pairs": [{{"topic": "<short label>", "entry_1": {{"framing": "{field_1}", "statement": "<statement>"}}, "entry_2": {{"framing": "{field_2}", "statement": "<statement>"}}}}, ...]}}
\end{Verbatim}

User message (first call per subfield pair):
\begin{Verbatim}[breaklines=true,breakanywhere=true,fontsize=\small]
Subfield 1: {field_1}
Subfield 2: {field_2}

Connection between the two fields: {connection}

Produce {n_pairs} pairs.
\end{Verbatim}

User message (later calls, which pass back the topics already produced):
\begin{Verbatim}[breaklines=true,breakanywhere=true,fontsize=\small]
Subfield 1: {field_1}
Subfield 2: {field_2}

Connection between the two fields: {connection}

The following topics have already been written for this subfield pair:
{covered}

Produce {n_pairs} further pairs. Prefer topics that are not in that list; where the shared idea is unavoidable, write a genuinely different statement of it (different objects, different hypotheses, different phrasing).
\end{Verbatim}

\paragraph{The MELD attack: the rewritten prompt of the $84/16$ control.}
\begin{Verbatim}[breaklines=true,breakanywhere=true,fontsize=\small]
You are preparing translation exercises for a graduate reading course that runs two seminars in parallel.

Students in seminar A work in {field_1}; students in seminar B work in {field_2}. Write exactly {n_pairs} exercises. Each exercise states one fact twice, once as seminar A would record it and once as seminar B would record it, so that a student attending only one seminar still meets the same mathematics.

For each exercise:
- entry_1 is the seminar-A version, phrased as that seminar phrases things;
- entry_2 is the seminar-B version, phrased as that seminar phrases things;
- both versions must be true and must say the same thing;
- the content should sit within a basic graduate curriculum.

Give each exercise a short "topic" label naming the fact being recorded, and make the {n_pairs} labels distinct.

Output ONLY a single valid JSON object, no code fences and no commentary:
{{"pairs": [{{"topic": "<short label>", "entry_1": {{"framing": "{field_1}", "statement": "<statement>"}}, "entry_2": {{"framing": "{field_2}", "statement": "<statement>"}}}}, ...]}}
\end{Verbatim}

\paragraph{The MELD attack: the validity and equivalence screen.}
MELD's three-question check, run through Qwen3-32B-AWQ; we gate on validity and equivalence and only record the dissimilarity answer.
\begin{Verbatim}[breaklines=true,breakanywhere=true,fontsize=\small]
You are an expert mathematician screening candidate items for an evaluation set of mathematically equivalent but lexically different statement pairs.

You will be given Statement 1 (written in one subfield's language) and Statement 2 (written in another's). Answer three questions:
(i)   is each statement, on its own, a valid and correct mathematical statement?
(ii)  are the two statements mathematically equivalent -- do they assert the same thing about the same underlying mathematics?
(iii) could the pair be rewritten to sound LESS similar to each other without breaking the equivalence?

Judge (ii) strictly. Sharing a topic is not equivalence: a changed hypothesis, a dropped finiteness assumption, a different quantifier, or a statement that merely implies the other in one direction makes them NOT equivalent.

Return ONLY a single valid JSON object, no code fences and no commentary:
{"entry_1_valid": true|false, "entry_2_valid": true|false, "equivalent": true|false, "could_be_less_similar": true|false, "confidence": 0.0-1.0, "reason": "one sentence"}
\end{Verbatim}

User message of the screen:
\begin{Verbatim}[breaklines=true,breakanywhere=true,fontsize=\small]
Statement 1 ({framing_1}):

{statement_1}

Statement 2 ({framing_2}):

{statement_2}
\end{Verbatim}

\paragraph{The SABER-Math attack: SABER's own solution-summary prompt.}
Taken byte for byte from SABER's repository (\texttt{annotate/ideas/prompts.py}) and run through Qwen3-32B-AWQ in place of GPT-OSS-120B; the summaries it writes are the only text the attack trains on.
\begin{Verbatim}[breaklines=true,breakanywhere=true,fontsize=\small]
# INSTRUCTION
You are an expert mathematical anotator tasked with identifying the *core idea* - the central mathematical insight, from a math problem.

# GOALS
1. Identify what makes the solution work conceptually, not how to carry it out. Capture the untrial step or idea that is the greatest hint for the solution.
2. Never include any multi-step reasoning, equations or numeric computations. Don't include any anotations that are in the solution, but not in the original problem statement.
3. Never try to solve the problem on your own and don't include your reasoning or thoughs.
4. Output a sinlge valid JSON object matching the schema below.
5. Structure the ideas imperatively so they look like you are giving a hint to someone.
5. If the problem seems too easy, straightforward or you can't identify a core idea, store its value as 'null' and set the 'noCoreIdea' to 'true'

# SCHEMA
```json
{
    "noCoreIdea": <true|false>,
    "coreIdea": "<string - one short sentence (up to 30 words) naming the main insight to the problem>",
    "supporingIdeas: ["<strings - 0-3 short techniques phrases>"],
    "keywords": ["<strings - 1-2 word phrases summarizing the ideas, theorems, etc. in the solution"],
    "confidence": <0.0-1.0>
}

Here are the problem statement and solution:

Statement: {{statement}}

Solution: {{solution}}
\end{Verbatim}

\end{document}